\documentclass[a4paper,11pt]{article}
\pdfoutput=1 

\usepackage{jcappub} 

\usepackage[T1]{fontenc} 

\usepackage{graphicx}
\usepackage{dcolumn}
\usepackage{bm}
\usepackage{color}
\usepackage{subfigure}

\newcommand{\be}{\begin{equation}}
\newcommand{\ee}{\end{equation}}
\newcommand{\bea}{\begin{eqnarray}}
\newcommand{\eea}{\end{eqnarray}}

\newcommand{\bseq}{\begin{subequations}}
\newcommand{\eseq}{\end{subequations}}

\allowdisplaybreaks[1]
\newcommand{\ms}{$M_{\odot}$\,}

\title{Neutron stars in the theory of gravity with nonminimal derivative coupling and realistic equations of state}

\author[a]{P.E. Kashargin}
\author[a]{A.A. Lebedev}
\author[a,1]{S.V. Sushkov\note{Corresponding author.}}

\affiliation[a]{Kazan Federal University, Kremliovskaya str. 16a, Kazan 420008, Russia}

\emailAdd{pkashargin@mail.com}
\emailAdd{lebedev.aleks2012konnor@yandex.ru}
\emailAdd{sergey\_sushkov@mail.ru}

\abstract{We numerically construct compact stars in the scalar-tensor theory of gravity with non-minimal derivative coupling of a scalar field to the curvature and nonzero cosmological constant. 
There are two free parameters in this model of gravity: the non-minimal derivative coupling parameter $\ell$ and the cosmological constant parameter $\xi$. We study the relationship between the model parameters and characteristic of the neutron star, what allowed us to limit the permissible range of $\xi$ and $\ell$. 
In particular, in the case $\xi=-1$ the external geometry of the neutron star coincides with the Schwarzschild anti-de Sitter geometry, while the internal geometry of the star differs from the case of the standard gravity theory. Plenty realistic equations of state of neutron star matter were considered. In general the neutron star model in the theory of gravity with a non-minimal derivative coupling does not contradict astronomical data and is viable.}

\begin{document}
\maketitle
\flushbottom

\section{Introduction}
\label{sec:intro}

\subsection{Mass and radius of a neutron star}
The concept of neutron stars was proposed by Baade and Zwigi in 1932 \cite{BaadeZwicky1934} and was discovered many years later in 1967 as rapidly rotating stars with strong magnetic fields, now known as pulsars \cite{Hewish1968}. Since the discovery of neutron stars, methods for observing them have been improved, and knowledge about their nature have been changed. A detailed discussion of the properties of neutron stars can be found in the following monographs and reviews \cite{ShapiroTeukolsky, Haensel2007, Potekhin2010, LattimerPrakash, Chandrasekhar, bookAstrNS2018, Lattimer2015}. 

Neutron stars belong to the class of compact stars, as the white dwarfs and  quark stars \cite{Weber2008}, which are so far undetected. It is generally accepted that a canonical neutron star has a mass $M=1.4\, M_{\odot}$ 
and radius  $R=10$ km. Stars of such extreme compactness have central densities that are 5-10 times the nuclear saturation density $\rho_{sat} = 2.8 \times 10^{14}\, {\rm g\,cm^{-3}}$. The nature of such matter is still not clear, and there are various models to describe it.  The structure of relativistic stars and their mass-radius relation are determined by integrating the relativistic structure equations, also known as the Tolman-Oppenheimer-Volkoff equations \cite{Tolman1939, Oppenheimer1939}. The Tolman-Oppenheimer-Volkoff equations connect such macroscopic parameters of a star as mass and radius with microphysics, which is described by the equation of state of superdense matter. 

The maximum neutron star mass is very important to distinguish between black holes and massive neutron stars and is a crucial parameter in studying the equation of state of cold dense matter. Twenty years ago, estimates of the mass of radio pulsars showed a narrow distribution interval of about $1.35\pm0.04\, M_{\odot}$ \cite{Thorsett1999}. Thanks to more recent studies \cite{Ozel2016, Miller2017}, 
we know now that neutron star masses span a wide range, between 1.2 and at least 2 \ms. 
It was found evidence for neutron stars with more than two and almost three Solar masses. 
The review the most massive neutron stars presently known in light of recent findings can be found in \cite{Linares2019}. 

Typically obtained radii are in the range 10–14 km \cite{Ozel2016}. For example in the work \cite{Collin2020} one finds that the radius of a $1.4$ \ms neutron star is $R=11.0^{+0.9}_{-0.6}$ km (90\% credible interval). 


\subsection{Neutron stars in modified theory of gravity}
Alternative theories of gravity attracted significant attention in the past decades. Modifications are often introduced to build 
cosmological models and explain phenomena in the universe on large scales including  the dark mater or the accelerated expansion of the universe. Reviews on modified theories of gravity including the motivations to consider extensions of general relativity  can be found in \cite{Cliftona2012, Myrzakulov2013, Langlois2018, Berti2015, Review_Salvatore:2011, Review_Nojiri:2017}.

In context of modification of general relativity, one needs to consider not only the cosmological problems, but stellar structures too, especially compact relativistic objects (neutron stars and black holes). 
Due to the fact that neutron stars have a very strong gravitational field and are composed of super-dense matter, they provide information about both the theory of gravity and nuclear interactions that we cannot obtain from laboratory tests. Observations of neutron stars offer opportunities for measuring the effects of general relativity, as well as testing its modifications. 
Neutron star solutions have been constructed in various alternative theories of gravity \cite{Odintsov, Doneva:2019, Horbatsch:2011, Doneva:2020, Raissa,  OdintsovOikonomou, Rosca, Oliveira:2015, HarkoLoboSushkov}.
For a recent review of compact star models in modified theories of gravity, see \cite{Olmo2019, Lehebel2018, Doneva2018, JuttaKunz, PaoloPani2011} and references therein.

\subsection{Neutron stars in the theory of gravity with non-minimal derivative coupling}
Horndeski theory \cite{Horndeski1974, Charmousis2012, Deffayet2011} is the most general scalar-tensor theory of gravity having second-order field equations in four dimensions. 
There are four arbitrary functions in the theory. By taking these functions appropriately, one can reproduce any second-order scalar-tensor theory as a specific case, including  quintessence and k-essence, $f(R)$ gravity, Brans-Dicke theory, Einstein-dilaton-Gauss-Bonnet, etc.  Horndeski theory has been studied both in cosmological and astrophysical frameworks (for a comprehensive review see \cite{Kobayashi2019} and the references therein). Neutron stars in Horndeski gravity have been also widely discussed in the literature.   

The important subclass of Horndeski gravity is represented by models with non-minimal derivative coupling of a scalar field with the Einstein tensor with the action
\begin{equation}\label{action1}
	S=\int d^4x\sqrt{-g}\,
	\left[\frac{1}{2\kappa} (R-2\Lambda_0)-\frac12\left( \alpha g_{\mu\nu}+\beta G_{\mu\nu} \right)\nabla^{\mu}\phi\nabla^{\nu}\phi\right]+S^{(m)},
\end{equation}
where $R$ and $G_{\mu\nu}$ are the Ricci scalar and the Einstein tensor, respectively,  $\kappa=8\pi G/c^4$ is the Einstein gravitational constant, and $S^{(m)}$ is the action for ordinary matter fields, supposed to be minimally coupled to gravity in the usual way. Coefficients $\alpha$ and $\beta$ are real parameters, where $\alpha$ corresponds to the usual kinetic term of the scalar field $\phi$, while $\beta$  determines its modified part. $\Lambda_0$ is a `bare' (i.e. unobserved) cosmological constant. As we will see later, an observed cosmological constant $\Lambda_{AdS}$ appears as a certain combination of $\Lambda_0$ and the parameter of nonminimal derivative coupling $\beta$.  The theory (\ref{action1}) has very interesting cosmological properties \cite{Sushkov:2009, Sushkov:2010, Sushkov:2012, Sushkov:2016, Sushkov:2020},  provides black hole \cite{Hui2013, Rinaldi:2012, Minamitsuji:2013, Anabalon:2014, Babichev:2014, Kobayashi:2014, Babichev:2015} and wormhole \cite{Sushkov:2012b, Sushkov:2014} solutions. 

Solutions describing black holes were considered in Horndeski theory. Hui and Nicolis \cite{Hui2013} presented a no-hair theorem\footnote{In this case the no-hair means a trivial  scalar field solution.} which is valid for shift-symmetric Horndeski gravity, i.e., the subclass of the Horndeski action which remains invariant under a transformation $\phi\to\phi+const$ of the scalar field. The
theorem is applicable to vacuum, static, spherically symmetric and asymptotically flat black holes. In this case it is proved that the scalar field of the black hole solution must be constant, and by exploiting the shift symmetry we can set its value to zero. Consequently black holes cannot sustain non-trivial scalar field profiles in this case. 
However hairy solutions can be obtained by relaxing some of the assumptions that enter the proof of the no-hair theorem. 
In particular, for the nonminimal derivative coupling theory it was found hairy solutions with $\phi(t,r)=qt+F(r)$ \cite{Babichev:2014}. 
Among the solutions constructed in this way, the non-minimally coupled theory with $\beta=\Lambda_0=0$  admits a “stealth”
solution, where a Schwarzschild black hole metric supports a nontrivial, regular scalar field configuration which does not backreact on
the spacetime. 
 A nontrivial scalar field are also possible in the case of not asymptotically flat black holes solutions \cite{Rinaldi:2012, Anabalon:2014, Minamitsuji:2013}. In paper \cite{Rinaldi:2012}, a solution was found that describes black holes in special case $\beta=\Lambda_0=0$. This solution was later generalized to the case arbitrary $\beta,\Lambda_0$. Black holes solutions with AdS asymptotics was also obtained \cite{Babichev:2014}, where the effective cosmological constant is given by geometric coupling constants in the action.

Neutron stars have been also explored within the model with nonminimal derivative coupling \cite{Rinaldi:2015, Rinaldi:2016, Silva:2016, Maselli:2016, Eickhoff:2018}. 
An overview of solutions describing neutron stars can be found in the literature \cite{Silva:2016, Lehebel1}. In work \cite{Lehebel1} a no-hair theorem was presented for the spherically symmetric and static star configurations in the shift-symmetric Horndeski theories with minimal matter coupling. It was noted \cite{Lehebel1} that the regular spherically symmetric and static solution with the asymptotically flat spacetime in the shift symmetric Horndeski theory with a minimally coupled matter sector with the action that is analytic around a trivial scalar field configuration has a constant scalar field, and in particular, star solutions are identical to their general relativity counterpart. But there is possibilities to escape the assumption of the no-hair theorem and consider neutron stars in the  an asymptotically de Sitter universe, with timedependent scalar field or etc \cite{Rinaldi:2015, Rinaldi:2016, Maselli:2016}.

Spherically symmetric neutron star solutions in the theory (\ref{action1}) with $\alpha=\Lambda_0=0$ with the scalar field is linear in times $\phi=qt+F(r)$ and usual equation for the matter fluid with polytrope equation of state was constructed in \cite{Rinaldi:2015}; 
in this case the external geometry is identical to the Schwarzschild metric, but the interior structure is considerably different from standard general relativity.  
Subsequently this solution was generalized. Slowly rotating stars were considered in the theory with $\alpha=\Lambda_0=0$ using several realistic equations of state  \cite{Rinaldi:2016, Maselli:2016}. 

Previously we construct neutron star configurations with AdS asymptotic within the framework of the full theory (\ref{action1}) without imposing any restrictions on the parameters $\Lambda_0$ and $\alpha$ \cite{Kashargin2022}. However the previous work considered the simplest polytrope equation of state. Now our goal is to explore neutron star configurations in the full theory (\ref{action1}) and realistic equation of state of neutron star matter.
The paper is organized as follows. In Section \ref{sec2} we derive general equations describing an external and internal configuration of a neutron star in the theory of gravity with nonminimal derivative coupling. The external vacuum solution is analyzed in Section \ref{sec3}. A detail analysis of interior of the star and constructing of a complete solution joining internal and external configurations is provided in the section \ref{sec4}.

\section{Basic equations} \label{sec2}

\subsection{Action and field equations}
In this section we will present the basic equations. We will use the notation of our previous work \cite{Kashargin2022}. A detailed calculations can be found in the mentioned article, here we will briefly present the equations in a form convenient for numerical integration. A gravitational theory (\ref{action1}) can be represented as
\begin{equation}\label{action} 
S=\int d^4x\sqrt{-g}\,
\left[\frac{1}{2\kappa} (R-2\Lambda_0)-\frac12\left( \varepsilon_1 g_{\mu\nu}+\varepsilon_2 \ell^2 G_{\mu\nu} \right)\nabla^{\mu}\phi\nabla^{\nu}\phi\right]+S^{(m)},
\end{equation}
where $\ell$ is a characteristic length which characterizes the nonminimal derivative coupling between the scalar field and curvature and $\varepsilon_{1,2}=\pm 1$. 
As the matter we will consider a perfect fluid with the energy-momentum tensor 
\be 
T_{\mu\nu}^{(m)}=(\epsilon + p)u_{\mu}u_{\nu}+p g_{\mu\nu},
\ee
where $u_{\mu}$ is a unit timelike 4-vector, $u_\mu u^\mu=-1$, $\epsilon$ is an energy density, and $p$ is an isotropic pressure. 
We will focus on static spherically symmetric configurations:
\be
ds^2=-A(r)d(ct)^2+\frac{dr^2}{B(r)}+r^2\left(d\theta^2+\sin^2\theta d\varphi^2\right),
\ee
where $A(r)$ and $B(r)$ are functions of the radial coordinate $r$. Assume also that the scalar field $\phi$, the energy density $\epsilon$, and the pressure $p$ depend only on $r$, i.e. $\phi=\phi(r)$, $\epsilon=\epsilon(r)$, and $p=p(r)$. 
Nonzero independent components of the gravitational field equations and scalar field equation for the action (\ref{action}) can be represent in the following form:
\begin{eqnarray}
\nonumber\label{_eqB_dl}
&& \frac{dB}{dx} = -\frac{1}{\Delta}
\bigg[\Big((1+\varepsilon \xi)x^4 +(\varepsilon-5\xi) x^2 +2\Big)B +\Big((1-\varepsilon x^2) {\cal E}
+2 (3-\varepsilon x^2) {\cal P} \Big) x^2B
\\ 
&&~~~~~~~~~~~~ 
-(1-\varepsilon x^2)^2 \,
\big(2 -(\varepsilon+\xi) x^2 -x^2 {\cal E}  \big)\bigg],
\\ \label{_eqP_dl}
&& \frac{d{\cal P}}{dx} =-\frac{({\cal E} + {\cal P}) (1-B-\varepsilon x^2) }{2xB},
\\ \label{_eqA_dl}
&& \frac{dA}{dx}=\frac {A(1-B-\varepsilon x^2)}{xB},
\\ \label{_eqPsi_dl}
&& \Psi^2=- \frac{ x^2 (\varepsilon-\xi +{\cal P})} {\varepsilon_2 B\, (1-\varepsilon x^2)},
\end{eqnarray}
where $\varepsilon=\varepsilon_1/\varepsilon_2$,
$$
\Delta =x(1-\varepsilon x^2 ) \big(2 -(\varepsilon +\xi) x^2 +x^2 {\cal P}\big),
$$
and was introduced the dimensionless values as follows
\be\label{dlvalues}
\xi=\Lambda_0\ell^2, \quad
x=\frac{r}{\ell},\quad 
{\cal E}=\kappa\ell^2 \epsilon, \quad
{\cal P}=\kappa\ell^2 p, \quad
\Psi^2=\kappa\ell^2 \left(\frac{d \phi}{dr}\right)^2.
\ee
Note that in order to provide a regularity of solutions of (\ref{_eqB_dl})--(\ref{_eqPsi_dl}) on the entire interval of varying the radial coordinate, $x\in[0,\infty)$, we need to make a choice $\varepsilon=\varepsilon_1/\varepsilon_2=-1$, because in this case $(1-\varepsilon x^2)=(1+x^2)>0$, and denominators in  (\ref{_eqB_dl}) and (\ref{_eqPsi_dl}) do not go to zero. The choice $\varepsilon=-1$ means that $\varepsilon_1$ and $\varepsilon_2$ have different signs and, ultimately, it means that the usual kinetic term $\alpha g_{\mu\nu} \nabla^\mu\phi\nabla^\nu\phi$ and the modified term $\beta G_{\mu\nu} \nabla^\mu\phi\nabla^\nu\phi$ enter into the Lagrangian (\ref{action1}) with different signs. Note also that the sign $\varepsilon_2=\pm1$ in (\ref{_eqPsi_dl}) is still undefined. To be determined, $\varepsilon_2$ should provide the positivity of $\Psi^2$, that is $\varepsilon_2(1+\xi-{\cal P}) \ge 0$. 


\subsection{Equation of state}
To make the system of four field equations (\ref{_eqB_dl})--(\ref{_eqPsi_dl}) complete, one needs to add an equation of state relating the pressure and the energy density. 
In this work we will use polytropic  and 31 realistic equation of state. 
\subsubsection{Polytropic equation of state}
In the previous work \cite{Kashargin2022} we considered the polytropic equation of state 
to construct neutron stars in the model (\ref{action})
\be\label{eos}
p=K \rho_0^{\Gamma}, 
\quad \epsilon = \rho_0 c^2 + \frac{p}{\Gamma-1},
\ee 
where $\rho_0$ is a baryonic mass density, 
$\Gamma = 1 + 1/n$ is the adiabatic index, $n$ is the polytropic index, and $K$ is the polytropic constant. 
In particular we used $\Gamma=2$ and $K=1.79\times 10^5$ cgs.
This equation of state is widely used to model neutron star configurations in general relativity \cite{Tooper:1965, ShapiroTeukolsky}  
and in various modified theories of gravity \cite{Olmo:2020, Babichev:2009, Babichev:2010, Cooney:2010, Pace:2017, Ilijic:2018, Pani:2011, Horbatsch:2011}, 
including Horndeski theory \cite{Rinaldi:2015, Maselli:2016, Eickhoff:2018, Silva:2016}. 
However, Eq. (\ref{eos}) is the simplest model equation of state which is not able to describe all complexity of the neutron star structure. 
For this reason, in the present work we will consider a set of more realistic (though still model) equations of state.

\subsubsection{Realistic equation of state}
It is generally accepted that the equation of state has a one-parameter character. Many equations of state have been proposed by considering different kinds of interactions into account. Equations of state are usually given in the form of tables. However, using the table is not always convenient, especially for considering the modified theory of gravity. In some cases it is convenient to use a piecewise polytropic approach \cite{Read2009, Read2009_2}, or an analytical representations of equation of state \cite{Haensel2004, Potekhin2015, Pearson2018, Gungor2011}. 
Here we use the analytical representations of 31 unified equations of state, presented in the works \cite{Haensel2004, Potekhin2015, Pearson2018, Gungor2011}. Standard abbreviation is used for equations of state, detailed references can be found, for example, in review \cite{Latimer:2001}.

The equation of state can be represented as a function of density from pressure, i.e. $p=p(\rho)$. 
Analytical representations of FPS (Friedman-Pandharipande-Skyrme)  and SLy (Skyrme Lyon) equations of state \cite{Douchin2001, Pandharipande1989} are derived in the work \cite{Haensel2004}.
The parametrization reads 
\begin{eqnarray}\label{FPSSLy}
\zeta &=& \frac{a_1+a_2\xi+a_3\xi^3}{1+a_4\xi}f_0(a_5(\xi-a_6))
 +
(a_7+a_8\xi)f_0(a_9(a_{10} -\xi)) \\
&&~~~~~~~~
+ (a_{11} + a_{12}\xi) f_0(a_{13}(a_{14} -\xi))
+ (a_{15} +a_{16}\xi) f_0(a_{17}(a_{18}-\xi)),\nonumber
\end{eqnarray}
where 
\begin{eqnarray}
\xi=\lg (\rho/{\rm\mbox{g}\,\mbox{cm}^{-3}}),\quad \zeta=\lg(p/{\rm\mbox{din}\,\mbox{cm}^{-2}}),\quad f_0(x)=\frac{1}{1+e^x},
\end{eqnarray}
and $a_i$ -- given constants \cite{Haensel2004}. Here $\rho$ is a full energy density including the rest energies of the matter constituents divided by $c^2$, i.e. $\rho=\epsilon/c^2$. Function $p(\rho)$ fits the original tables \cite{Douchin2001, Pandharipande1989} in the density from $10^5\,{\rm g\,cm^{-3}}$ to $10^{16}\,{\rm g\,cm^{-3}}$ within typical error of 1–2\%. 
In the dimensionless form we have 
\be\label{eos_dl2}
\zeta=\log_{10}\left(\frac{ {\cal P} }{\kappa\ell^2} \right), \quad
\xi=\log_{10}\left(\frac{ {\cal E} }{\kappa c^2 \ell^2} \right).
\ee

We also use unified Brussels-Montreal-Skyrme equations of state in our work, labeled BSk19, BSk20, BSk21 \cite{Goriely2010, Pearson2011} and BSk22, BSk24, BSk25, BSk26 \cite{Potekhin2015}.  
At the much higher densities they differ greatly in their stiffness. 
 It was argued that the real equation of state can probably not be much stiffer than BSk26, and certainly not much softer than BSk22, BSk24 or BSk25. 
Analytic parametrization of $p(\rho)$ reads \cite{Potekhin2015, Pearson2018}
\begin{eqnarray}\label{BSk}
\zeta &=& \frac{a_1+a_2\xi+a_3\xi^3}{1+a_4\xi}f_0(a_5(\xi-a_6))
 +
(a_7+a_8\xi)f_0(a_9(a_{6} -\xi))+ (a_{10} + a_{11}\xi) f_0(a_{12}(a_{13} -\xi)) \nonumber\\
&&
+ (a_{14} +a_{15}\xi) f_0(a_{16}(a_{17}-\xi))
+
\frac{a_{18}}{1+a_{19}^2(\xi-a_{20})^2}+\frac{a_{21}}{1+a_{22}^2(\xi-a_{23})^2},
\end{eqnarray}
which fits the numerical tables \cite{Goriely2010, Pearson2011} for $6\leq \xi\leq 16$ with a typical error of 1\%. 
Eqs. (\ref{FPSSLy}) and (\ref{BSk}) differ in the numbering of parameters $a_i$ and numerical value of  $a_i$.

In the work \cite{Gungor2011} was presented an analytical unified representation for 22 equations of state of dense matter in neutron stars: 
AP1-4, engvik, gm1nph, gm2nph, gm3nph, mpa1, ms00, ms2, ms1506, pal2, pclnphq, wff1, wff2, wff3, wff4, schaf1, schaf2, prakdat, ps. 
The function is
\begin{eqnarray}\label{engvik}
\zeta &=& \zeta_{low}f_0(a_1(\xi-c_{11})) + \zeta_{high}f_0(a_2(c_{12}-\xi)), 
\end{eqnarray} 
where
\begin{eqnarray}
\zeta_{low} &=& [c_1+c_2(\xi-c_3)^{c_4}] f_0(c_5(\xi-c_6)) +
(c_7+c_8\xi) f_0(c_9(c_{10}-\xi)), \\
\zeta_{high} &=& (a_3+a_4\xi) f_0(a_5(a_{6}-\xi))+
(a_7+a_8\xi+a_9\xi^2) f_0(a_{10}(a_{11}-\xi)),
\end{eqnarray} 
describe the low and high density regimes, respectively, the values of the fit parameters $c_i$ and $a_i$ for $5\leqslant\xi\leqslant 16$ are given  in the tables \cite{Gungor2011}. 

Some equations of state becomes superluminal ($v\geqslant c$) above critical density $\rho\geqslant\rho_{caus}$, $\rho_{caus}$  named causality limits. For example in the case of BSk24 equation of state  $\rho_{caus}=2.69\times10^{15}\, g\, cm^{-3}$. Fig.~\ref{bsk24}d shows the dependence of the sound speed $v^2/c^2$ on the density $\rho$ for BSk24 equation of state.  Apparently in the case $\rho\geqslant\rho_{caus}$  equation of state do not give a complete description, causality breaks down, in particular it was discussed in \cite{Pearson2018}.

\subsection{Boundary conditions}
The equations (\ref{_eqB_dl})--(\ref{_eqA_dl})  form a closed system of ordinary differential equations for function $A(r)$, $B(r)$ and $p(r)$, 
where the pressure and the energy are related by the equation of state $p=p(\rho)$. The scalar field $\phi$ can be found using the equation (\ref{_eqPsi_dl}). 
Boundary conditions are usually determined at the center of a star $r=0$
\be
B(0)=1,\quad p(0)=p_c,
\ee
where  $p_c$ is the central pressure, and the value of $A(0)$ will be fixed after matching internal and external solutions at the star boundary. Therefore, the only free parameter is the value of the pressure in the center of the star, $p_c$.

\section{External vacuum solution} \label{sec3}
Outside the star one has a vacuum solution ($p=0$ and $\rho=0$)
\begin{eqnarray}
	\label{vacsol_B}
	&& B(x) =\frac{(x^2+1)^2}{\big((1-\xi)x^2+2\big)^2}\, F(x),
	\\ 
	&& A(x)=3 C_2\, F(x),
	\label{vacsol_A}
	\\ \label{vacPsi1}
	&& \Psi^2 = \frac{x^2(1+\xi)} {\varepsilon_2  B\, (1 +x^2)},
\end{eqnarray}
where 
$$
F(x)=(1-\xi)(3+\xi)+\frac{1}{x}\left((1+\xi)^2 \arctan x+C_1\right) +\frac{x^2}{3}(1-\xi)^2,
$$
and $C_1$ and $C_2$ are constants of integration. 
The asymptotical form of $B(x)$ and $A(x)$ at $x\to\infty$ shows that one has anti-de Sitter-Schwarzschild spacetime geometry outside the star:
\begin{eqnarray}
\label{asB}
B(x) &=& \frac{x^2}{3}+\frac{7+\xi}{3(1-\xi)} +\frac{C_1+\frac12(1+\xi)^2\pi}{(1-\xi)^2}\,\frac1x
+{\cal O}(x^{-2}),
\\
\label{asA}
A(x) &=& 3C_2(1-\xi)(3+\xi)\left[1+\frac{1-\xi}{3(3+\xi)}x^2 +\frac{C_1+\frac{\pi}{2}(1+\xi)^2}{(3+\xi)(1-\xi)}\,\frac1x\right]
+{\cal O}(x^{-2}),\\
\Psi^2(x) &=& \frac{3(1+\xi)}{\varepsilon_2 x^2} +{\cal O}(x^{-4}).
\end{eqnarray}
Assuming that $t$ is the time of a distant observer, one can fix the value of $C_2$ as follows $3C_2(1-\xi)(3+\xi)=1$.
Additionally, one has to demand $C_2>0$ in order to guarantee the same sign of the metric functions $B(x)$ and $A(x)$, hence one has $(1-\xi)(3+\xi)>0$, or 
\begin{equation}\label{range_xi}
	-3<\xi<1\,.
\end{equation}   
Returning to the dimensional radial coordinate $r=\ell x$ and choosing appropriately the constants of integration $C_1$ and $C_2$, we find that far from the star
\be\label{A_AdS}
A(r)\approx 1-\frac{r_g}{r} +\frac{|\Lambda_{AdS}|}{3}\,r^2,
\ee 
where
\be\label{C1}
r_g=\frac{2GM}{c^2}=-\ell\,\frac{C_1+\frac{\pi}{2}(1+\xi)^2}{(3+\xi)(1-\xi)},
\ee 
 and
\begin{equation}
	\label{Leff}
	\Lambda_{AdS}=-\frac{1-\xi}{3+\xi}\, \frac{1}{\ell^2},
\end{equation}
is the effective negative cosmological constant. Formula (\ref{C1}) gives us the so-called asymptotic mass of the neutron star.
Constant $C_1$ is found from the condition of continuity on $B(r)$
$$B_{in}(R)=B_{out}(R),$$ 
where $B_{in}(R)$ and $B_{out}(R)$ are the value of the interior and exterior function at the boundary of the neutron star $r=R$. 
Substituting the value of the constant $C_1$ into the formula (\ref{C1}), we find the asymptotic mass $M$. 

In the case $\xi=-1$ the solution represents the anti-de Sitter–Schwarzschild black hole:
\begin{eqnarray}
A(r) = B(r) = 1-\frac{r_g}{r}+\frac{|\Lambda_{AdS}|}{3}\, r^2,
\quad \Psi^2(r)=0,
\label{stealthSS}
\end{eqnarray}
where $\Lambda_{AdS}=-1/\ell^{2}$ and the integration constants fixed as $C_1=-4 r_g/\ell=-8MG/c^2\ell$ and $C_2=1/12$. 

In the classical theory of gravity, the nucleon and asymptotic mass coincide due to Einstein's equations. However, in this case, the nucleon mass $M_b$  will differ from the asymptotic one and is calculated using formula 
\be\label{M_0}
M_0=4\pi \int_0^R A^{1/2}B^{-1/2} \rho\,r^2\, dr.
\ee 

In the general case $\xi\not=-1$ one has to choose a proper sign $\varepsilon_2=\pm1$ in order to provide positivity of $\Psi^2$. From Eq. (\ref{vacPsi1}) we obtain:
\begin{eqnarray}
	&& \textrm{(i)}\quad -3<\xi\le-1,\quad \varepsilon_2=-1;
	\nonumber\\
	&& \textrm{(ii)}\quad -1<\xi<1,\quad \varepsilon_2=+1.
	\nonumber
\end{eqnarray}

\section{Internal solution} \label{sec4}

\subsection{Scheme of numerical integration}
We explore internal configurations of neutron stars for different sets of model parameters $\xi$ and $\ell$ using the following scheme of 
numerical integration of the system (\ref{_eqB_dl})--(\ref{_eqPsi_dl}), where pressure ${\cal P}$ and energy ${\cal E}$ are related by one of the equations of state (\ref{FPSSLy})-(\ref{engvik}). 
First, we find solutions for $B(x)$ and ${\cal P}(x)$ integrating Eqs (\ref{_eqB_dl}) and (\ref{_eqP_dl}) from the center $r=0$ to the boundary of the star $r=R$ with the following initial conditions: $B(0)=B_c=1$, ${\cal P}(0)={\cal P}_c$. 
In the case of polytropic equation of state ${\cal P}_c=\kappa\ell^2 K\rho_{0c}^\Gamma$, 
in the case of realistic equations ${\cal P}_c=\kappa\ell^2 10^{\zeta(\xi_c)}$, where $\xi_c=\log_{10}(\rho_c/\kappa\ell^2)$. 
The boundary of the star is defined as ${\cal P}(R)=0$. At the boundary the internal solution $B_{in}(x)$ is matching with the external vacuum solution $B_{vac}(x)$ given by (\ref{vacsol_B}), i.e. $B_{in}(R)=B_{vac}(R)$. From the matching condition we fix the constant of integration $C_1$ and, ultimately, the Schwarzschild mass $M$ given by Eq.~(\ref{C1}). Then, with solutions found for $B(x)$ and ${\cal P}(x)$, we obtain $A(x)$ from (\ref{_eqA_dl}), and $\Psi^2(x)$ from (\ref{_eqPsi_dl}). A constant of integration of Eq. (\ref{_eqA_dl}) is fixed by the matching condition at the boundary of star, $A_{in}(R)=A_{vac}(R)$, where the vacuum solution $A_{vac}(x)$ is given by (\ref{vacsol_A}).

\subsection{Results of numerical integration}

\subsubsection{The case $\xi=-1$ }
\begin{figure}[h]\begin{center}
\includegraphics[scale=0.25]{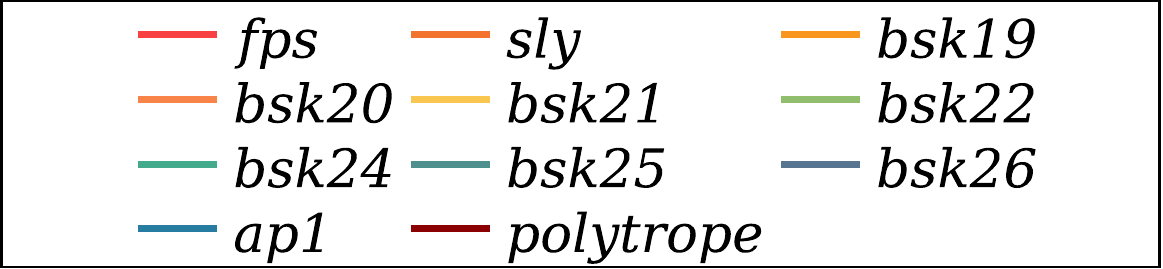}  
\includegraphics[scale=0.25]{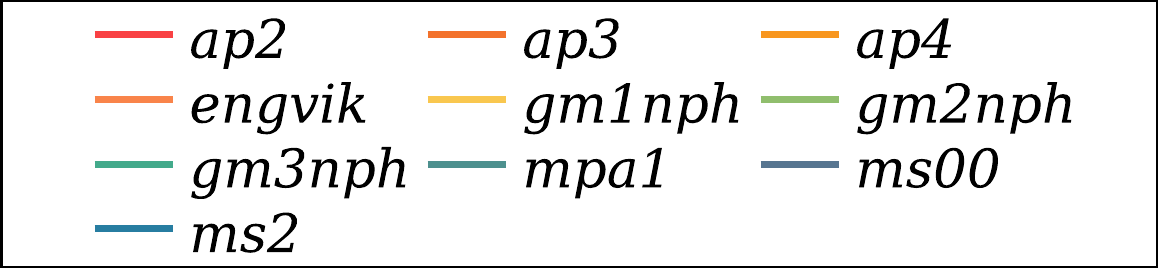}  
\includegraphics[scale=0.25]{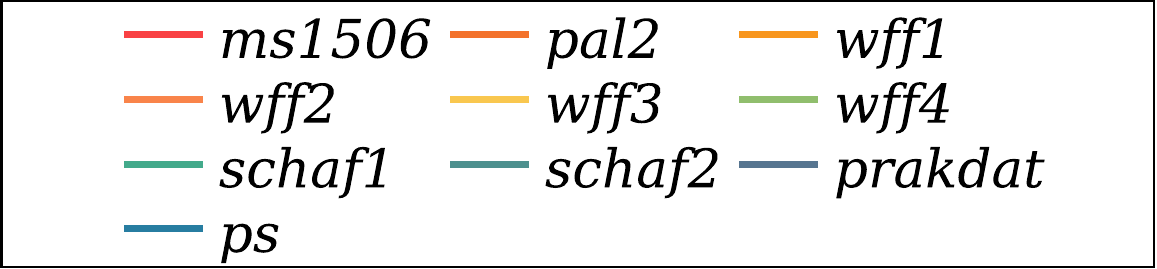}  \\
\includegraphics[scale=0.25]{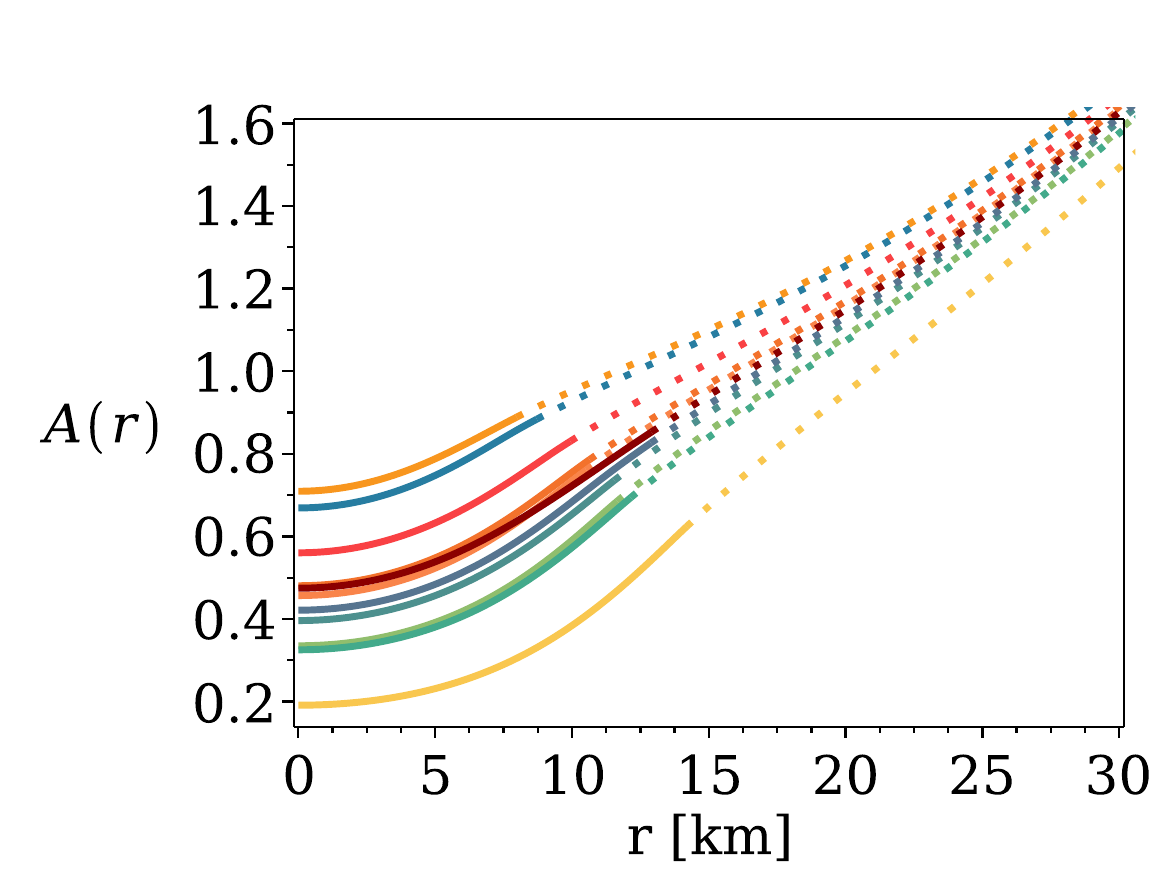} 
\includegraphics[scale=0.25]{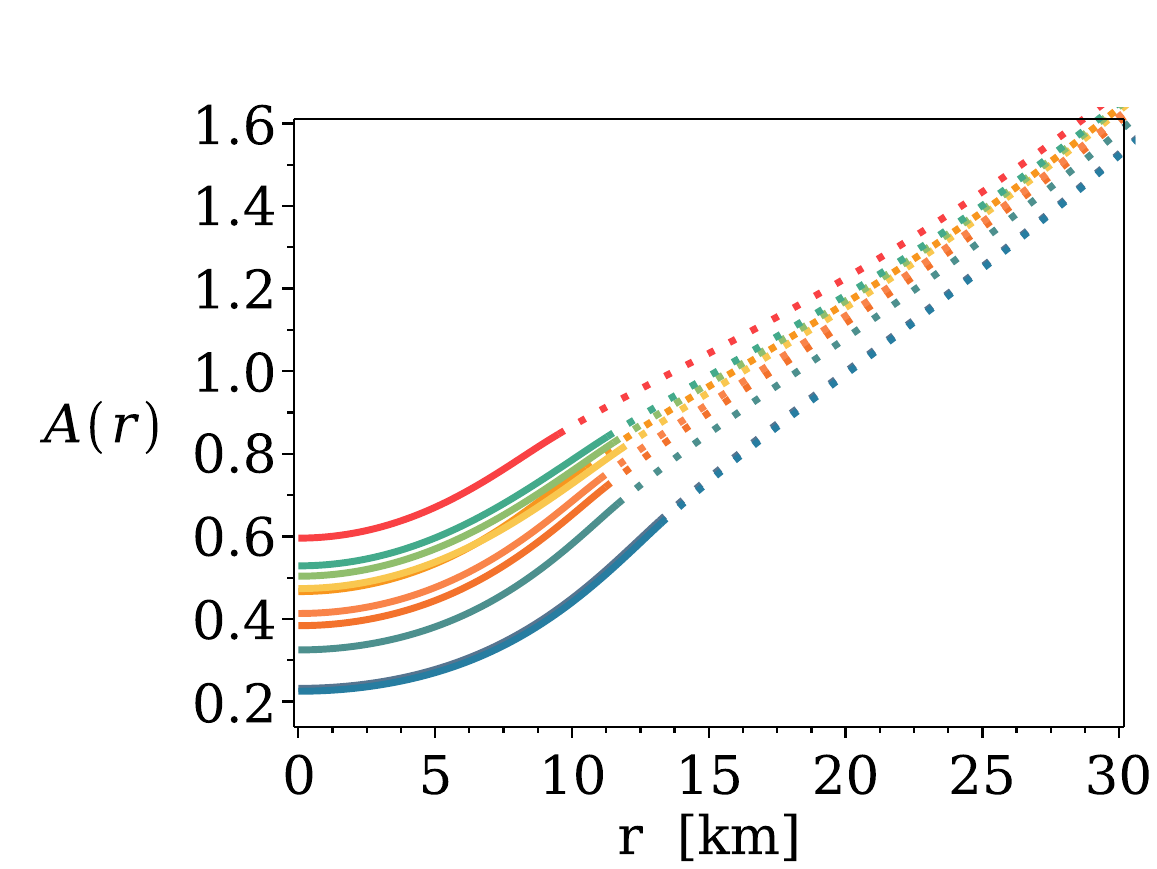} 
\includegraphics[scale=0.25]{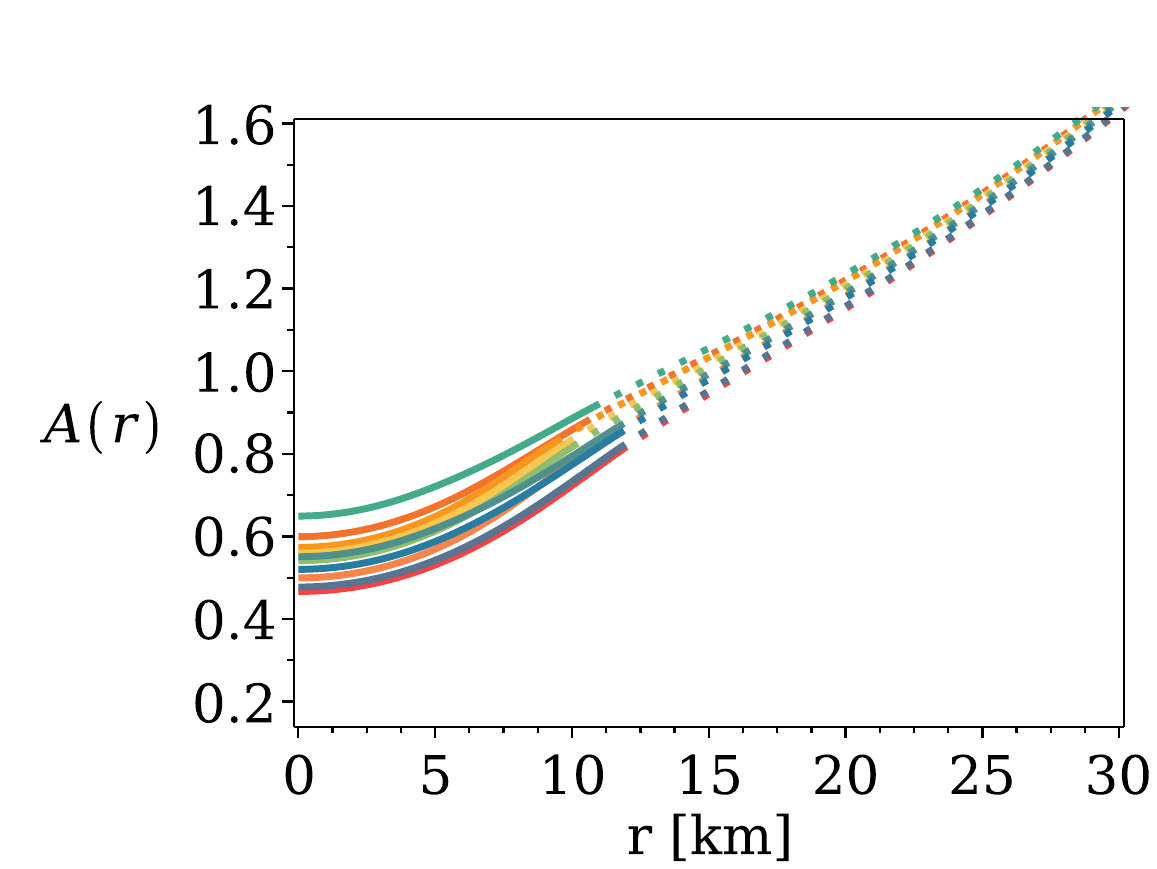} \\
\includegraphics[scale=0.25]{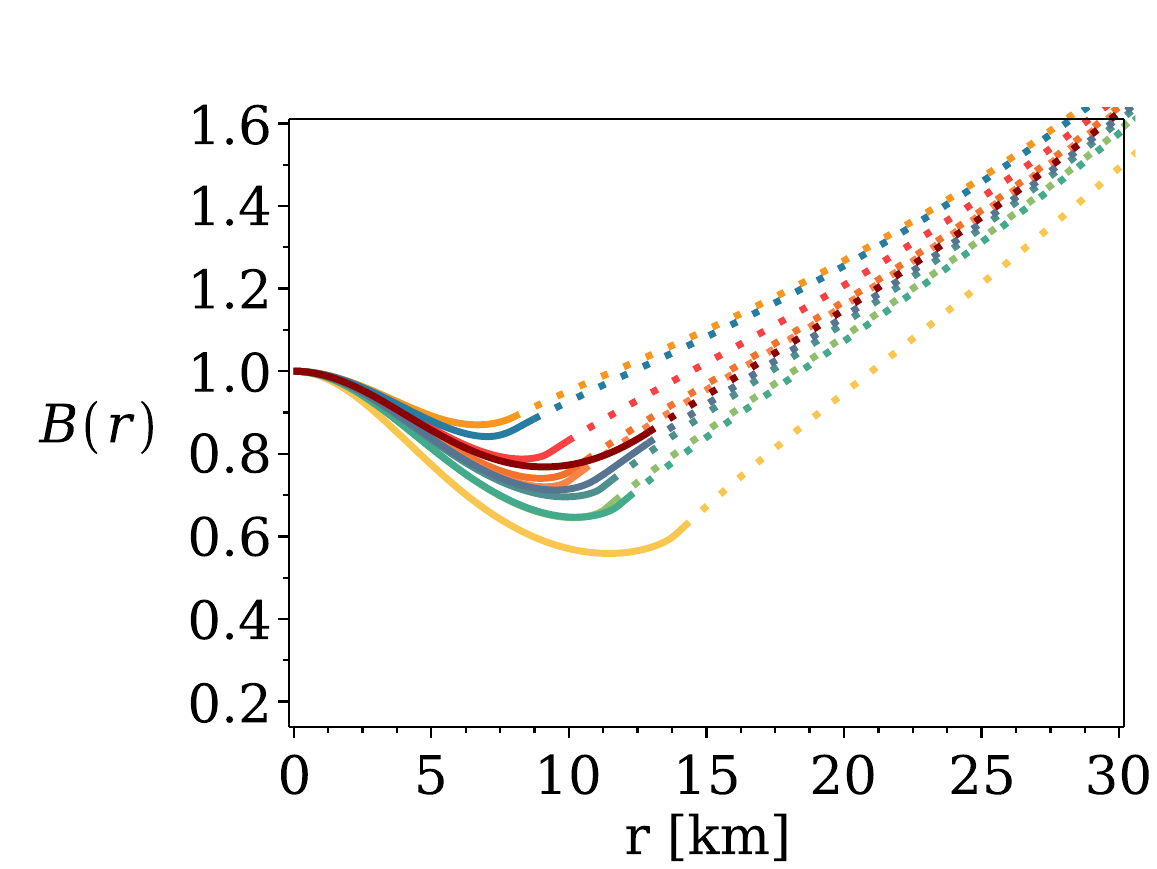} 
\includegraphics[scale=0.25]{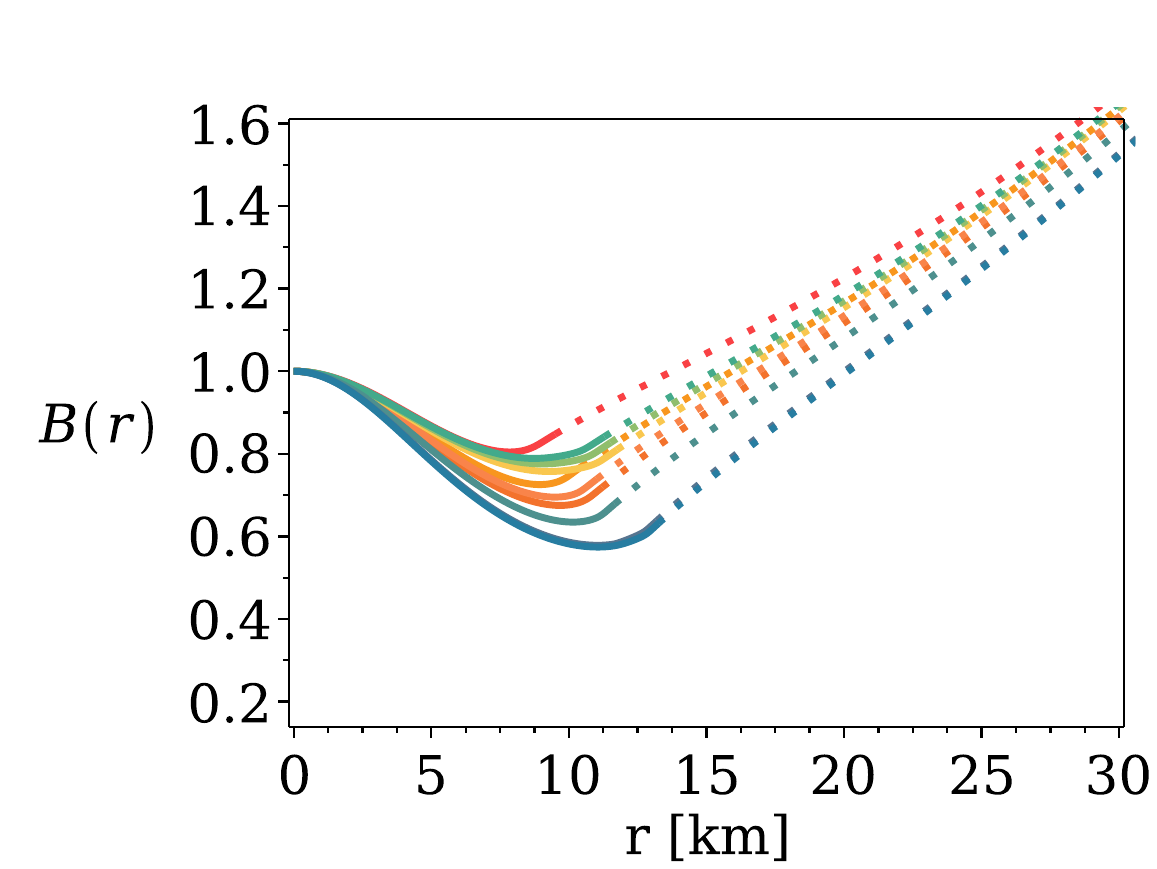} 
\includegraphics[scale=0.25]{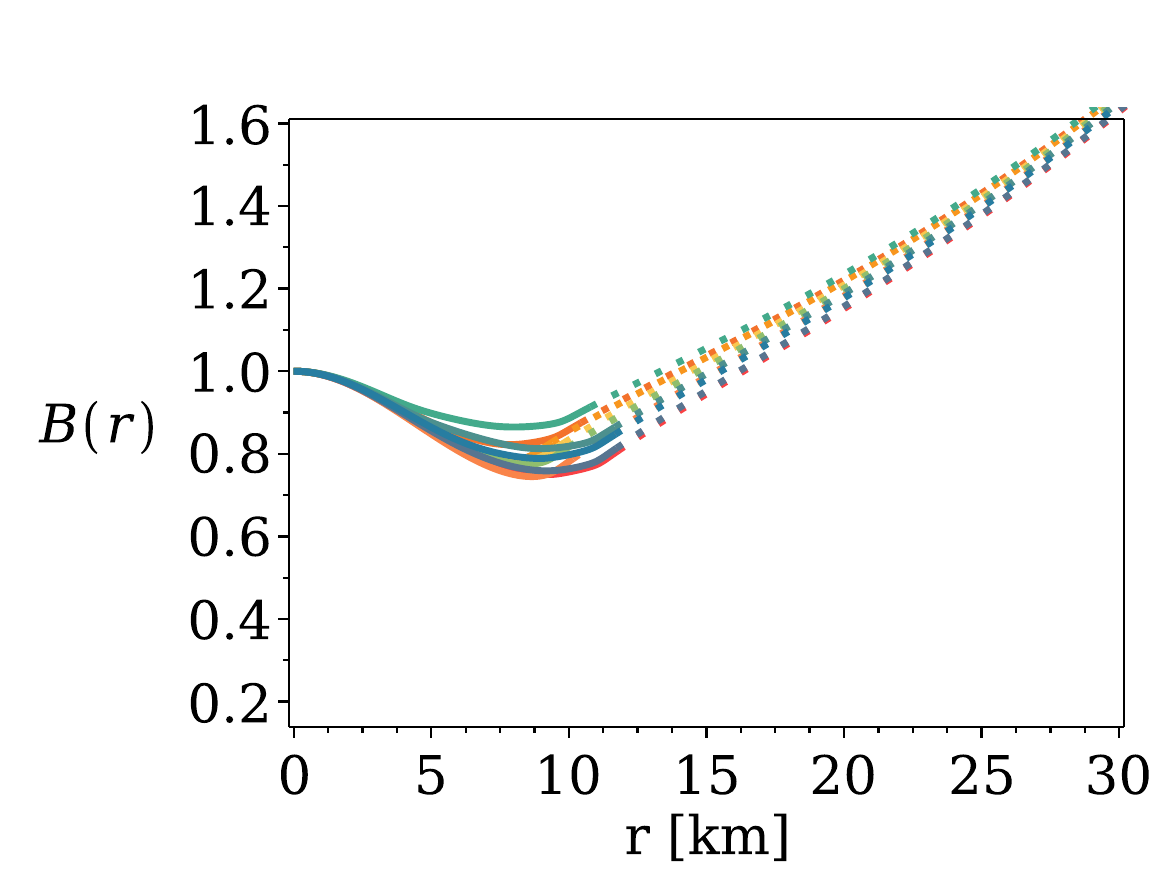} \\
\includegraphics[scale=0.25]{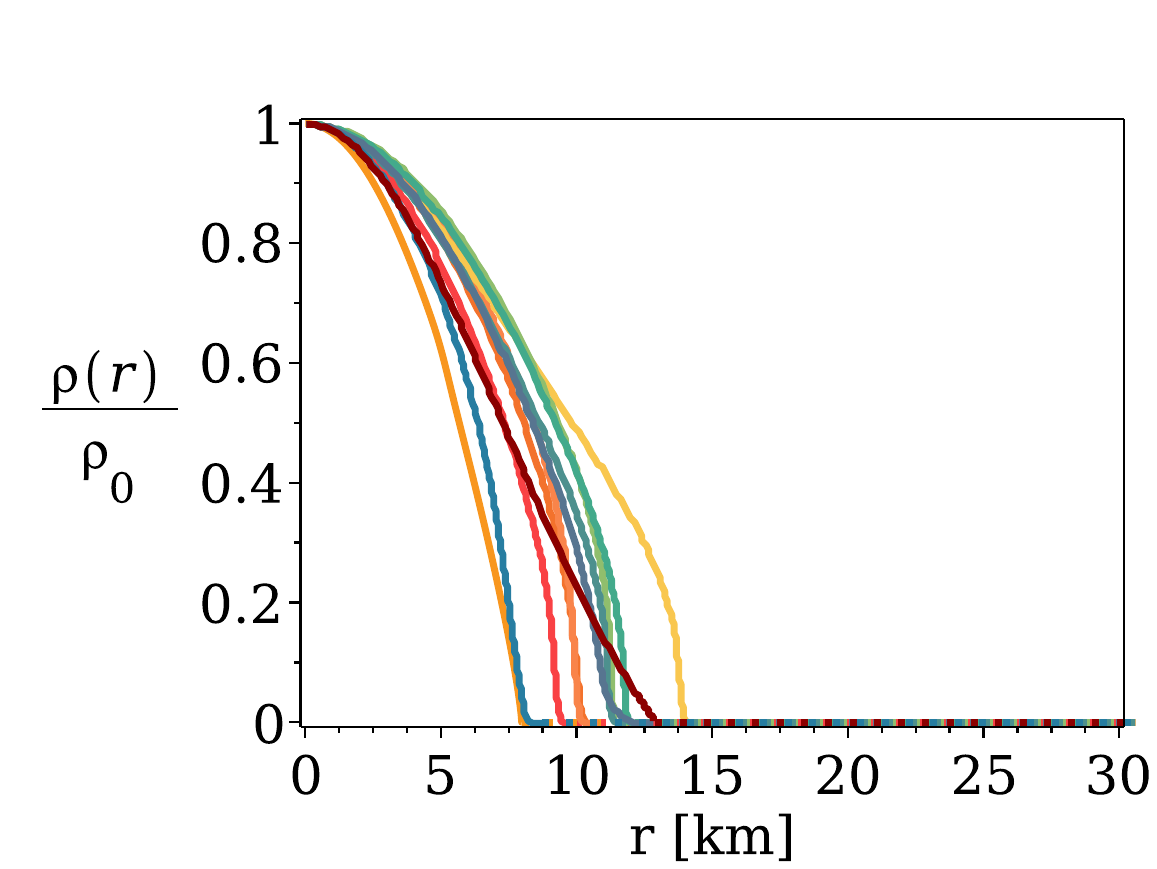}
\includegraphics[scale=0.25]{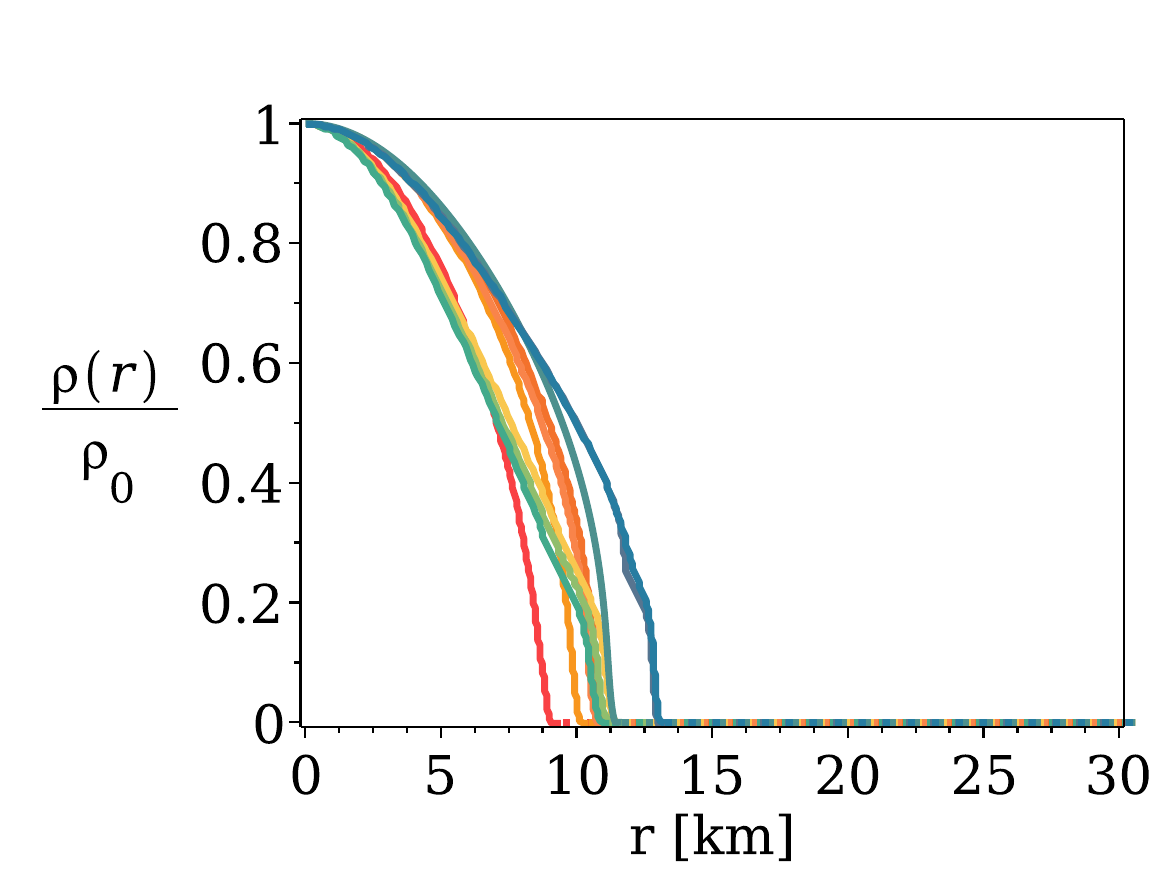}
\includegraphics[scale=0.25]{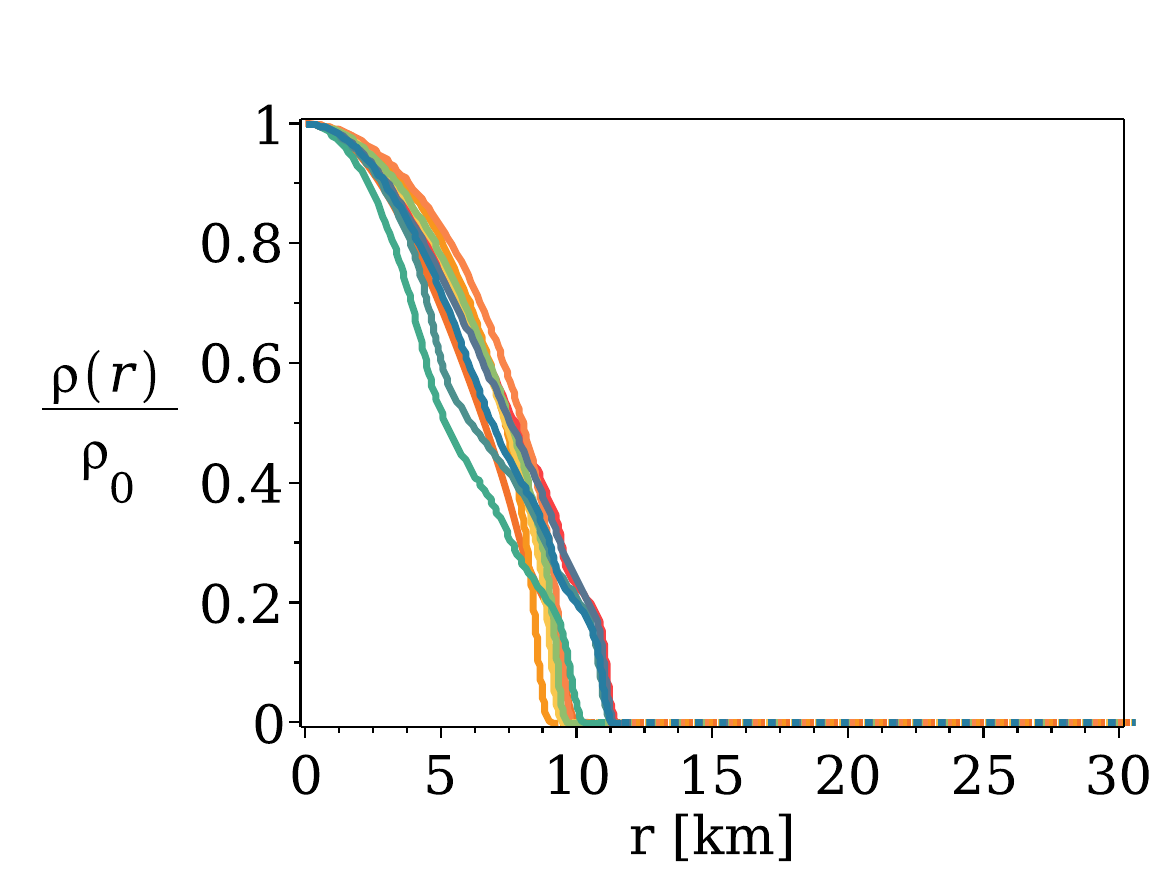}\\
\includegraphics[scale=0.25]{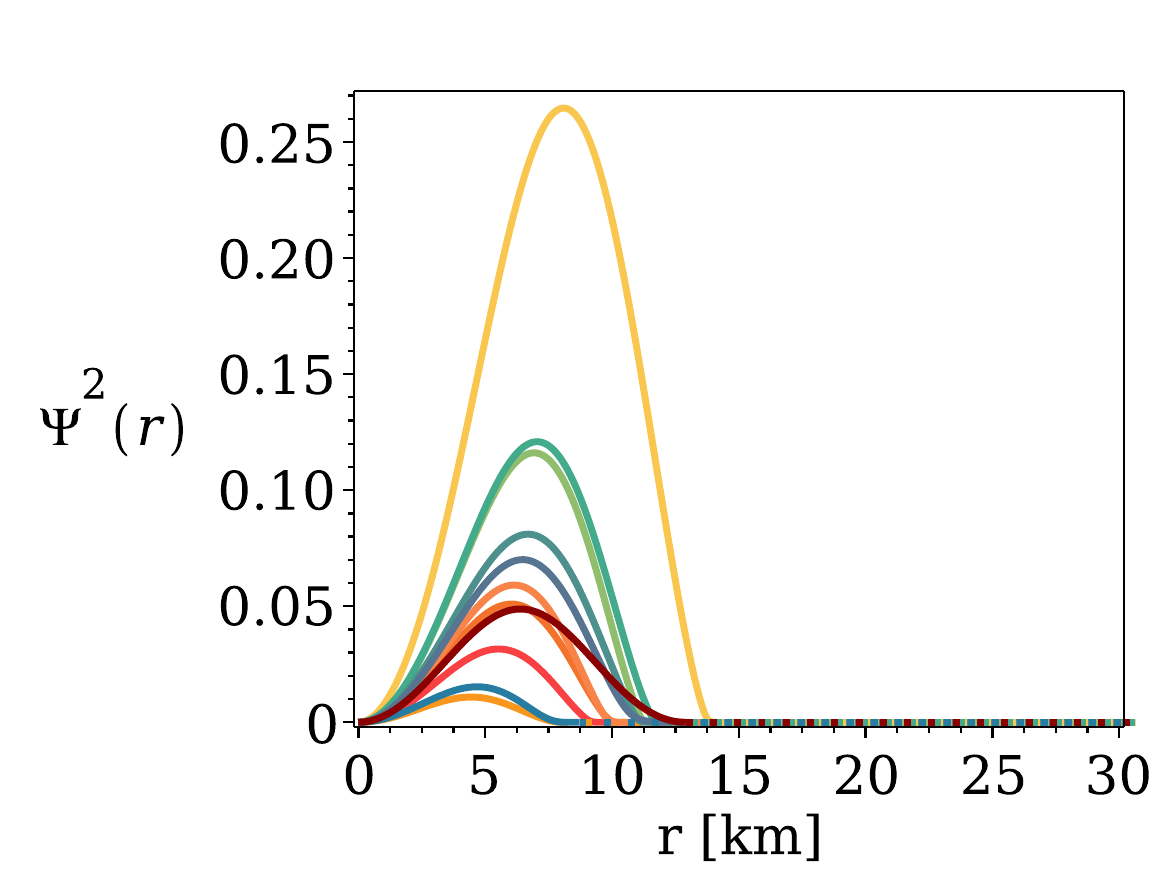} 
\includegraphics[scale=0.25]{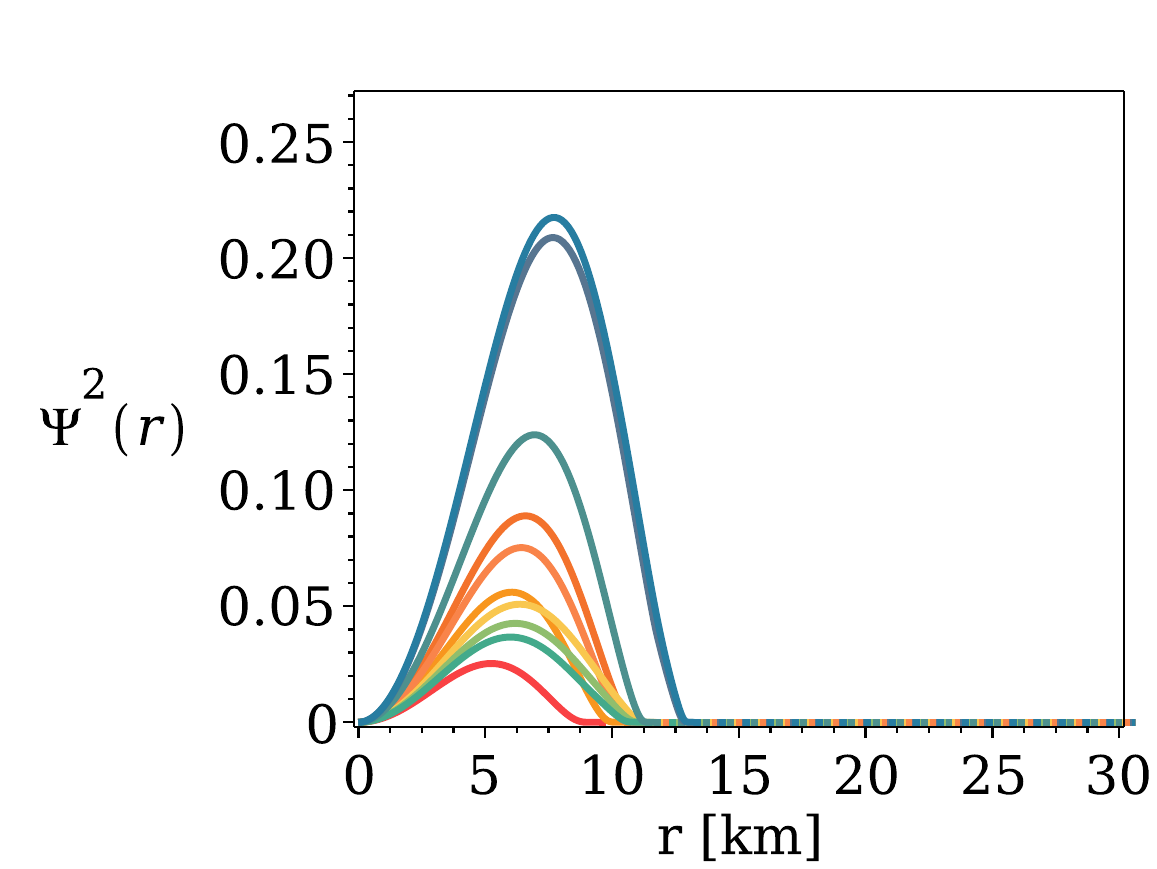} 
\includegraphics[scale=0.25]{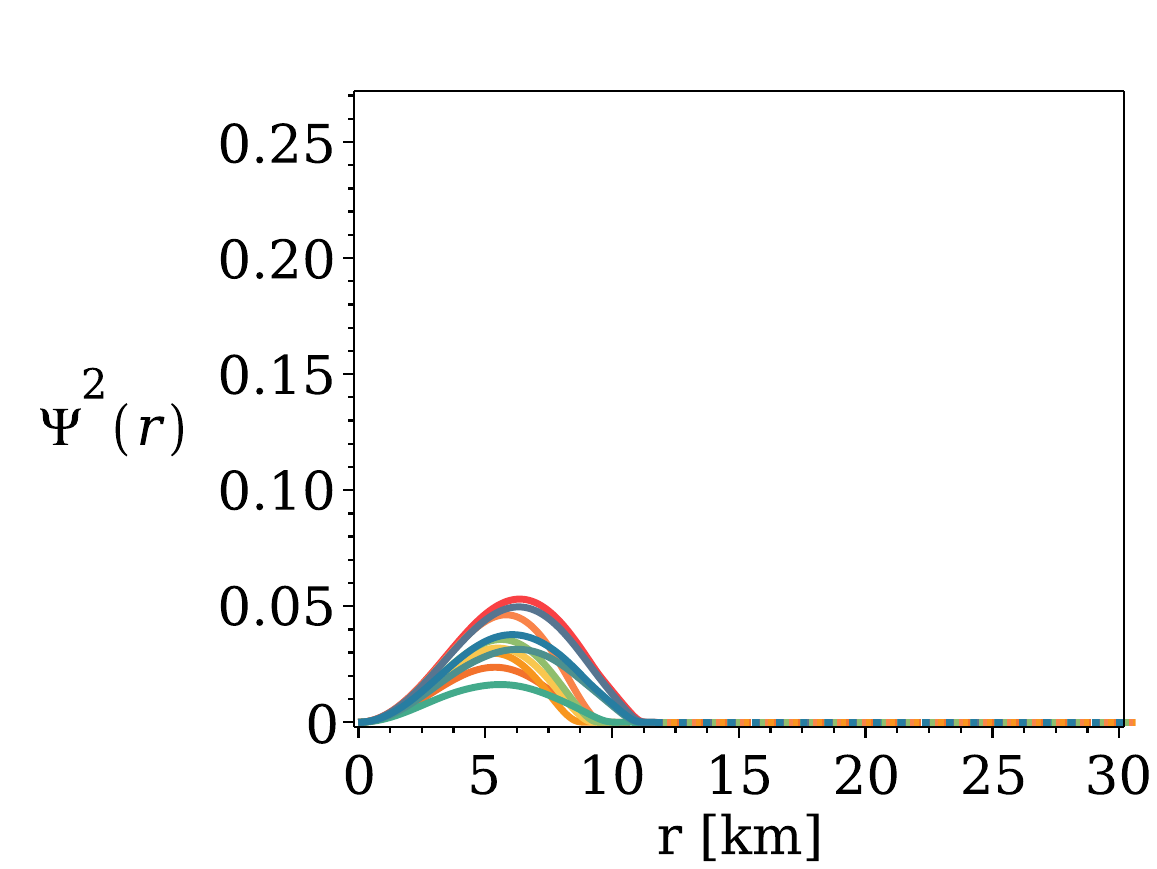}
\caption{Graphs of the functions  $A(r)$, $B(r)$, $\rho(r)$ and $\Psi^2(r)$ are shown for different equations of state in the case $\xi=-1$, $\ell=20$ km and the central mass density $\rho_{c}=1.0\times 10^{15}\ {\rm g/cm^3}$. Dotted line mark the vacuum solution.\label{ABrhoPsi1} }
\end{center}\end{figure}

\begin{figure}[t]\begin{center}
\includegraphics[scale=0.35]{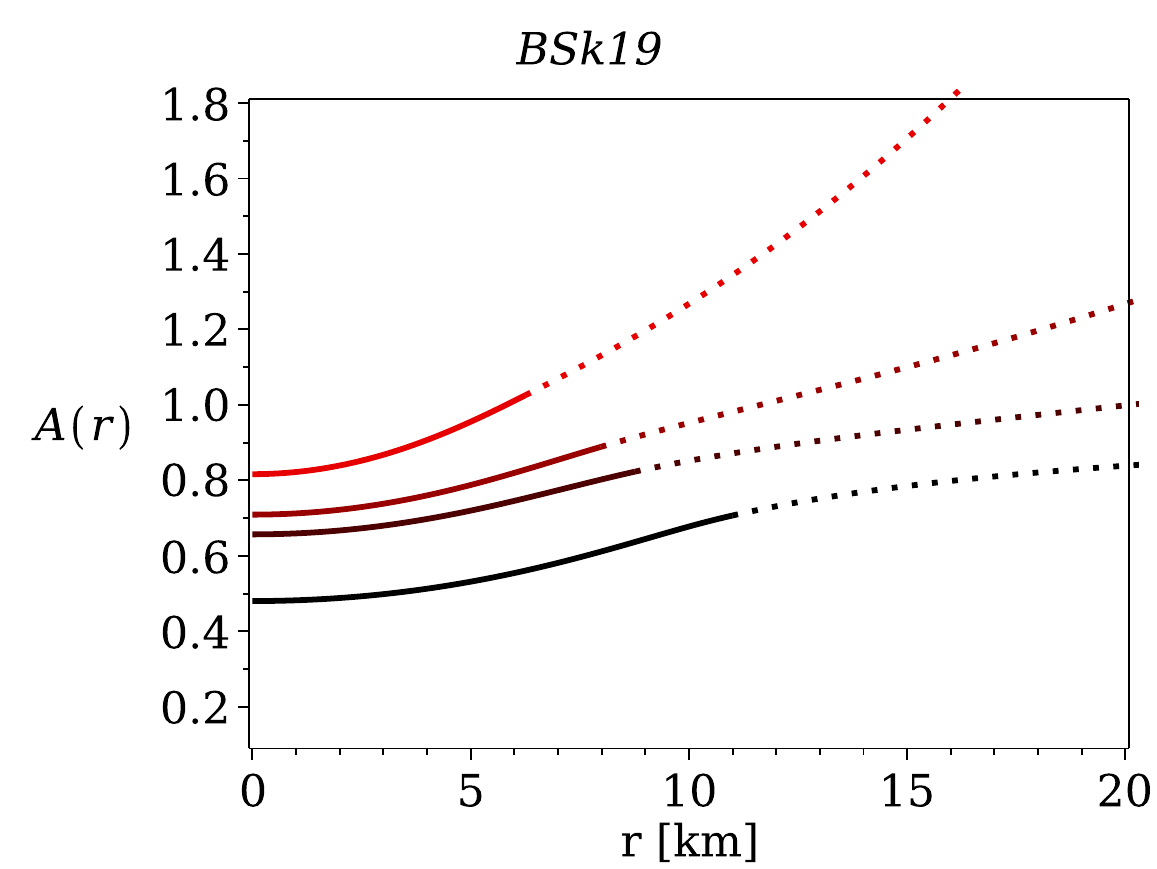}  
\includegraphics[scale=0.35]{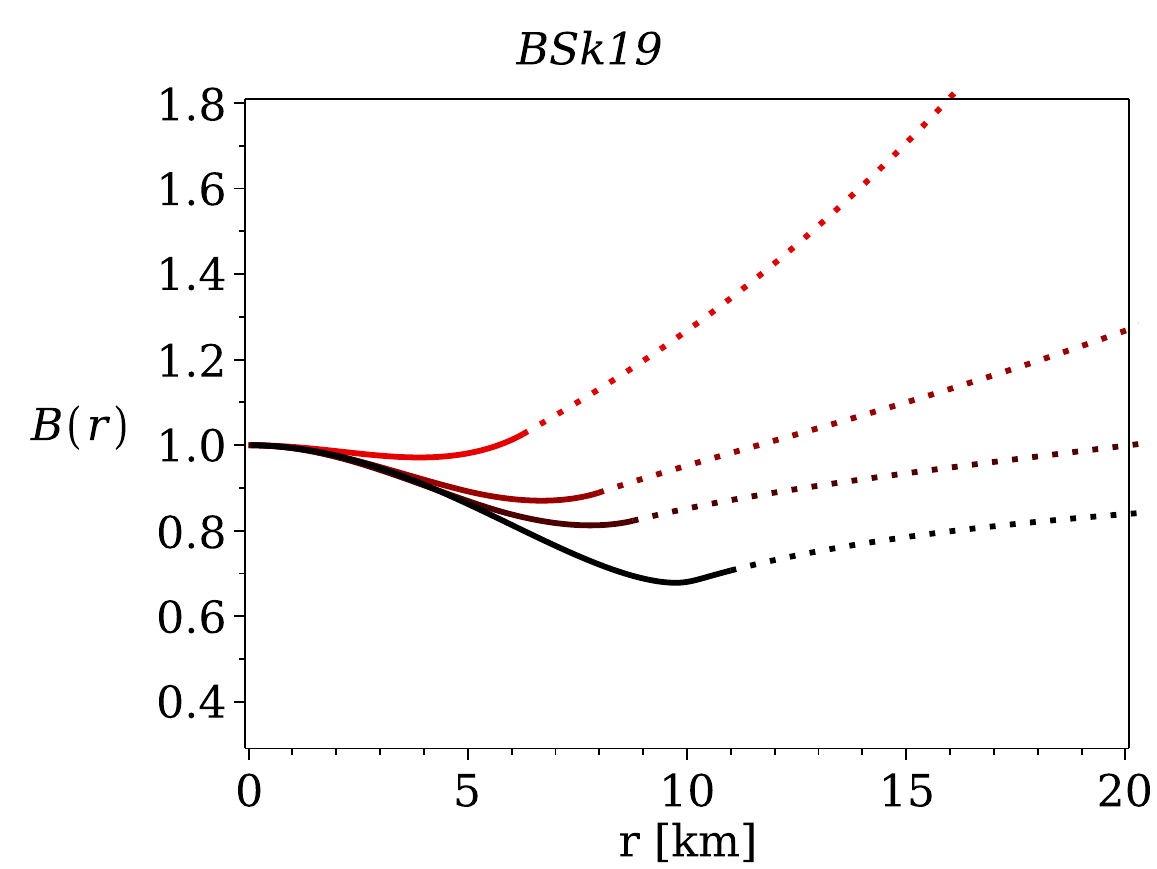} \\ 
\includegraphics[scale=0.35]{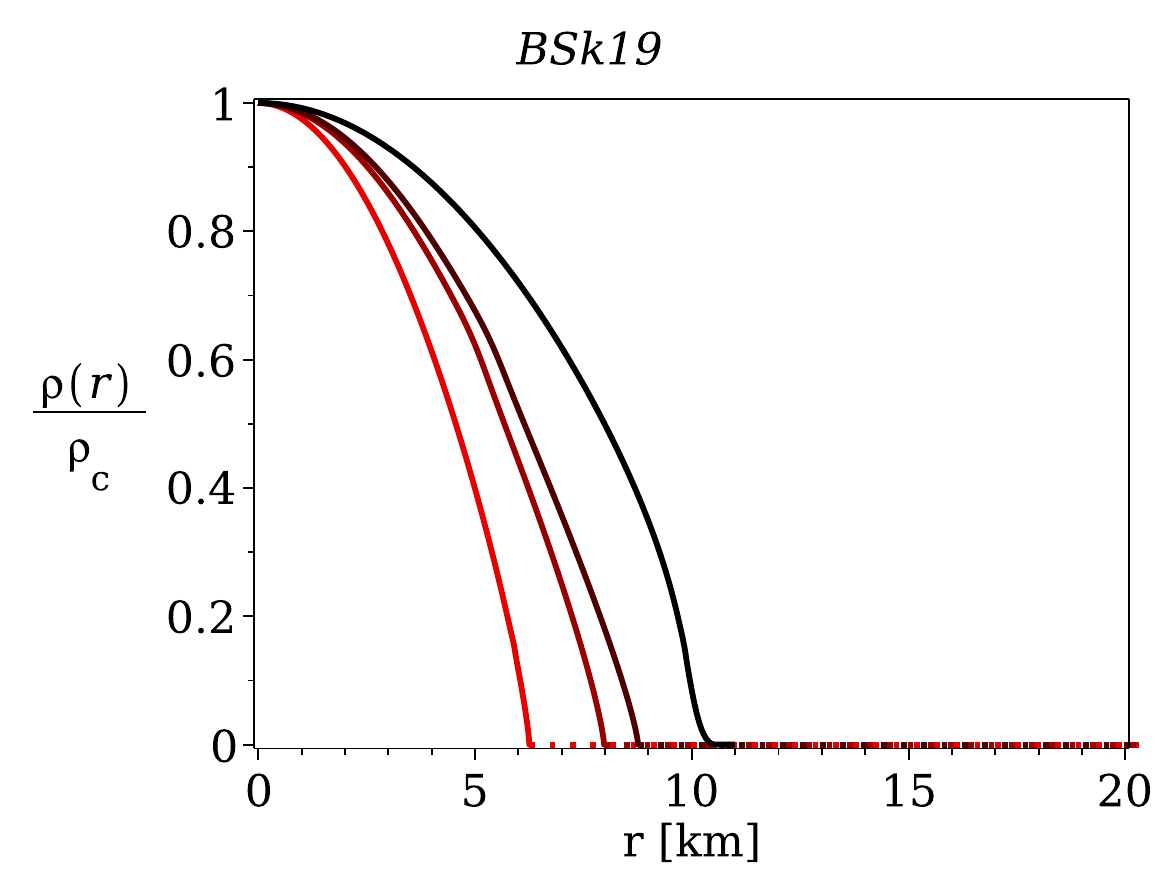}  
\includegraphics[scale=0.35]{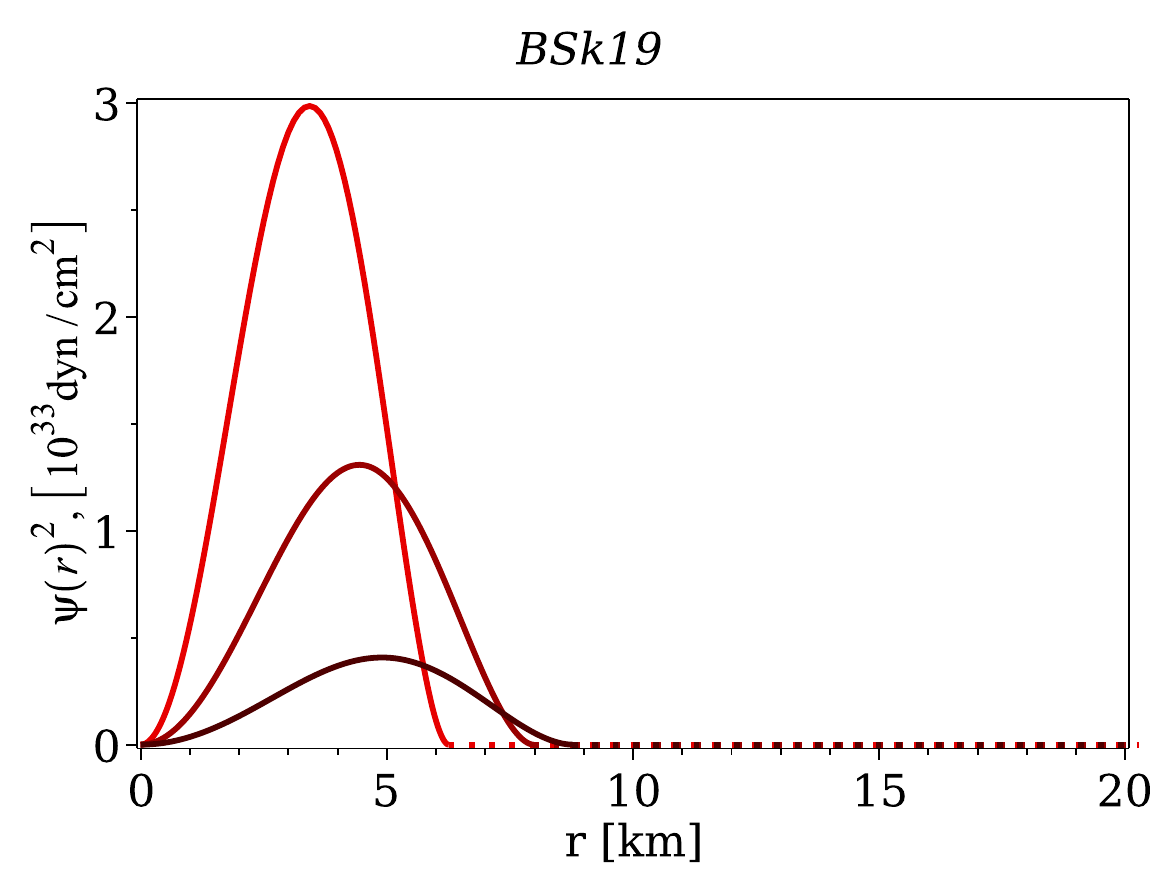} \\
\includegraphics[scale=0.35]{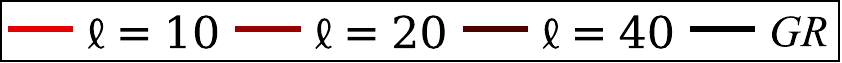}
\caption{Graphs of the functions  $A(r)$, $B(r)$, $\rho(r)$ and $\Psi^2(r)$ for BSk19 equation of state in the case $\xi=-1$,  the central baryonic mass density $\rho_{c}=1.0\times 10^{15}\ {\rm g/cm^3}$  are shown for three different values $\ell=10$, 20 and 40 km. Dotted line mark the vacuum solution. \label{ABrhoPsi_ell} }
\end{center}\end{figure}

\begin{figure}[t]
\begin{center}\begin{tabular}{cc}
 	\includegraphics[scale=0.4]{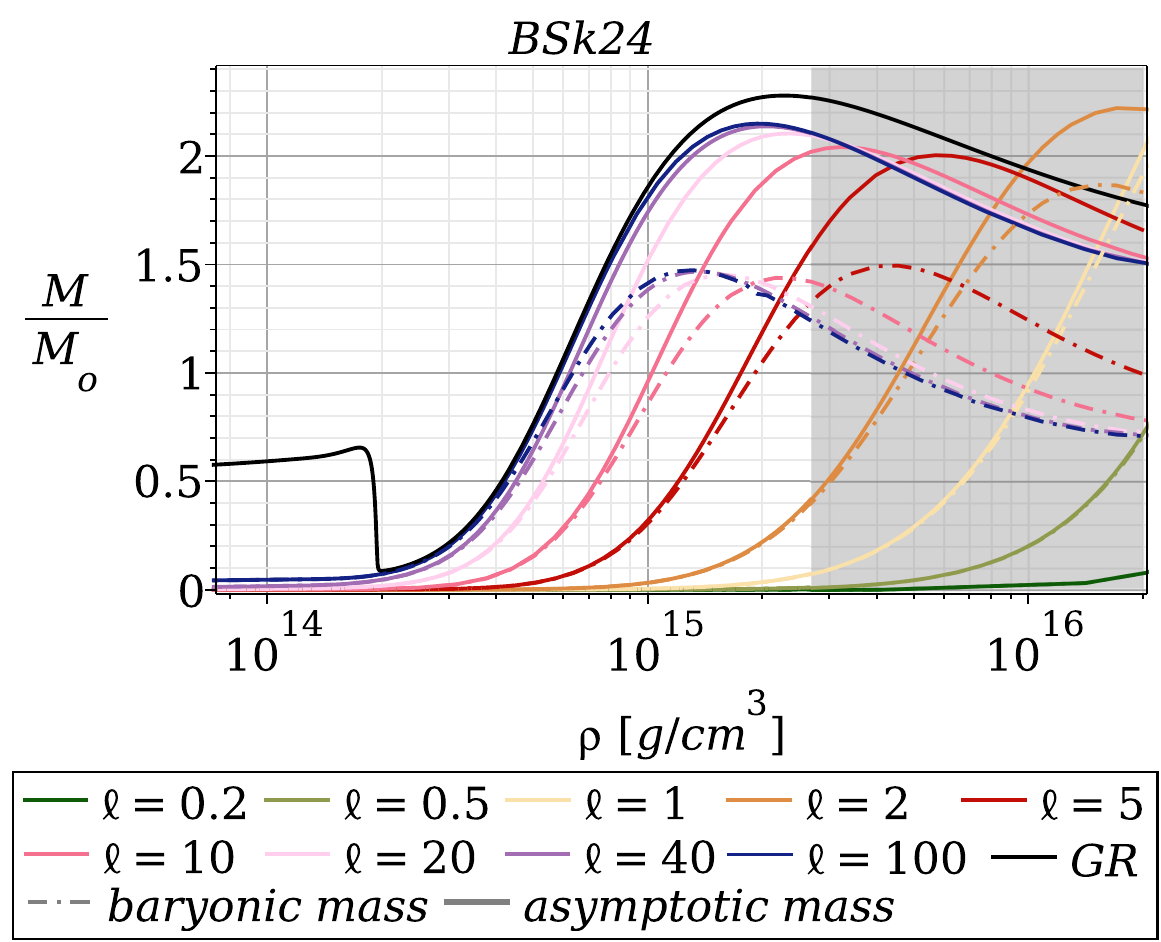} &
	\includegraphics[scale=0.4]{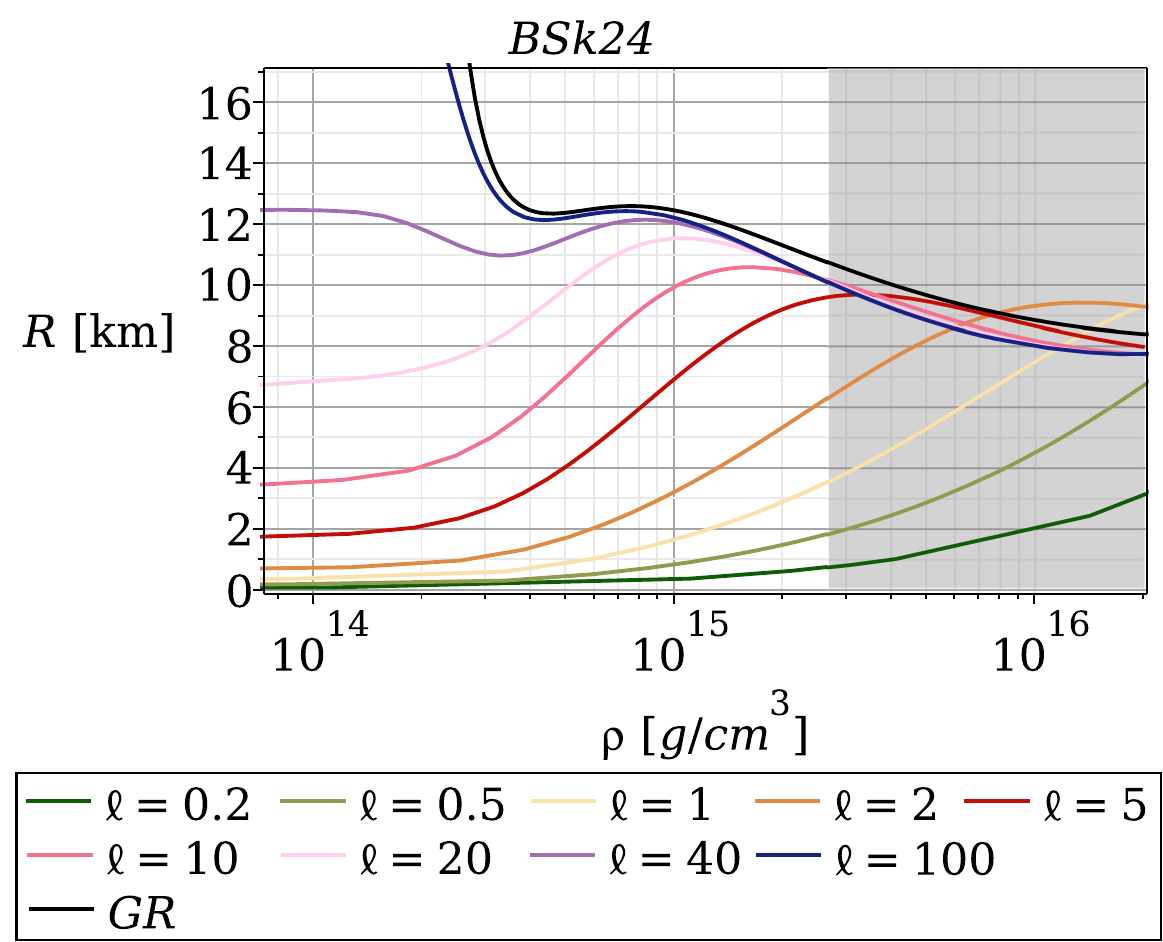} \\ a) & b) \\
\includegraphics[scale=0.4]{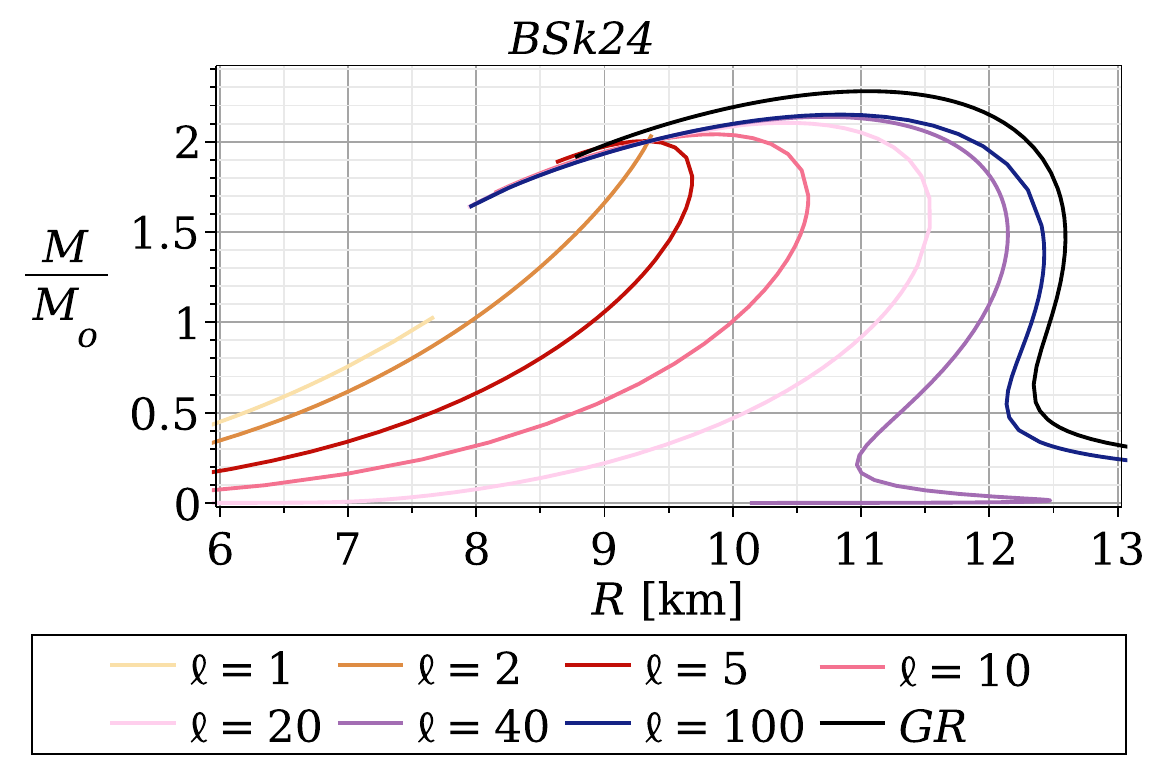} &
	\includegraphics[scale=0.4]{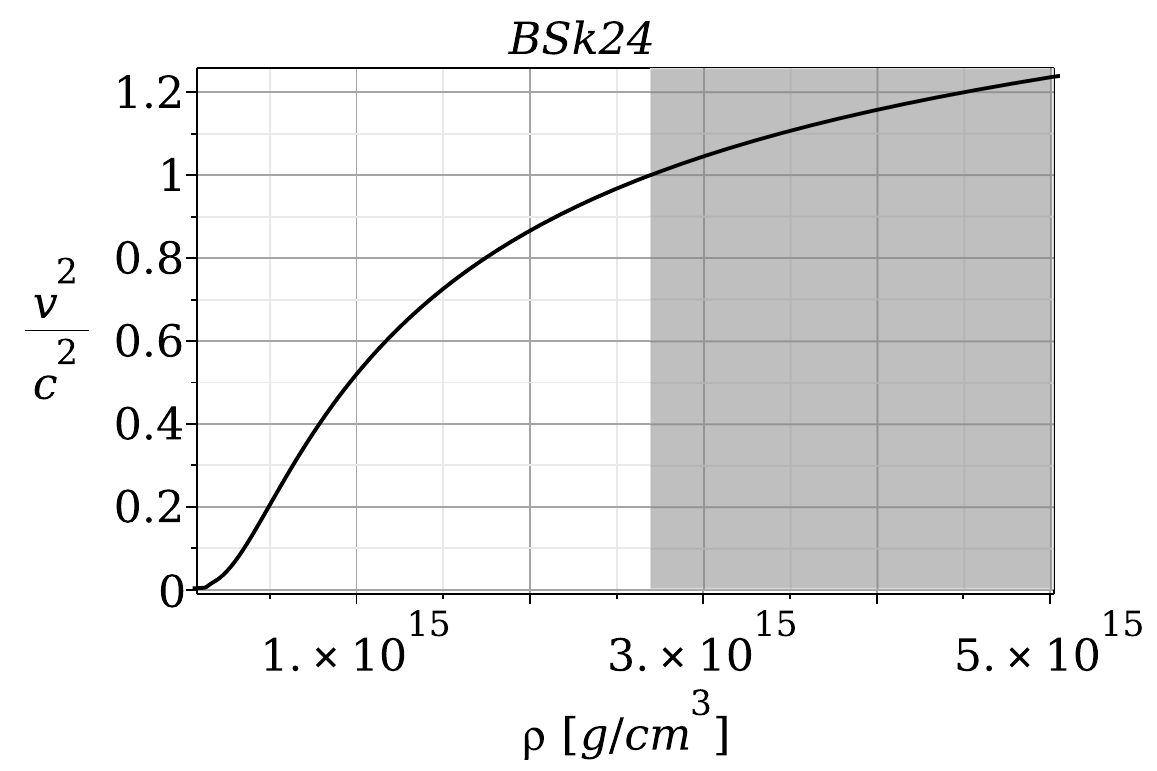} \\c) & d)  
	\end{tabular} \end{center}	
\caption{\label{bsk24} Parameters of the stars in the case $\xi=-1$, $\ell=$ 0.2, 0.5, 1, 2, 5, 10, 20, 100 km for  BSk24 equation of state: 
a) mass $M$ dependence from the central density $\rho_c$, the solid line corresponds to the asymptotic mass $M_a$, the dash-dot line --- to the baryonic $M_b$ mass;
b) the dependence of the star radius $R$ from the central density $\rho_c$; 
c) the mass-radius diagram in the range $10^{14} \ {\rm g/cm^3}\leqslant \rho_c\leqslant 10^{16} \ {\rm g/cm^3}$;
d)  the sound speed $v/c$ dependence from the density $\rho$. 
The correspondence between style line  and values $\ell$ is given at the legends, the black curve corresponds to Einstein's theory of gravity (GR). 
The gray area at the figures corresponds to the density at which the equation of state becomes superluminal.}
\end{figure}

\begin{figure}[h]
\begin{center}
\begin{tabular}{ccccc}  &$\ell=10$ km & $\ell=40$ km & $\ell=100$ km  & GR\\
 \includegraphics[scale=0.17]{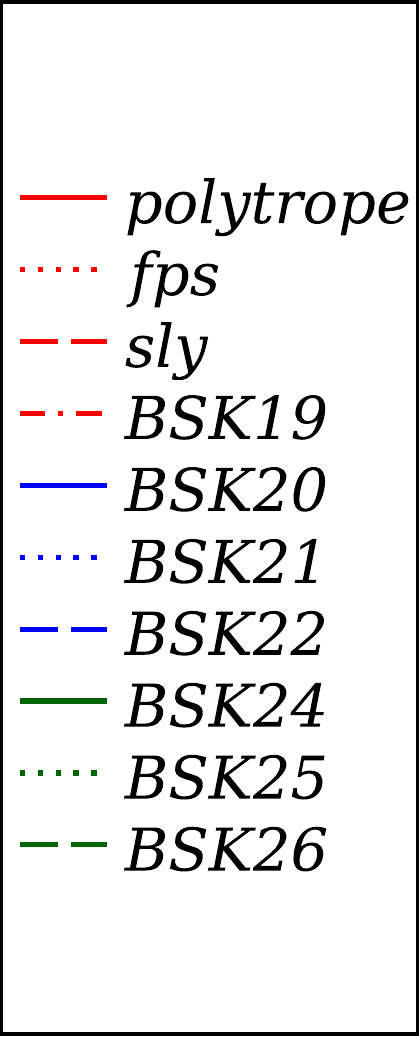} & \includegraphics[scale=0.17]{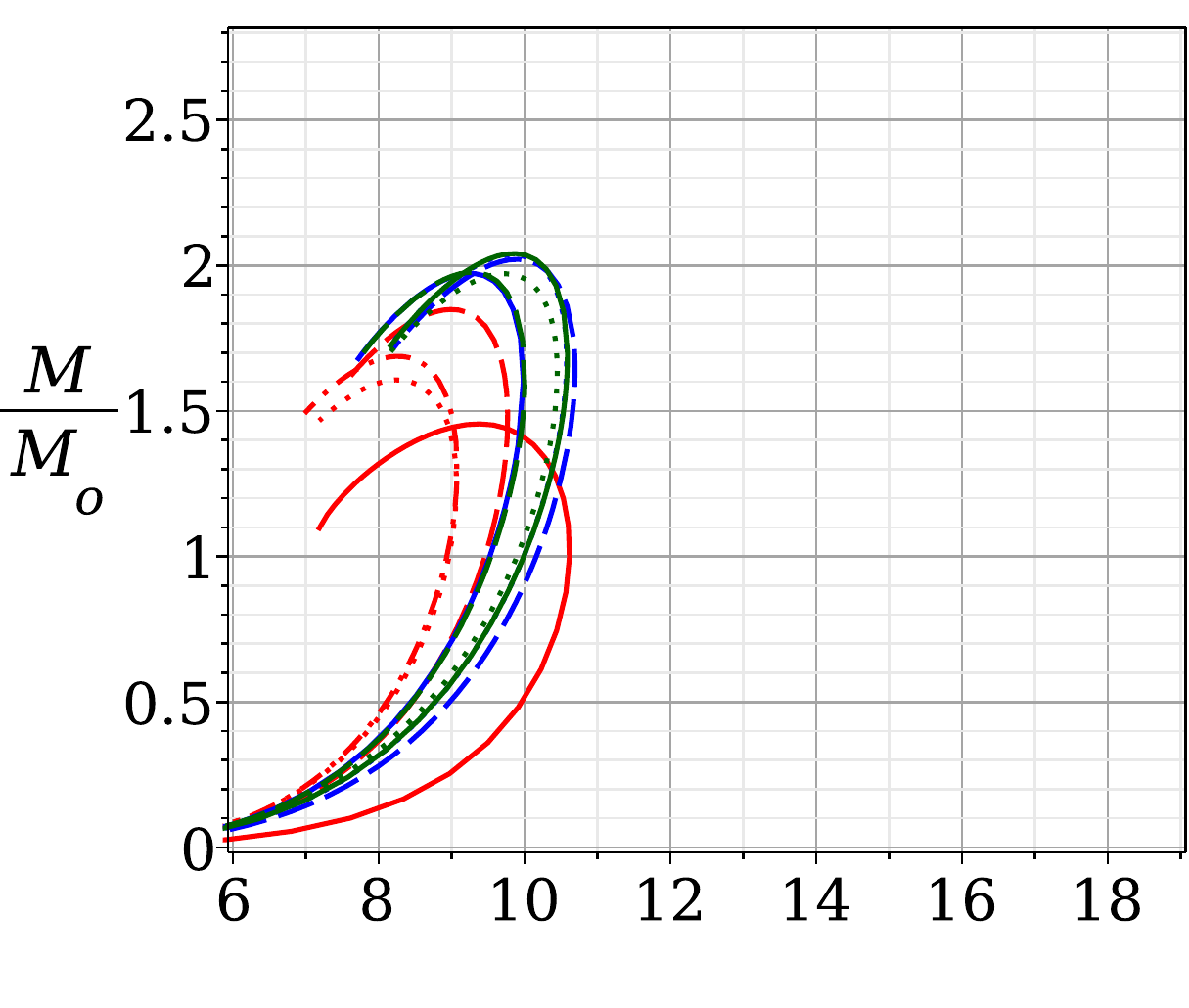} & 	\includegraphics[scale=0.17]{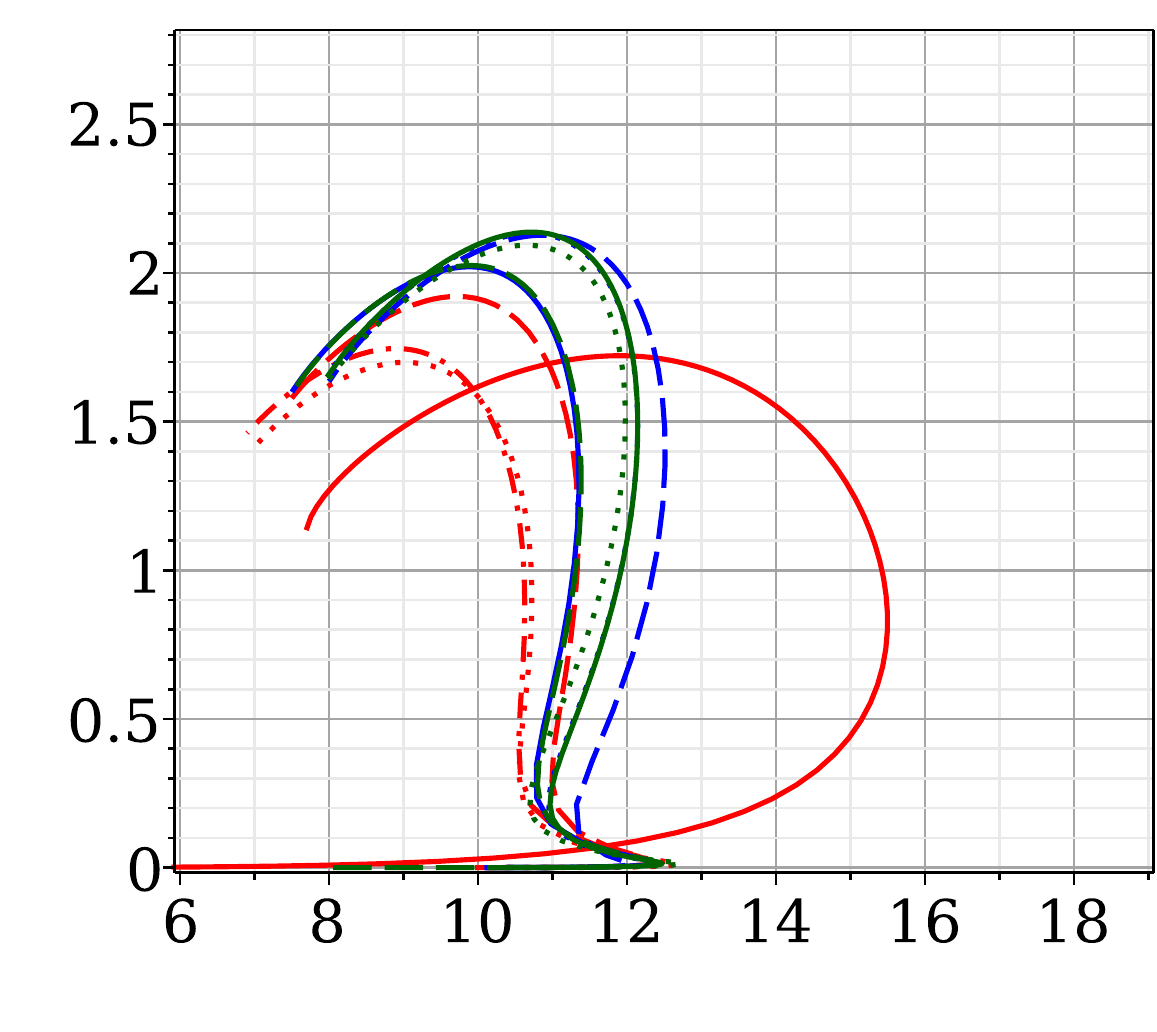} & 	\includegraphics[scale=0.17]{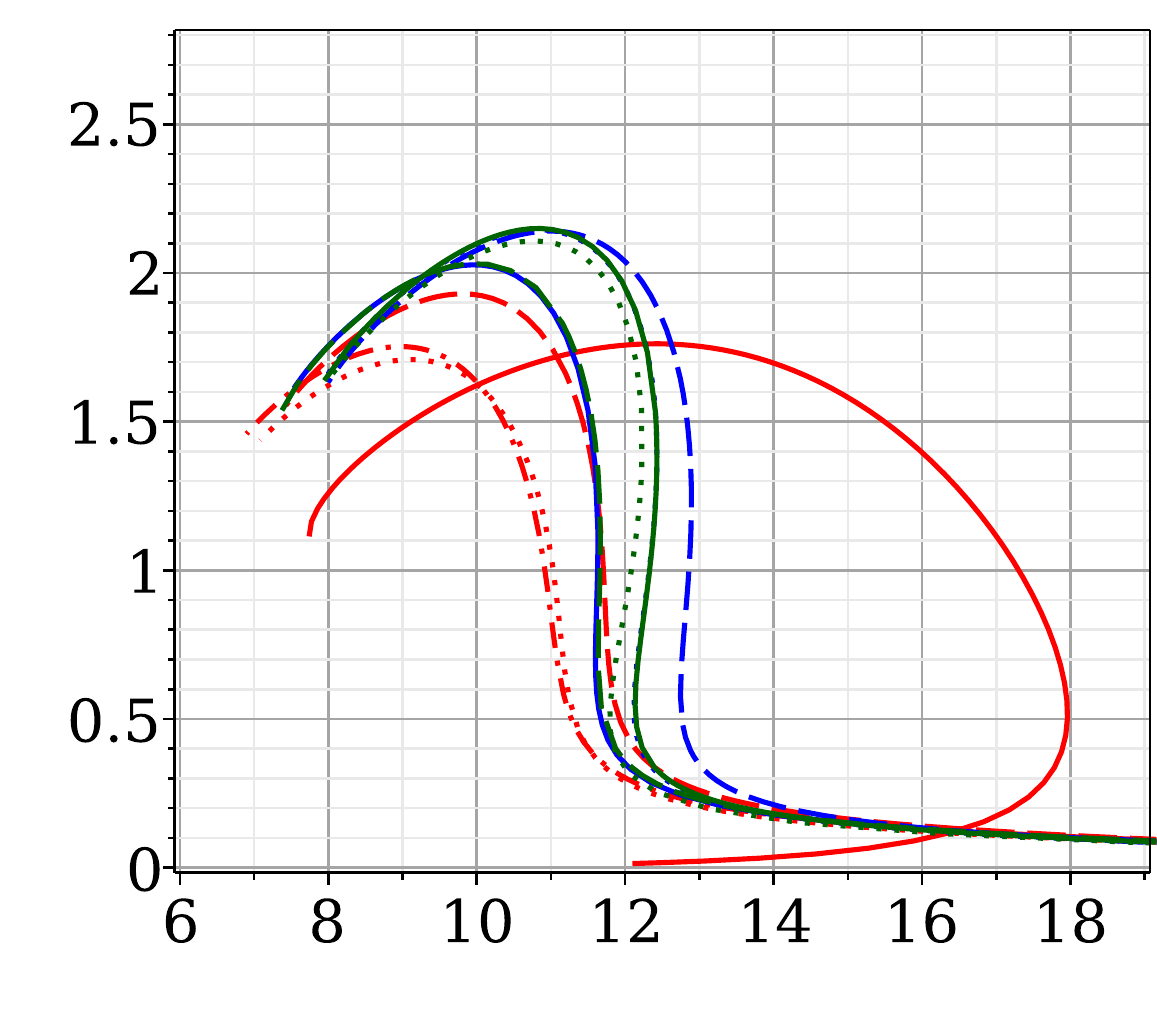} 
& 	\includegraphics[scale=0.17]{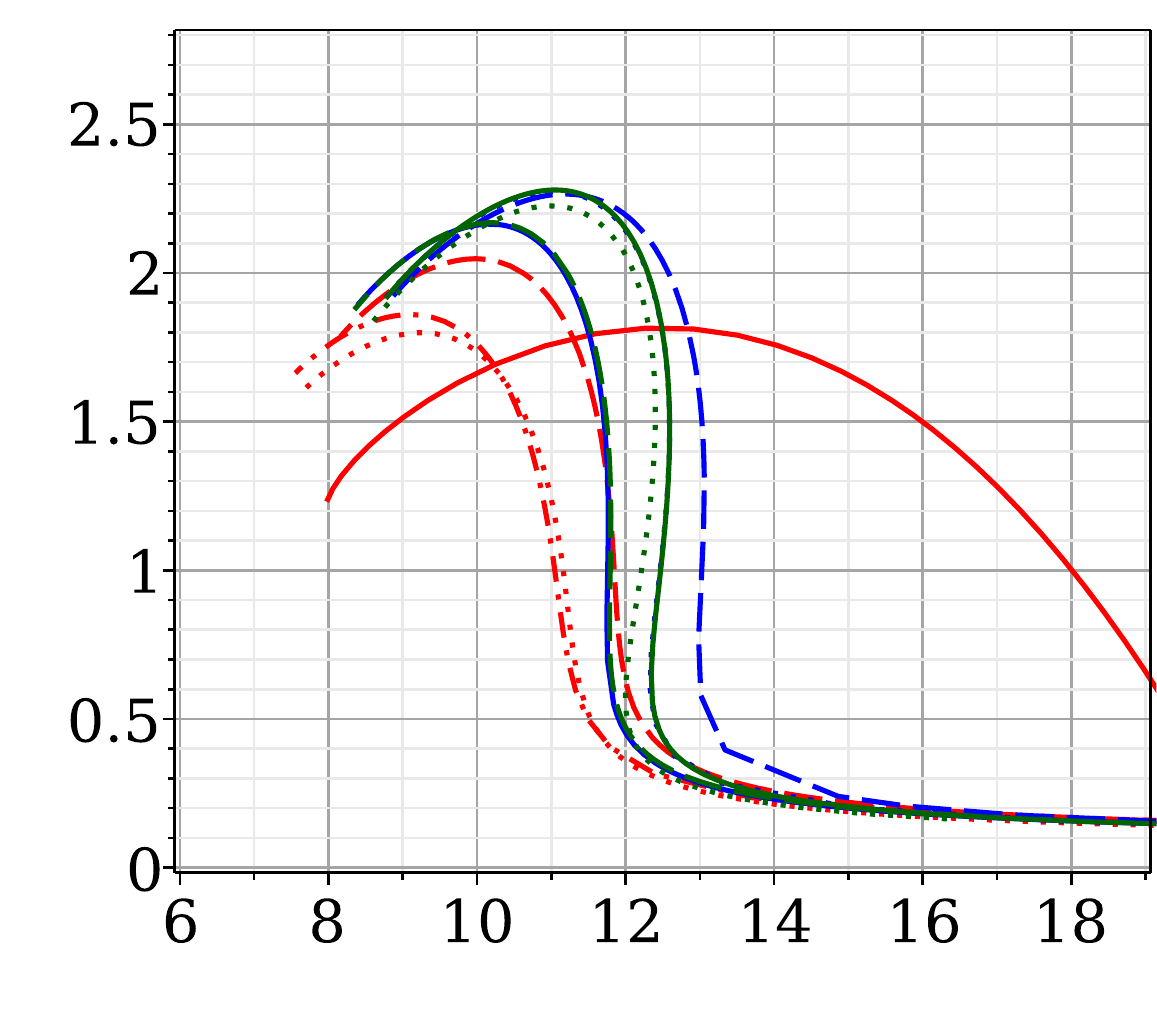}  \\
 \includegraphics[scale=0.17]{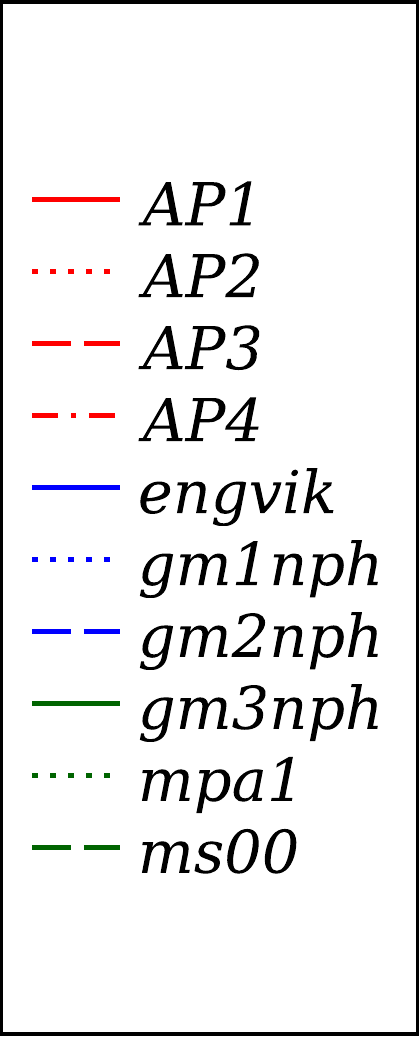}&\includegraphics[scale=0.17]{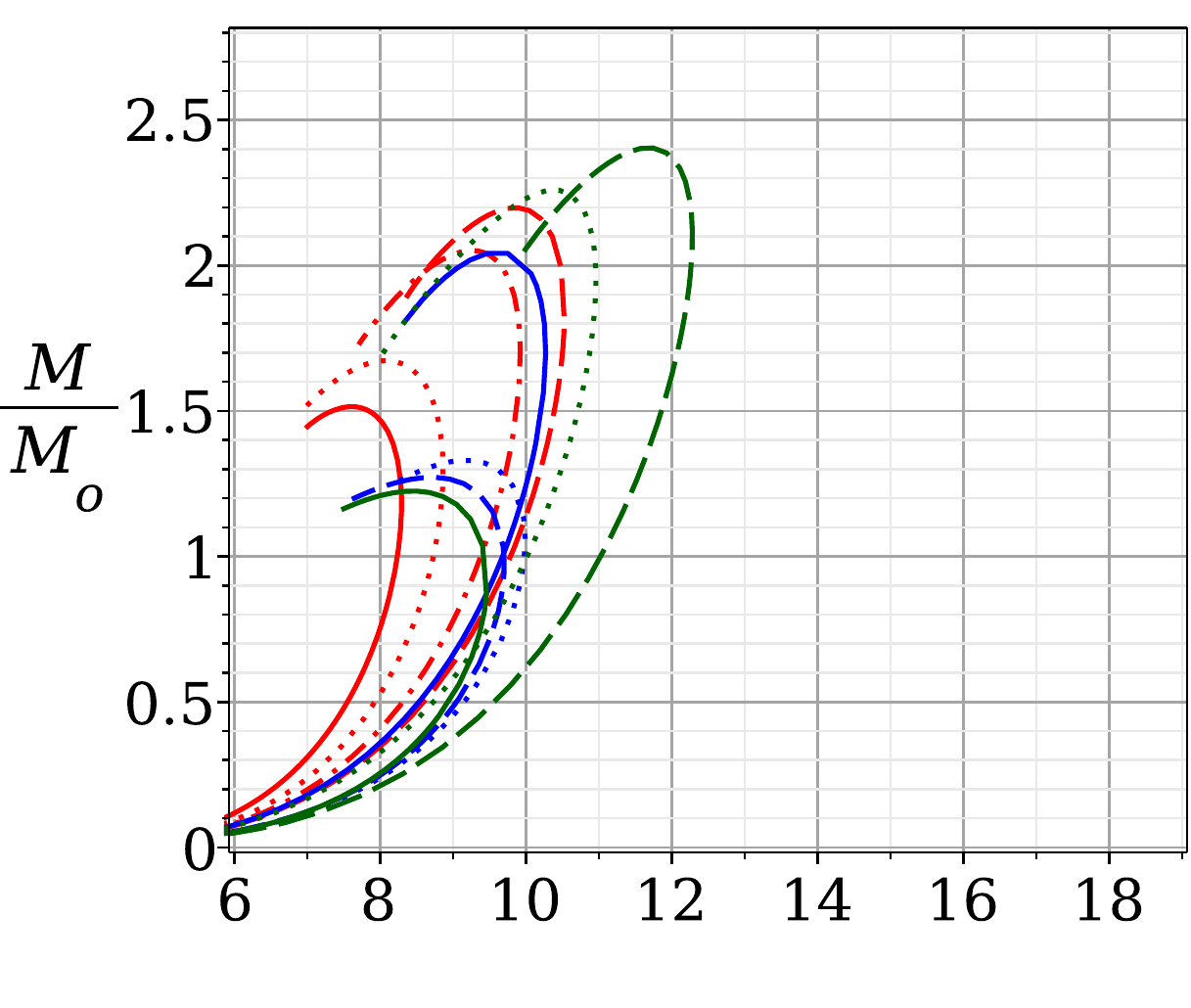} & 	\includegraphics[scale=0.17]{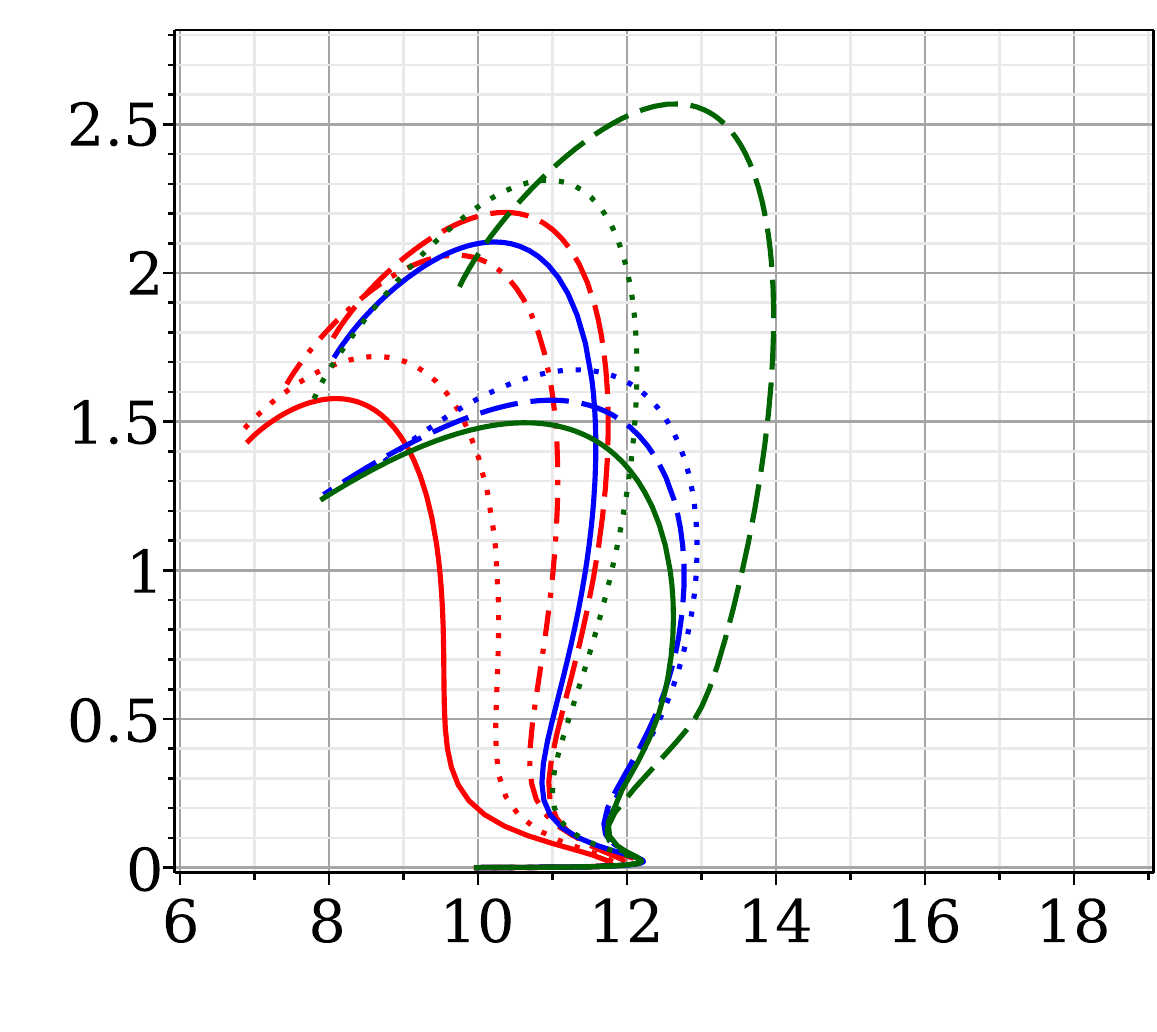} & 	\includegraphics[scale=0.17]{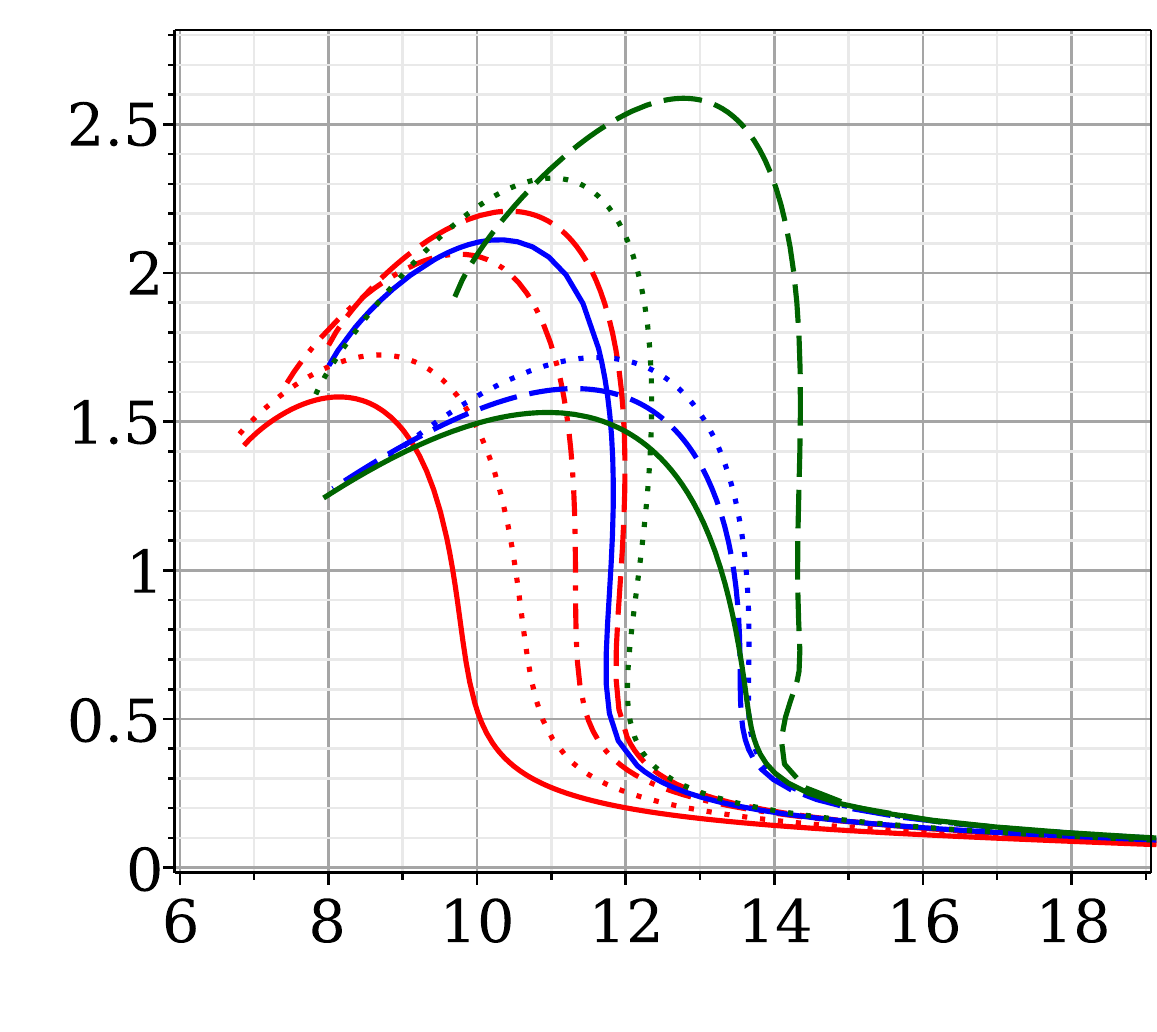}& 	\includegraphics[scale=0.17]{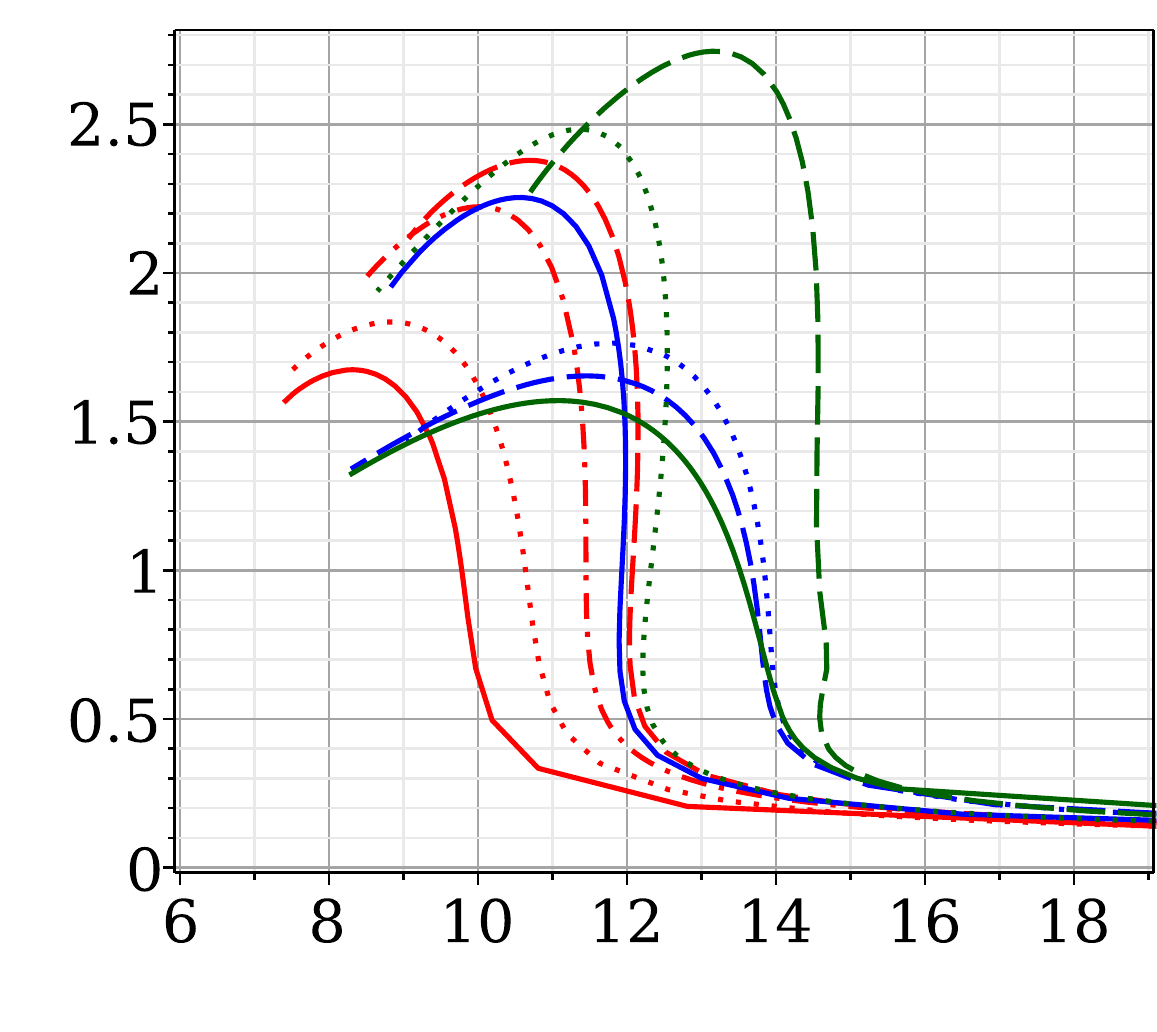}  \\
	\includegraphics[scale=0.17]{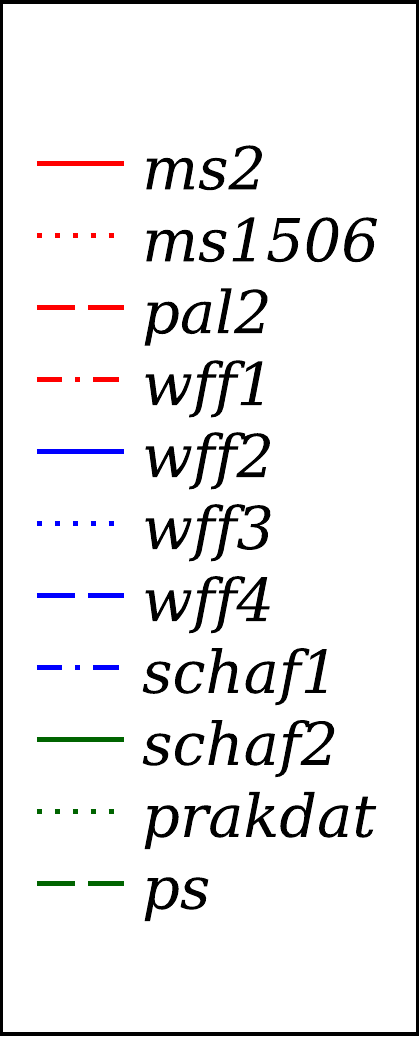}&\includegraphics[scale=0.17]{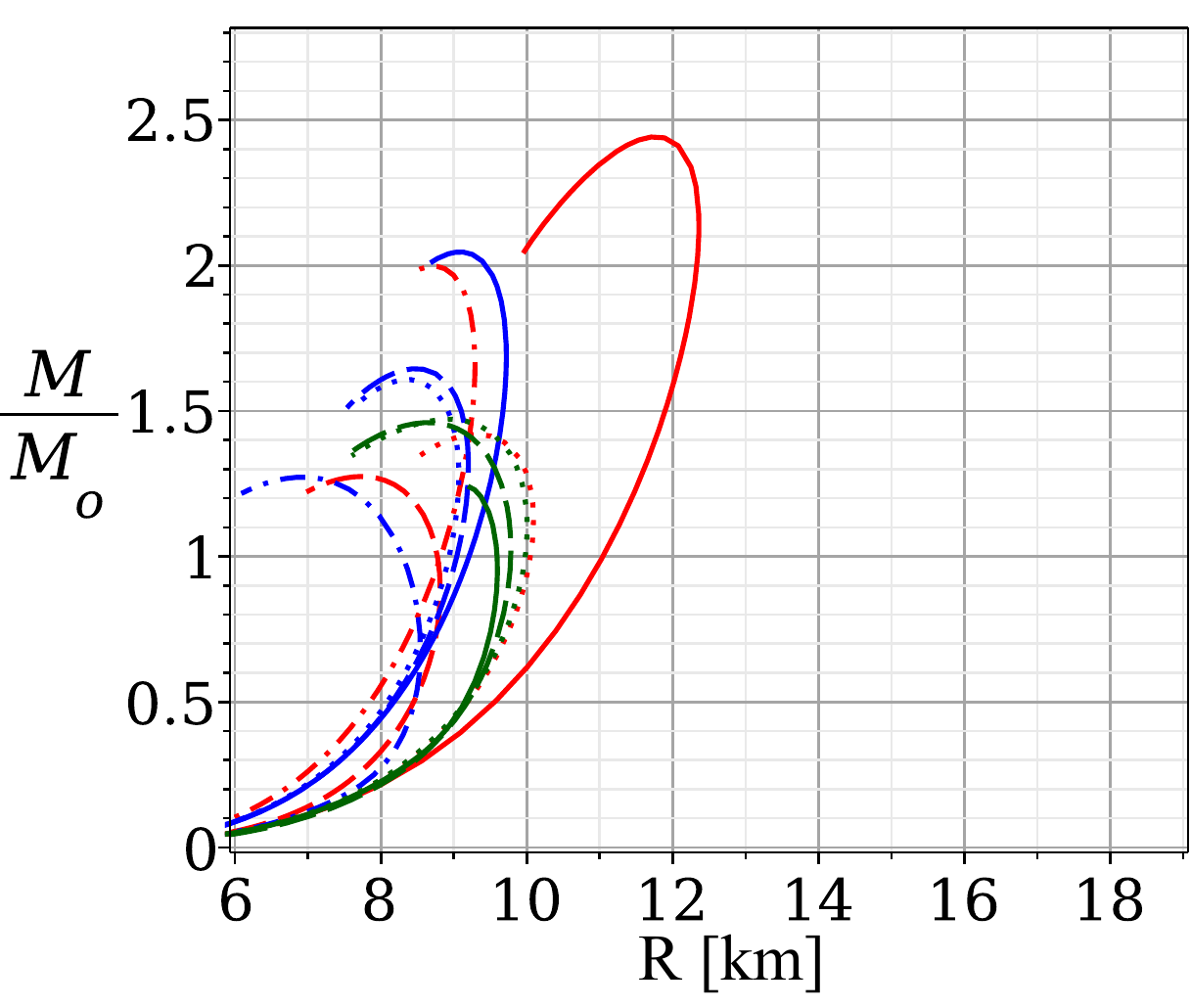} & 	\includegraphics[scale=0.17]{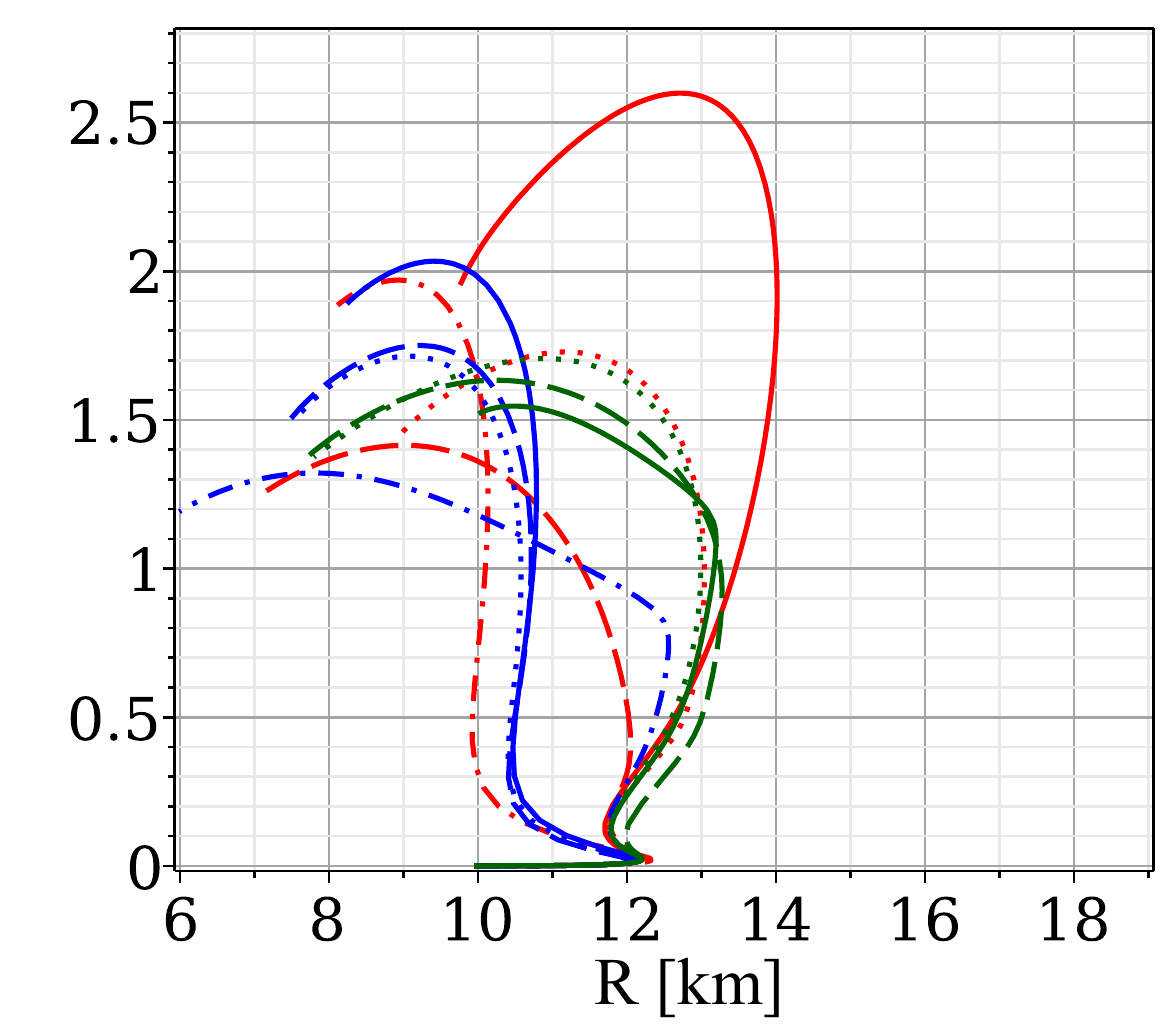} & 	\includegraphics[scale=0.17]{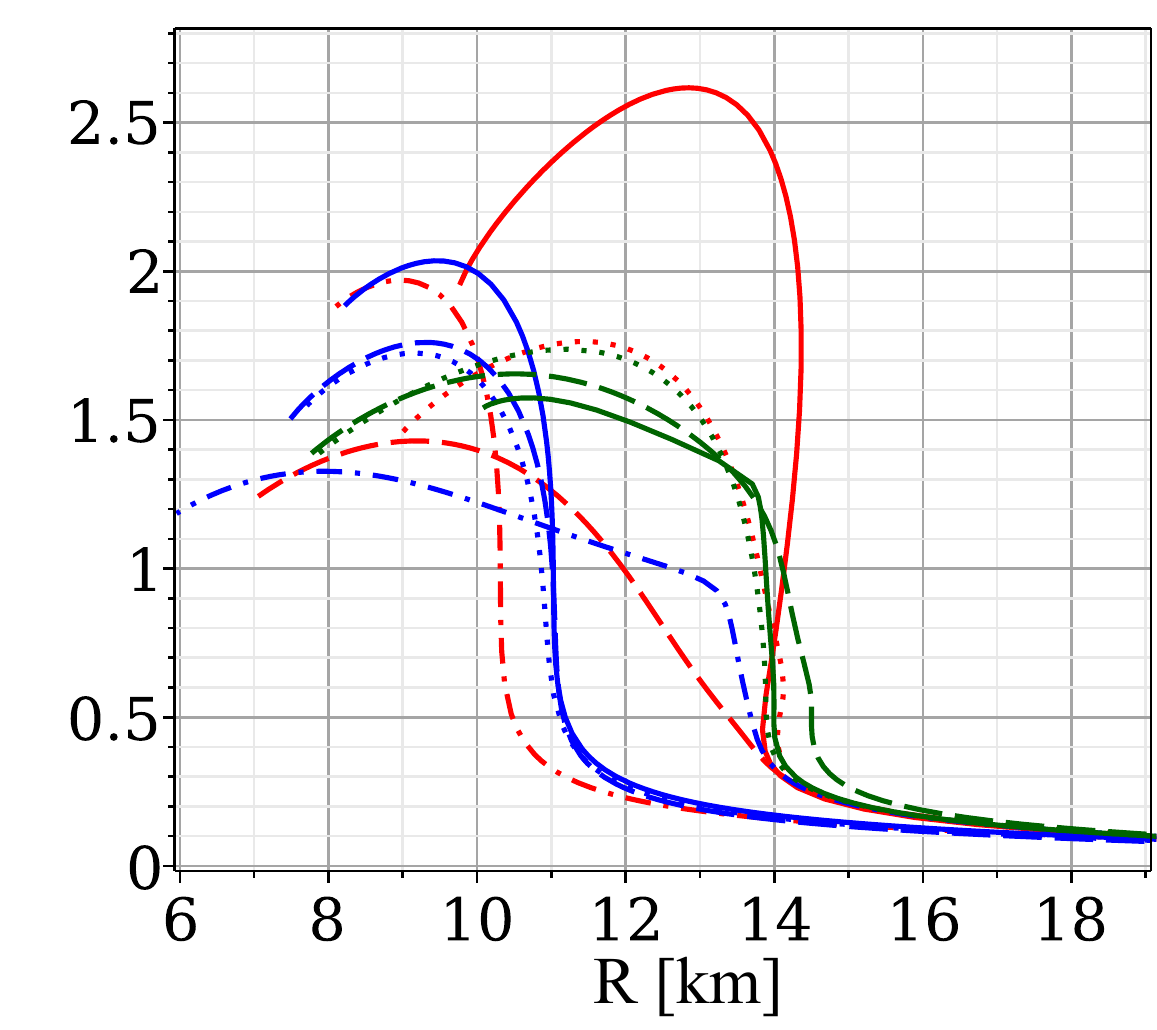} & 	\includegraphics[scale=0.17]{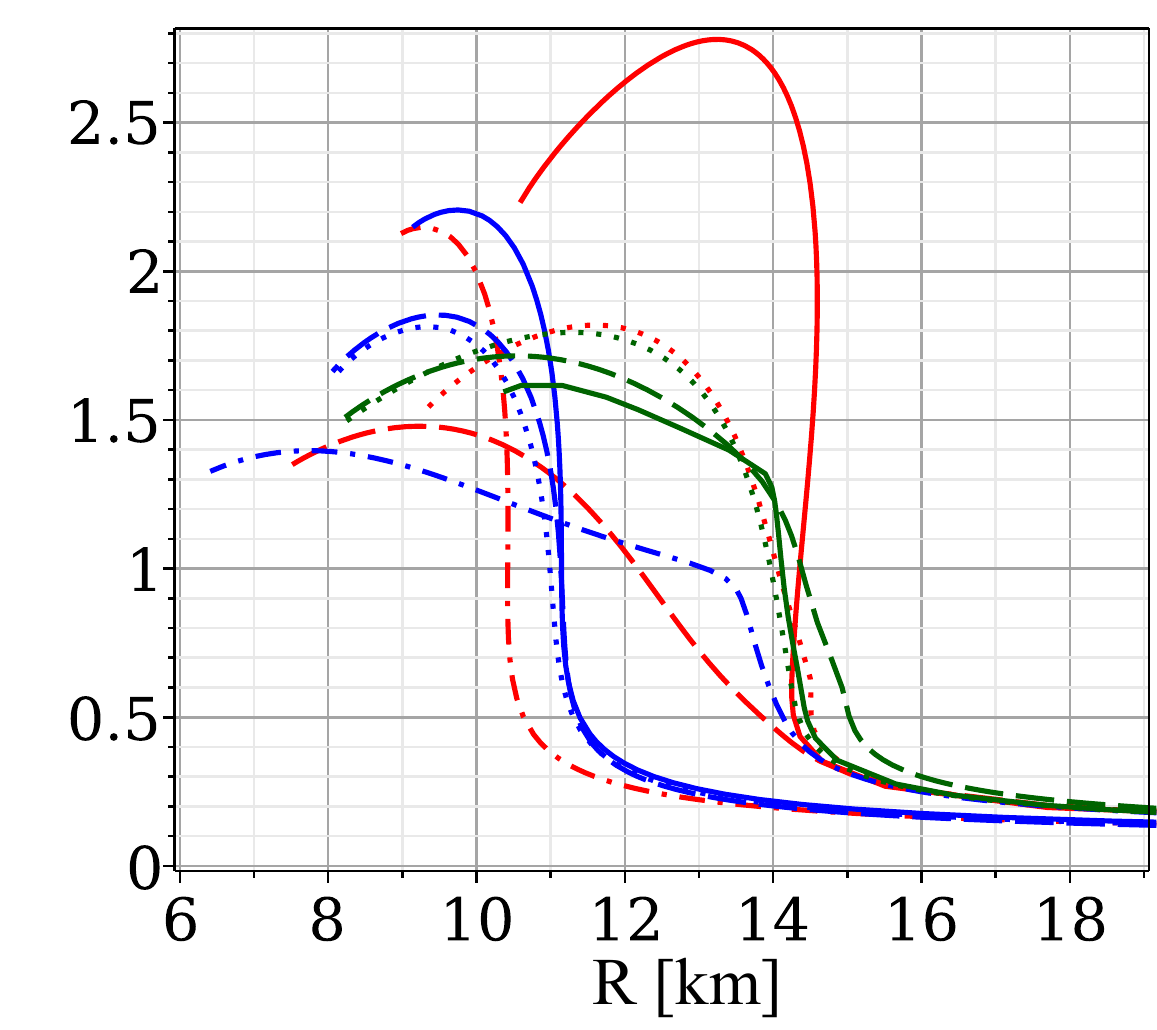} 
	\end{tabular} \end{center}	
\caption{\label{MRxi1}The mass-radius diagram for different equations of state in the case $\xi=-1$ and $\ell=10$, 40 and 100 km. 
The left column of pictures corresponds to $\ell=10$ km, the next column corresponds to $\ell=40$ km and the third column corresponds to  $\ell=100$ km, the right column  corresponds to unmodified theory of gravity (GR). 
The top row of figures corresponds to the polytrope equation of state, FPS, SLY, BSk19-22, BSk24-26  equations of state, 
the middle row corresponds to AP1-4, engvik, gm1nph, gm2nph, gm3nph, mpa1, ms00, 
the bottom row corresponds to ms2, ms1506, pal2, wff1, wff2, wff3, wff4, schaf1, schaf2, prakdat, ps. 
The correspondence between the equation of state and the curve style is indicated at the left.}
\end{figure}

\begin{figure}[h!]
\begin{center}
\begin{tabular}{l} 
\vspace{-0.4cm}\includegraphics[scale=0.15]{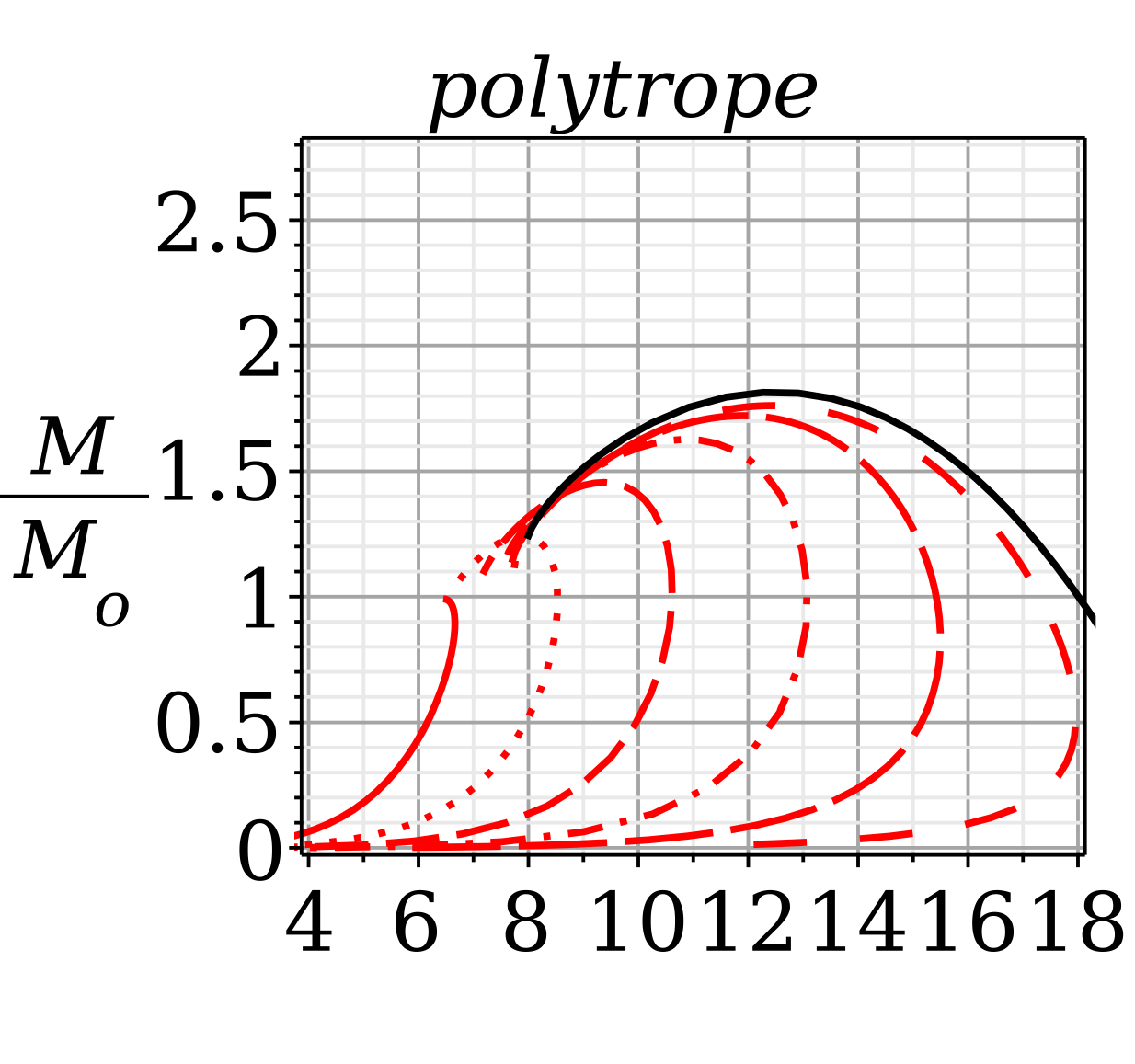} 	\includegraphics[scale=0.15]{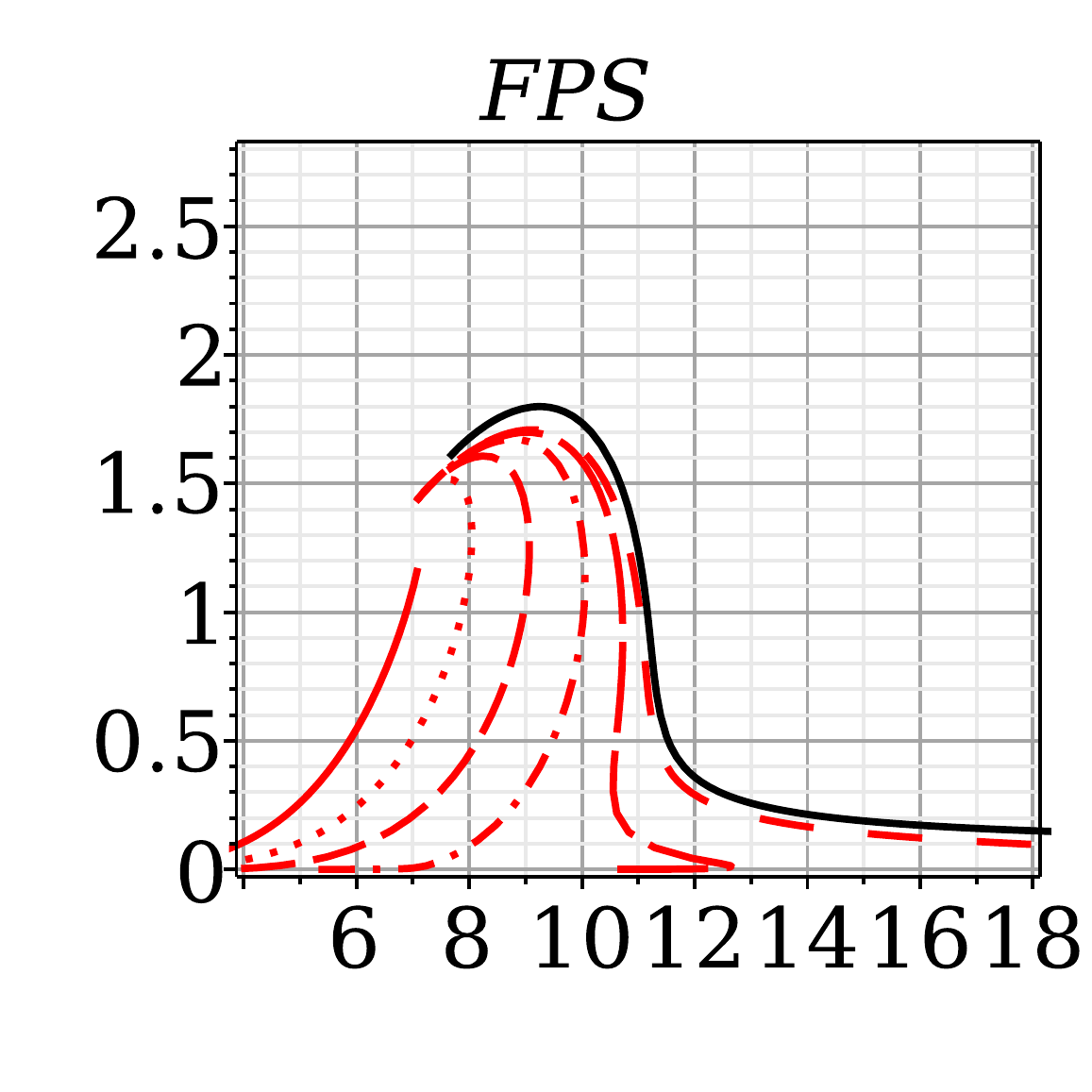}  	\includegraphics[scale=0.15]{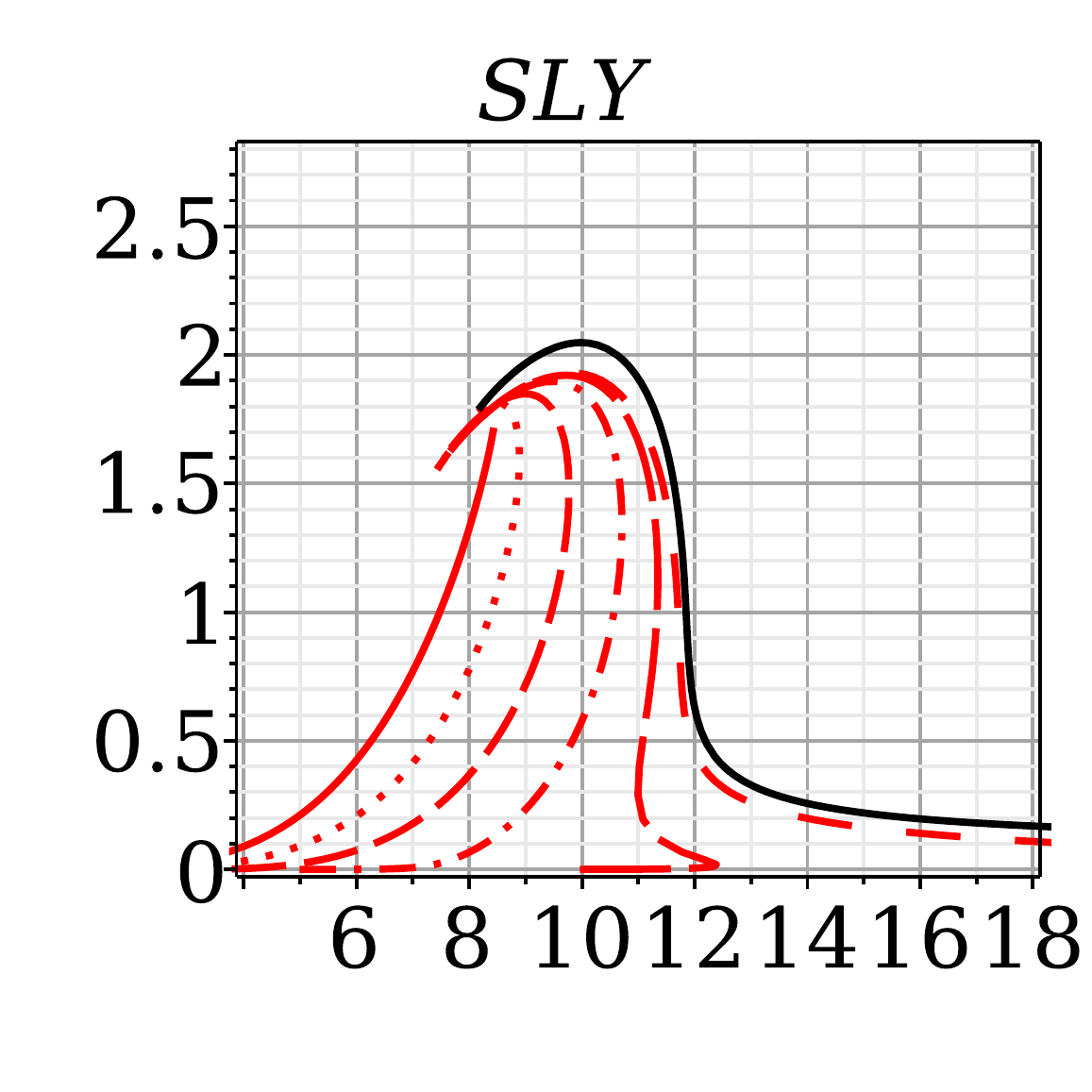}  
\includegraphics[scale=0.15]{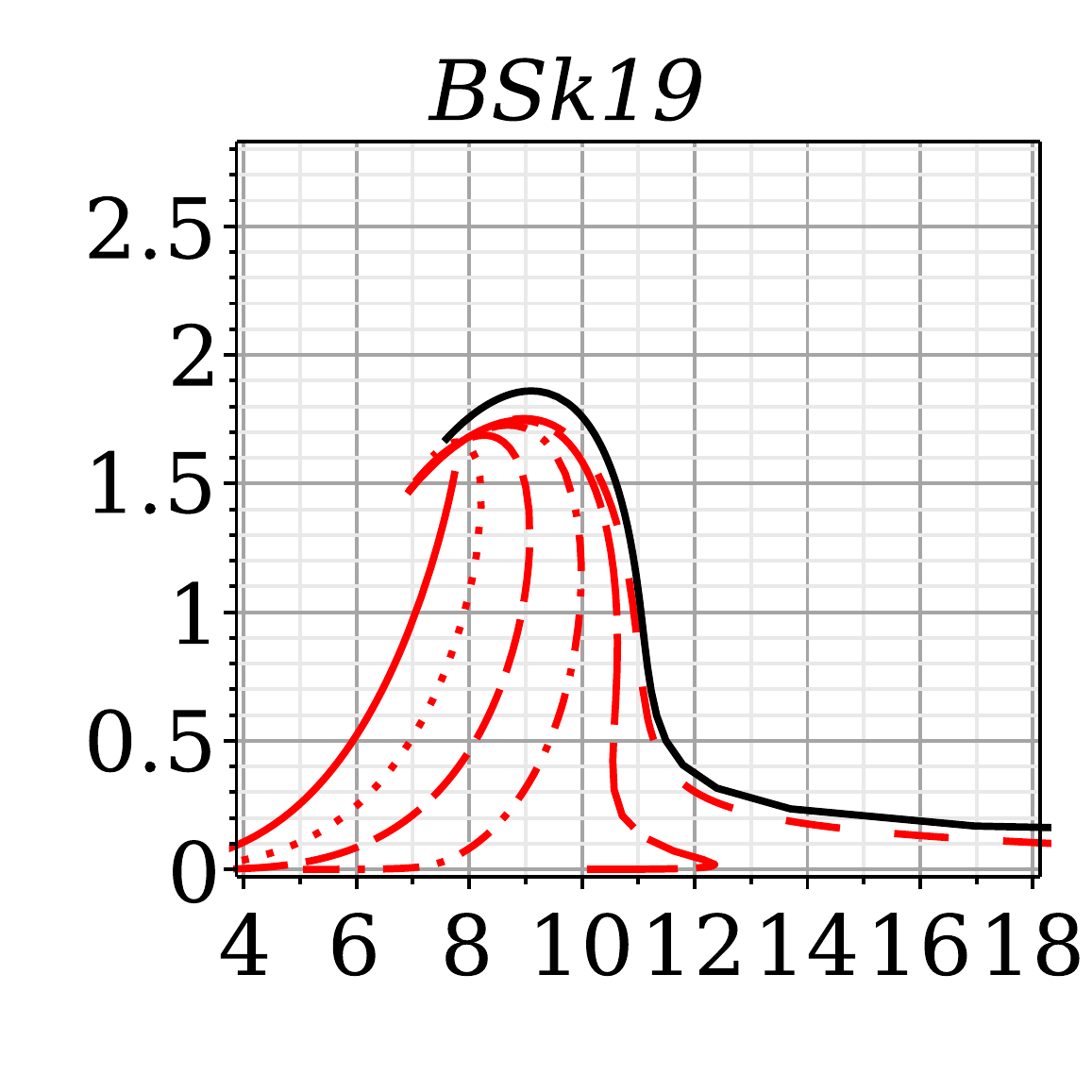}  \includegraphics[scale=0.15]{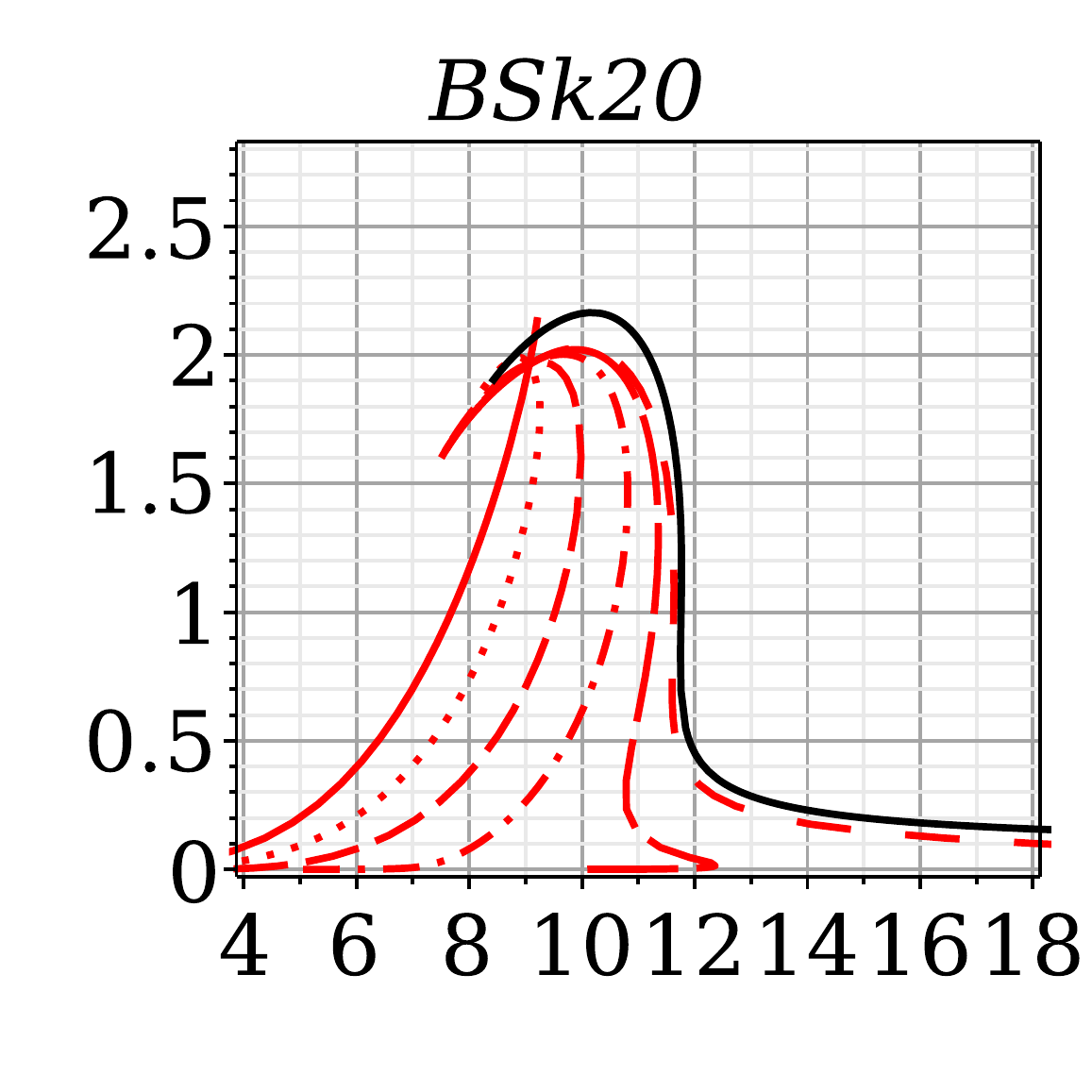} \\ \vspace{-0.4cm}
\includegraphics[scale=0.15]{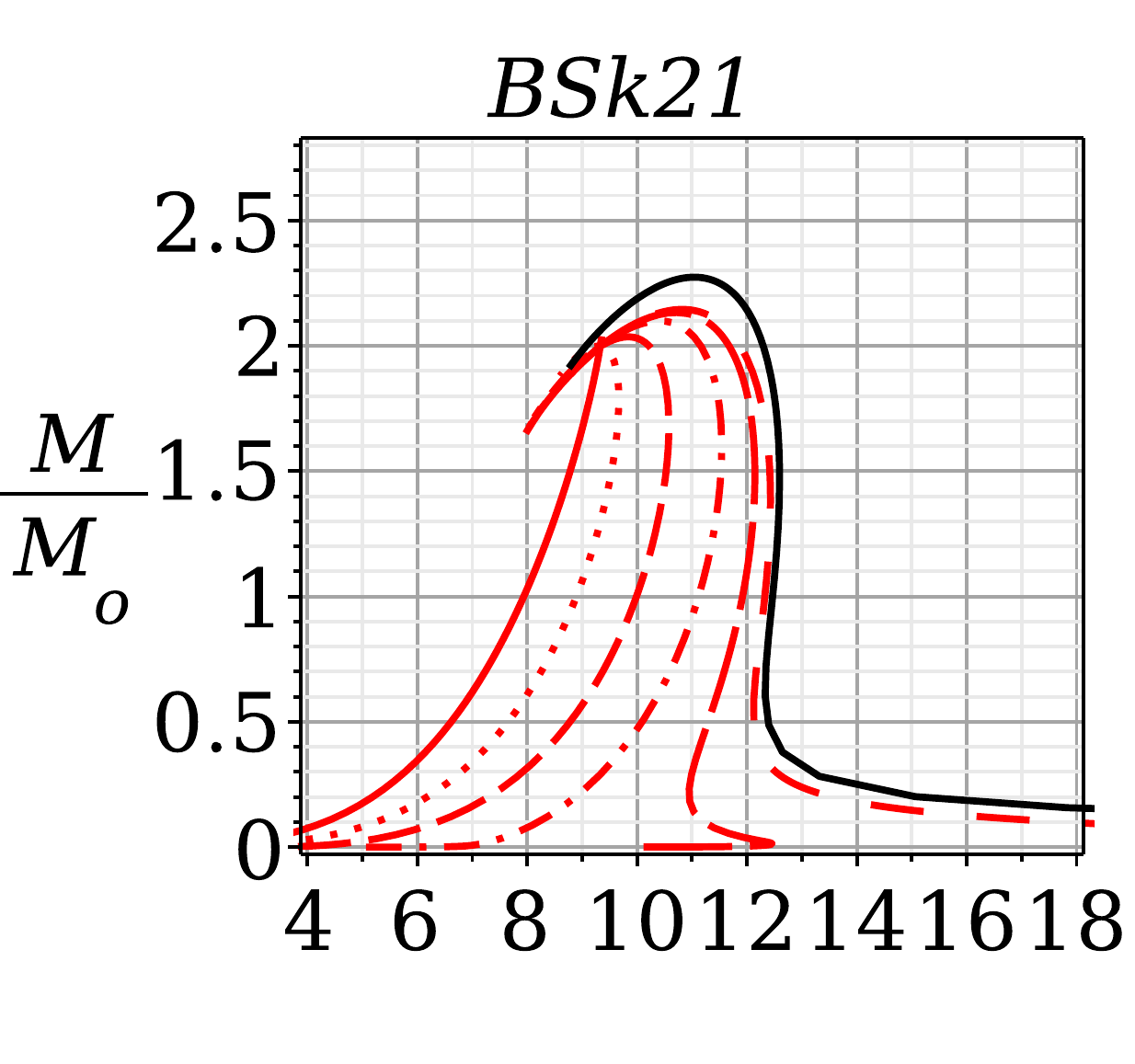} \includegraphics[scale=0.15]{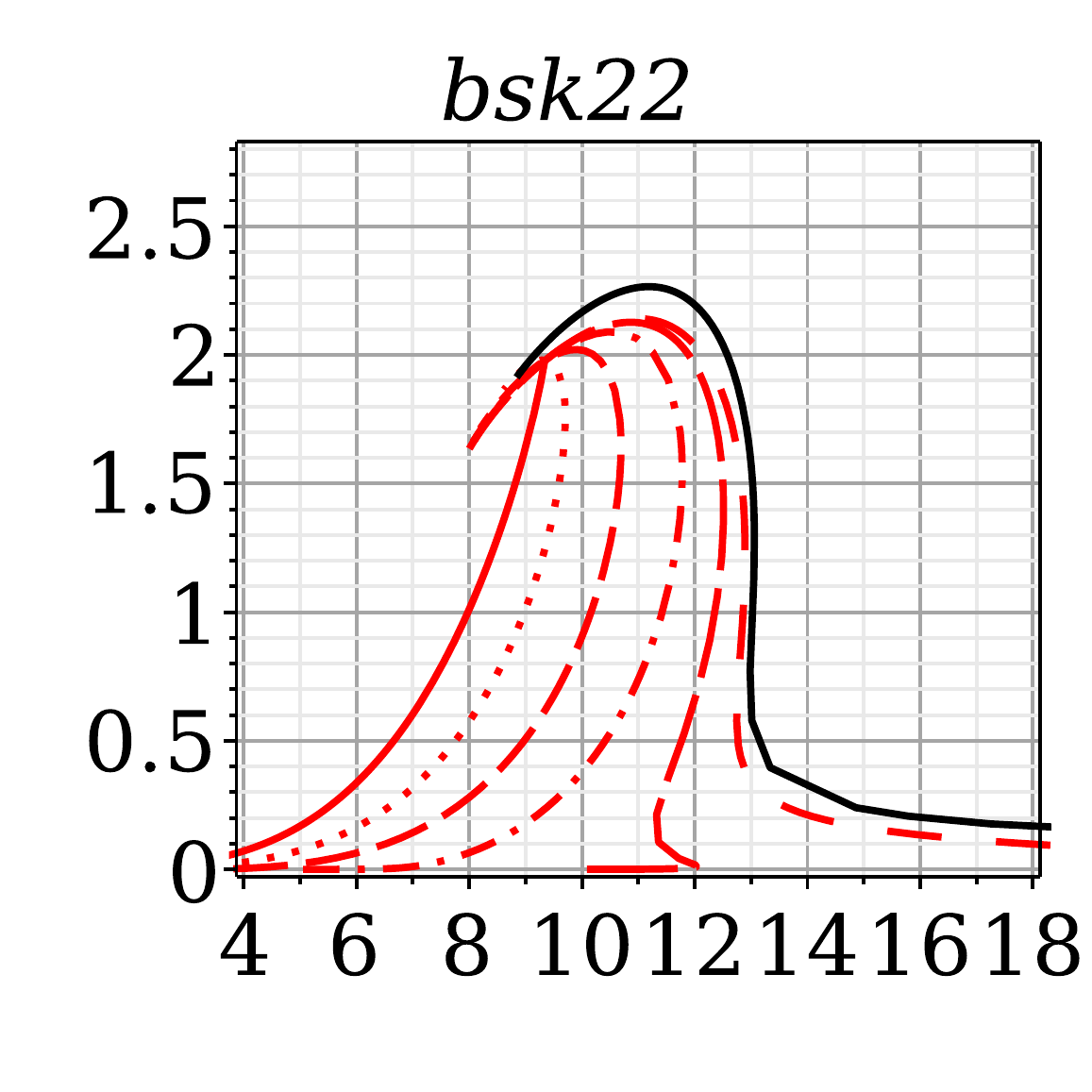} 	\includegraphics[scale=0.15]{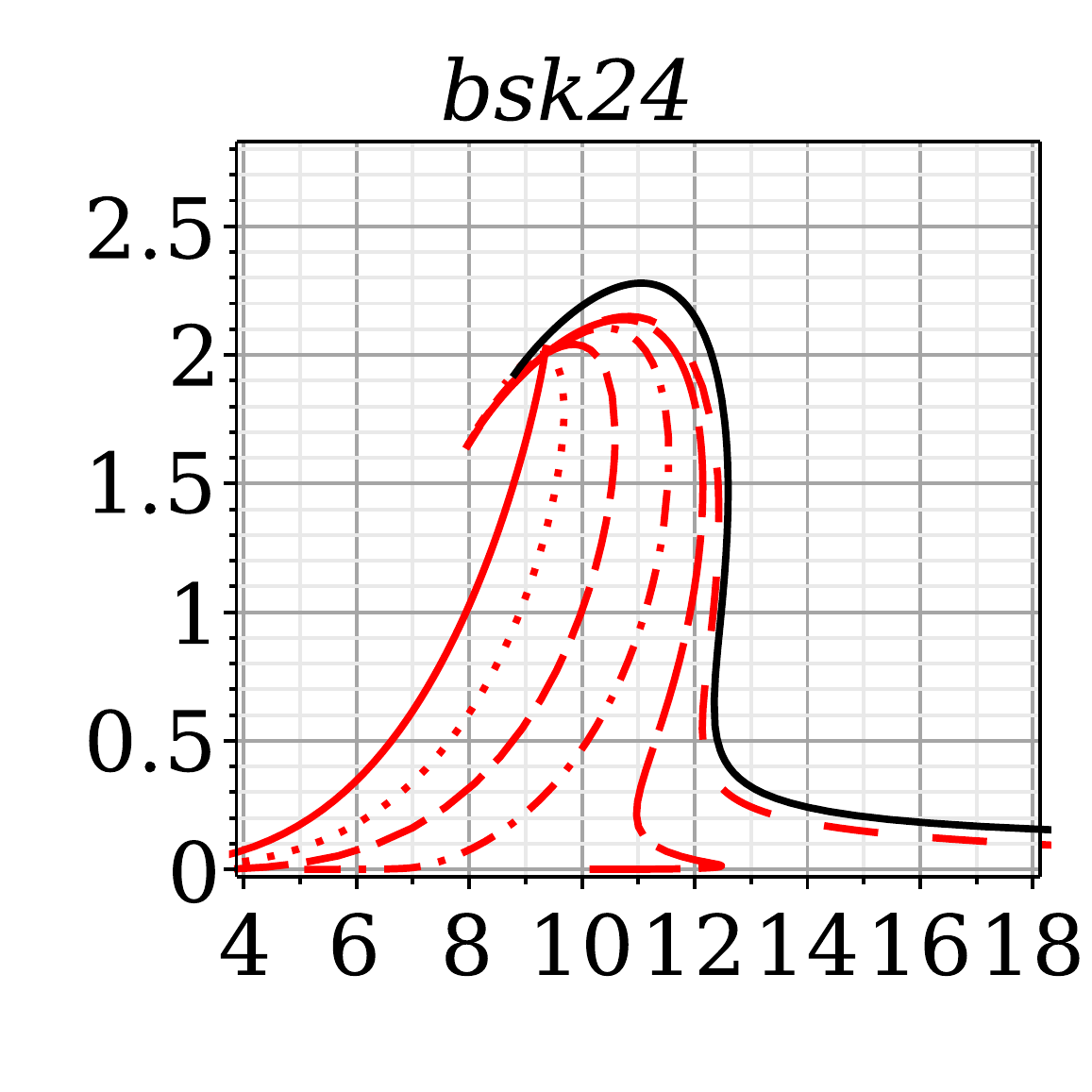}  
\includegraphics[scale=0.15]{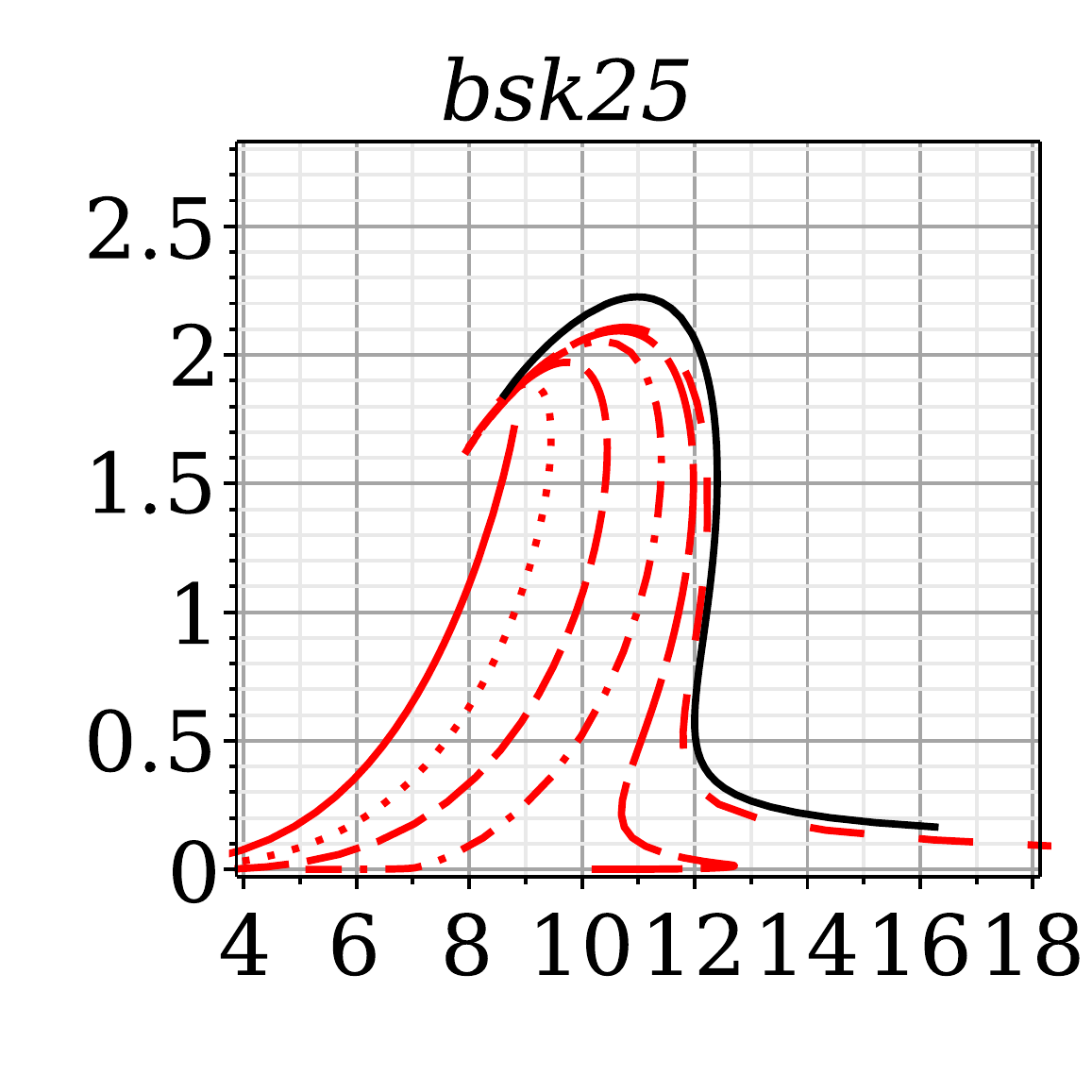}  \includegraphics[scale=0.15]{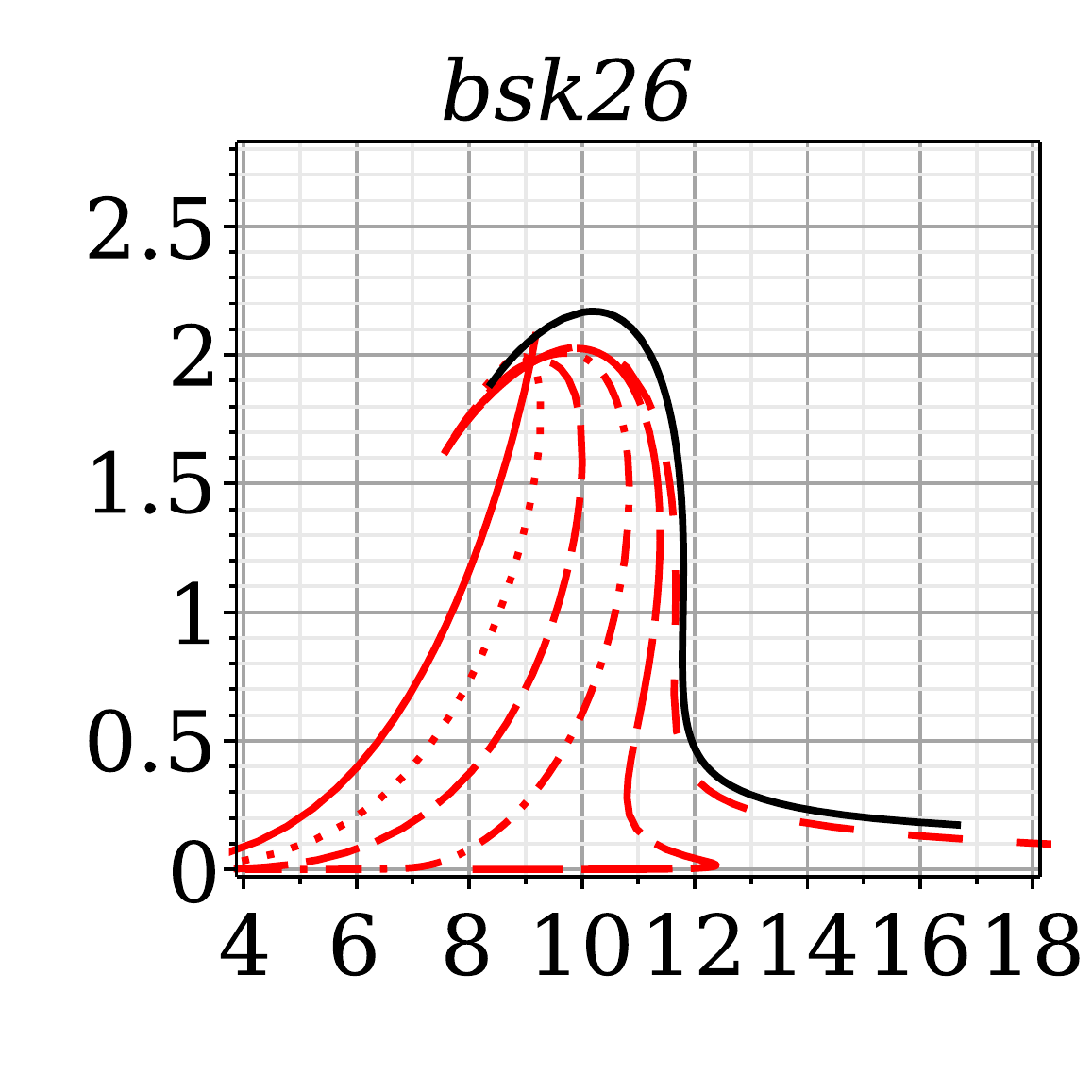} \\  \vspace{-0.4cm}
\includegraphics[scale=0.15]{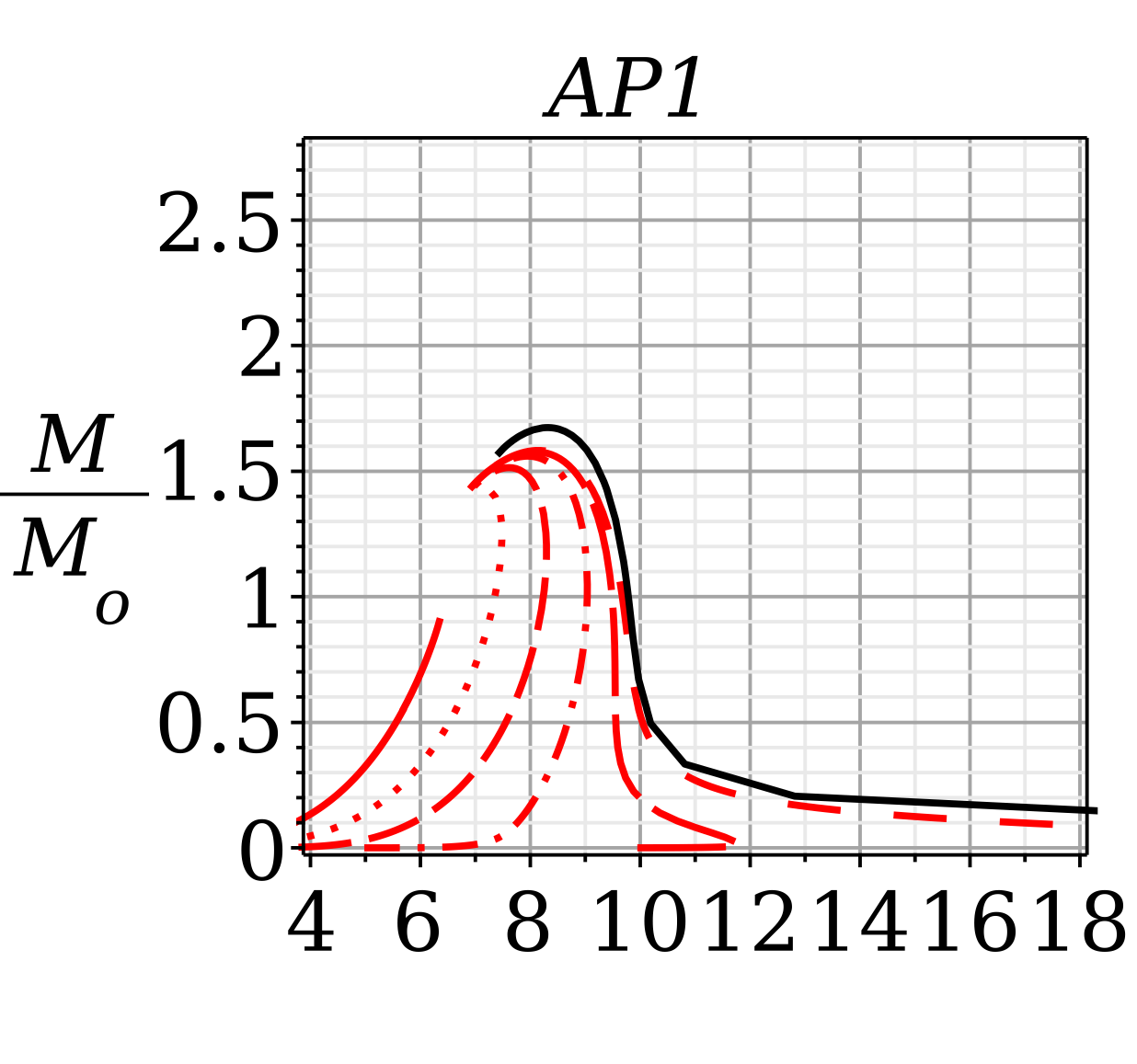}  	\includegraphics[scale=0.15]{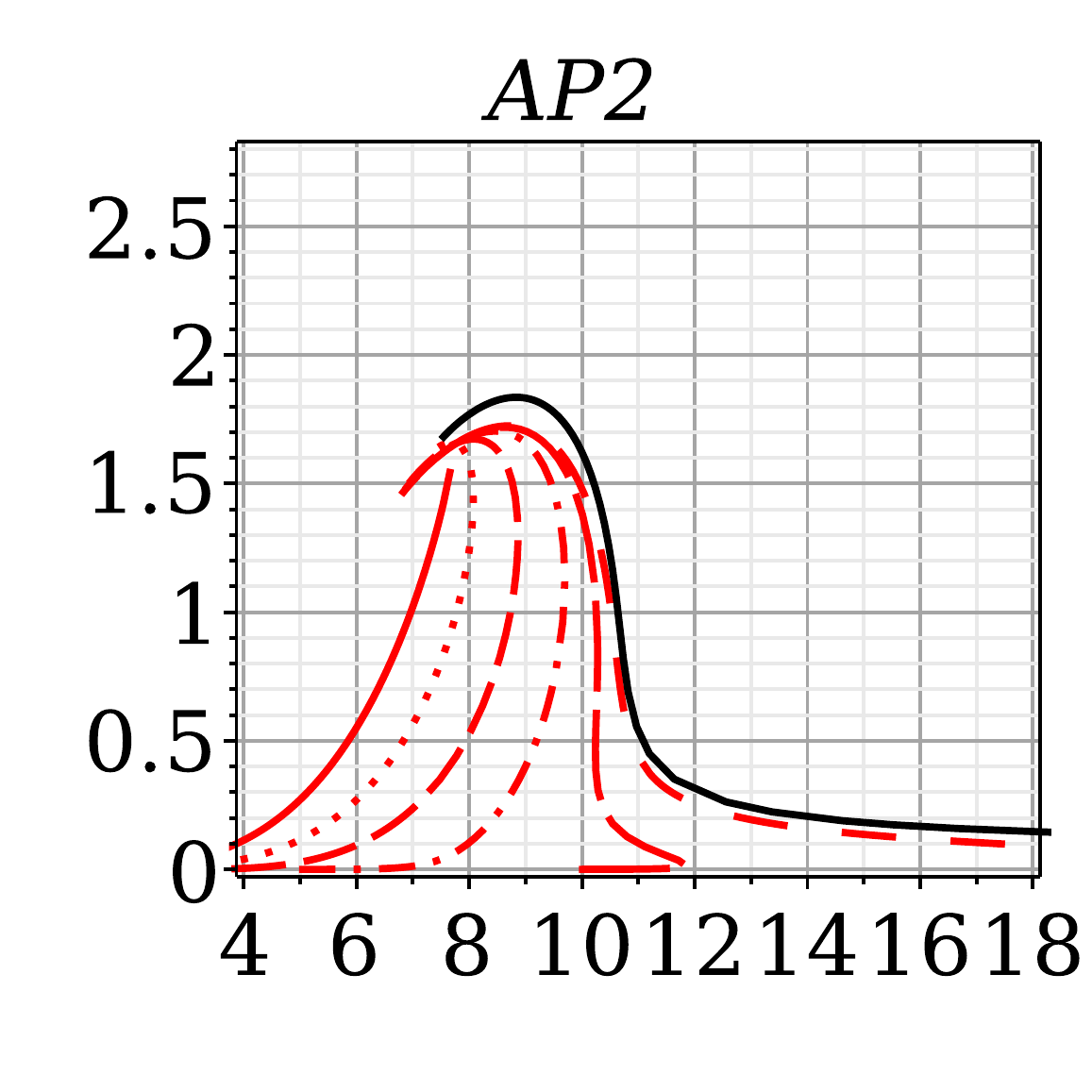} 	\includegraphics[scale=0.15]{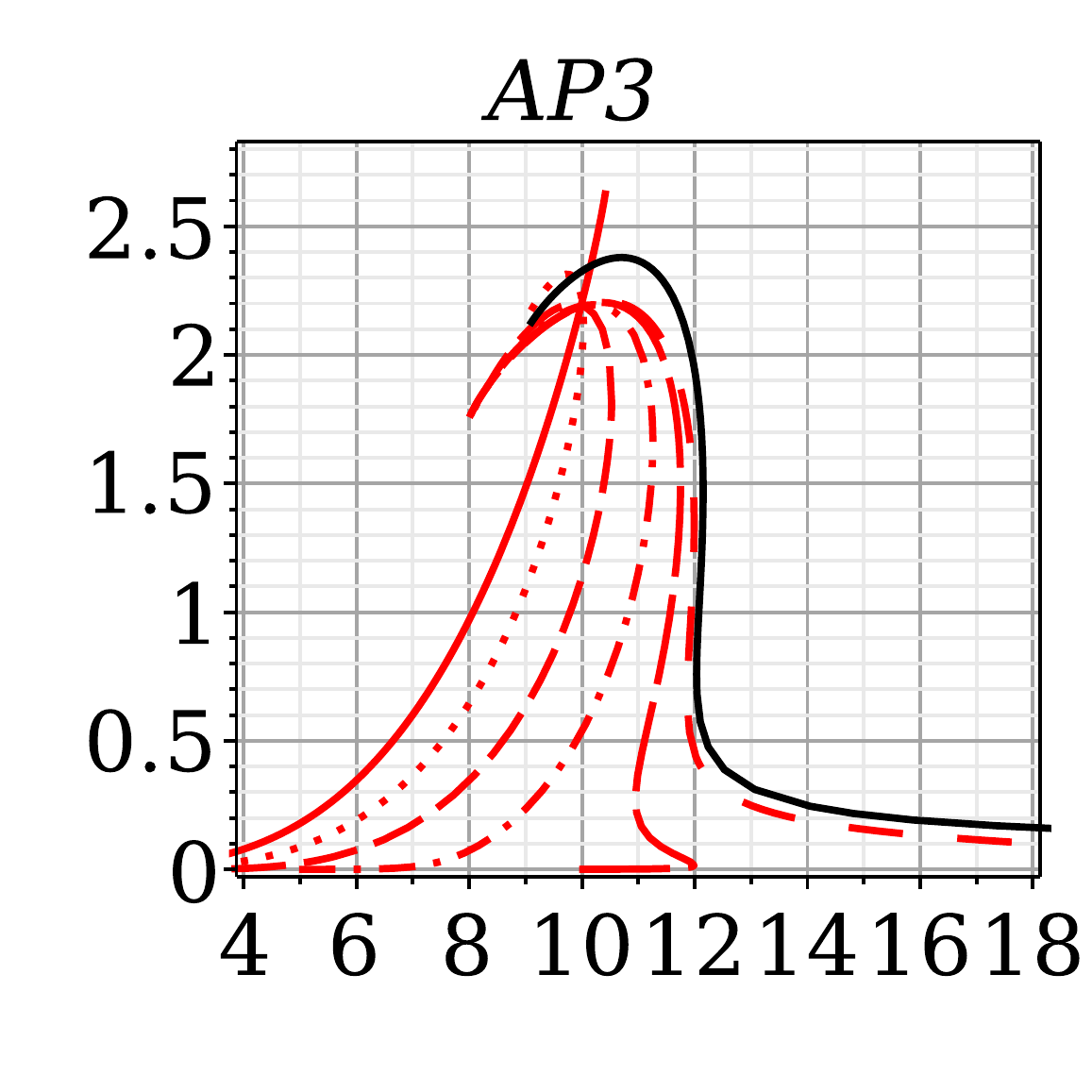}  
\includegraphics[scale=0.15]{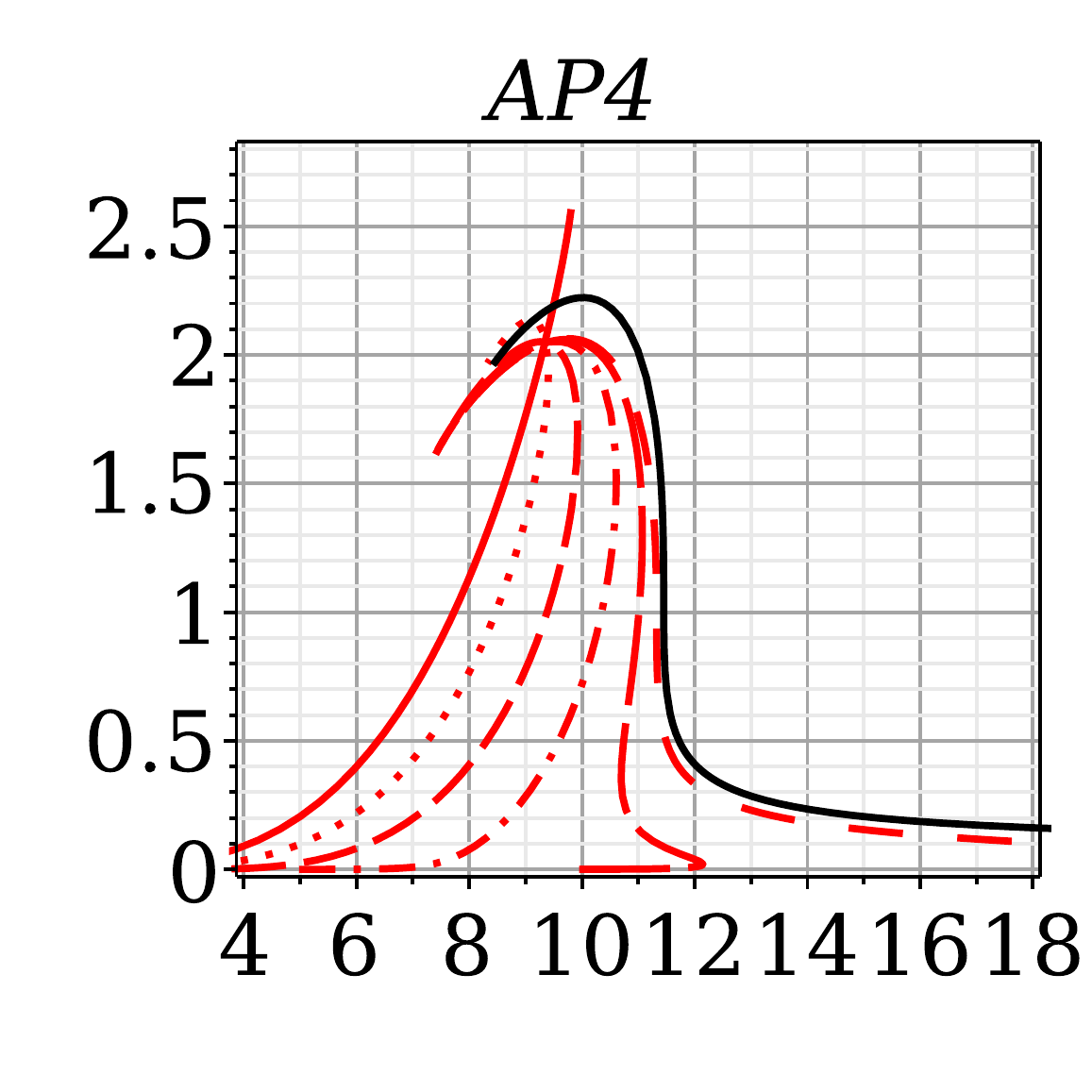}  \includegraphics[scale=0.15]{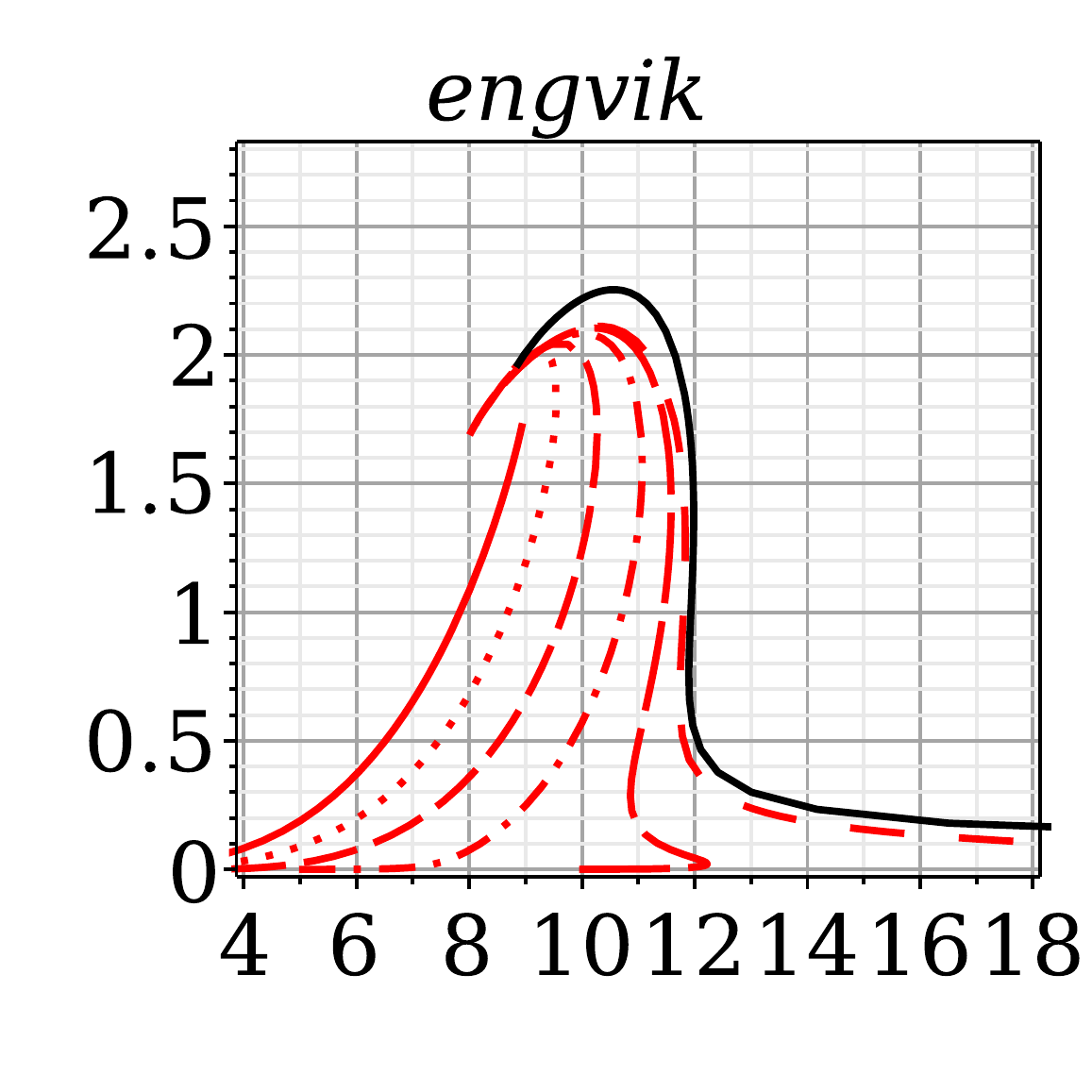} \\ \vspace{-0.4cm}
\includegraphics[scale=0.15]{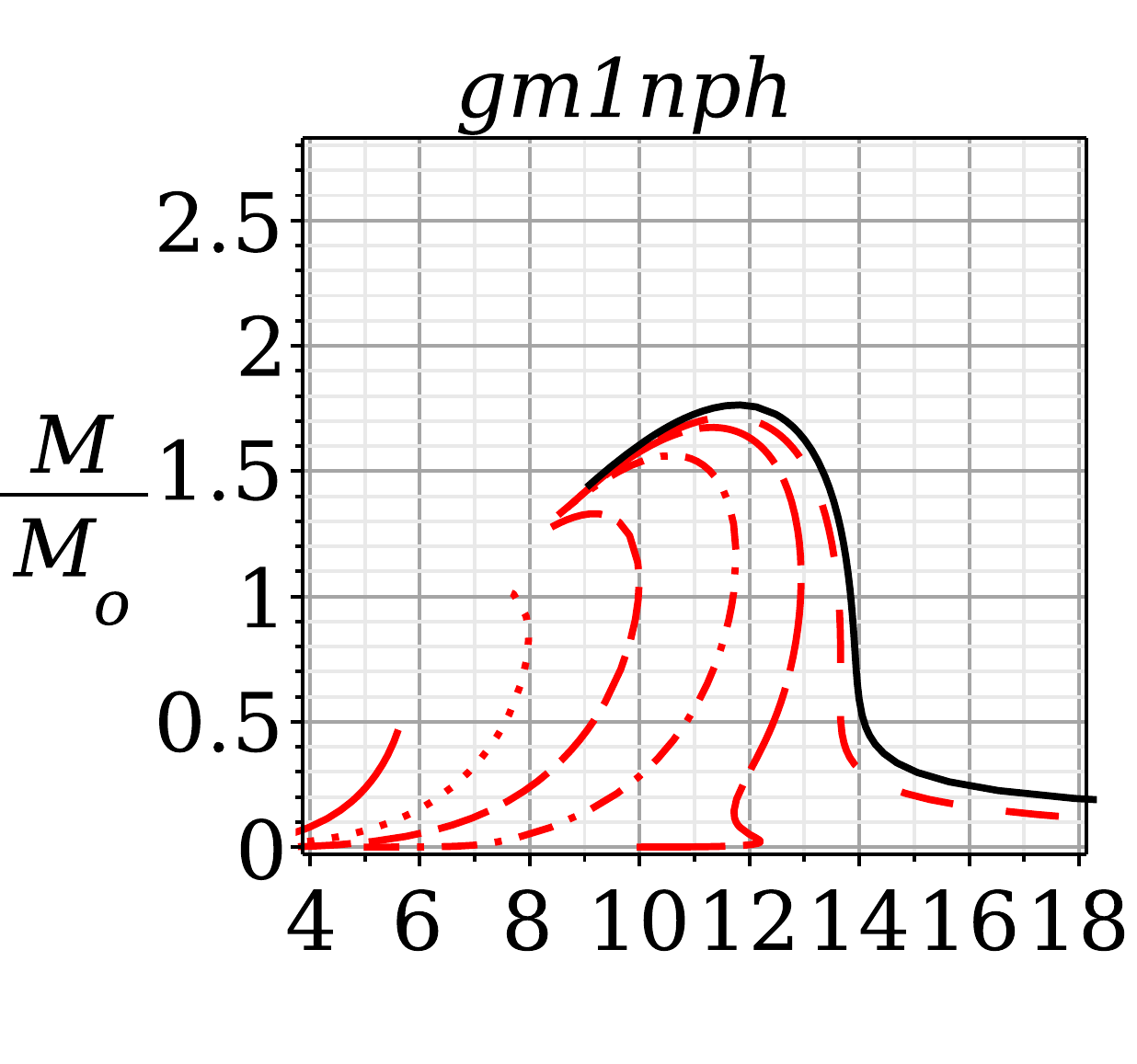} 	\includegraphics[scale=0.15]{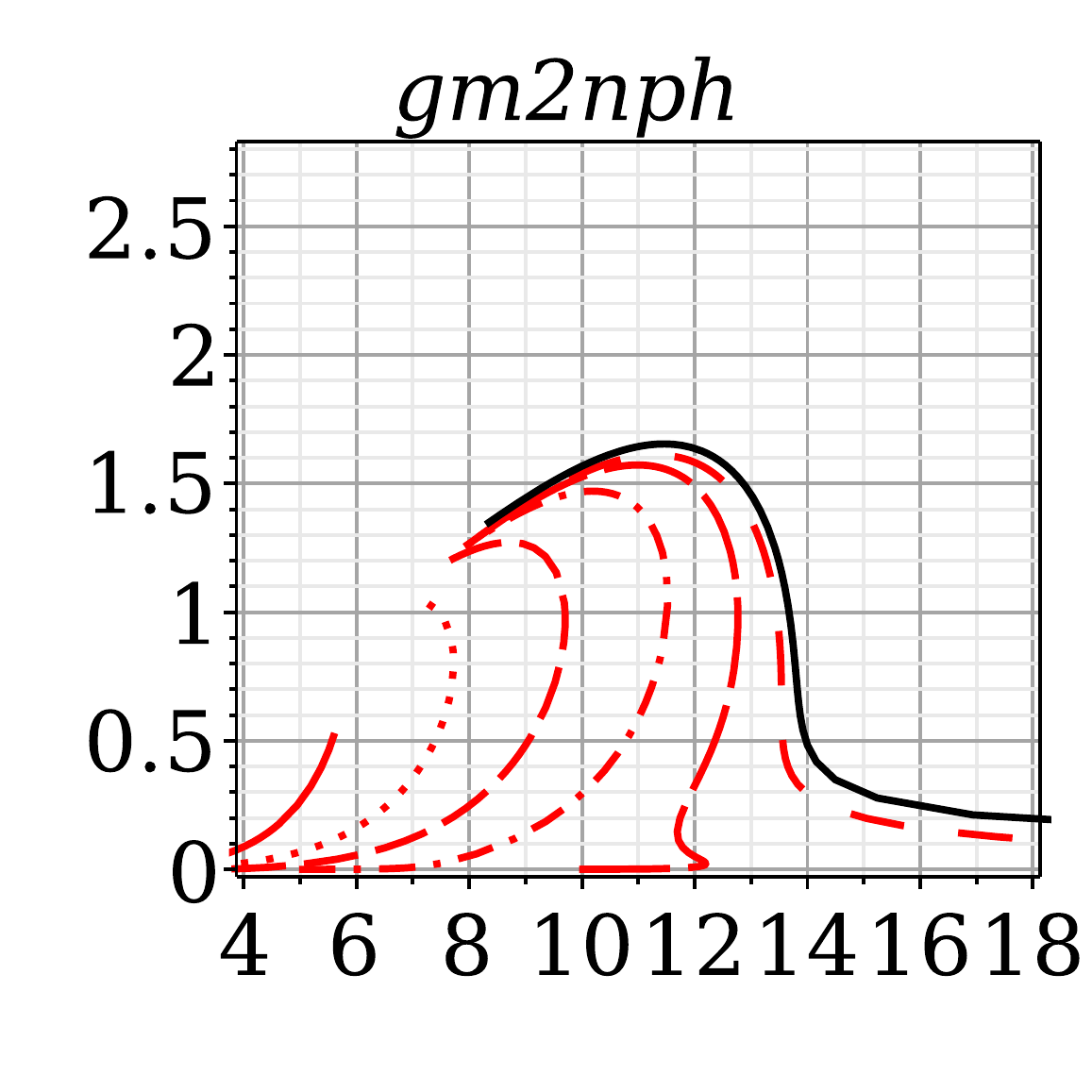} \includegraphics[scale=0.15]{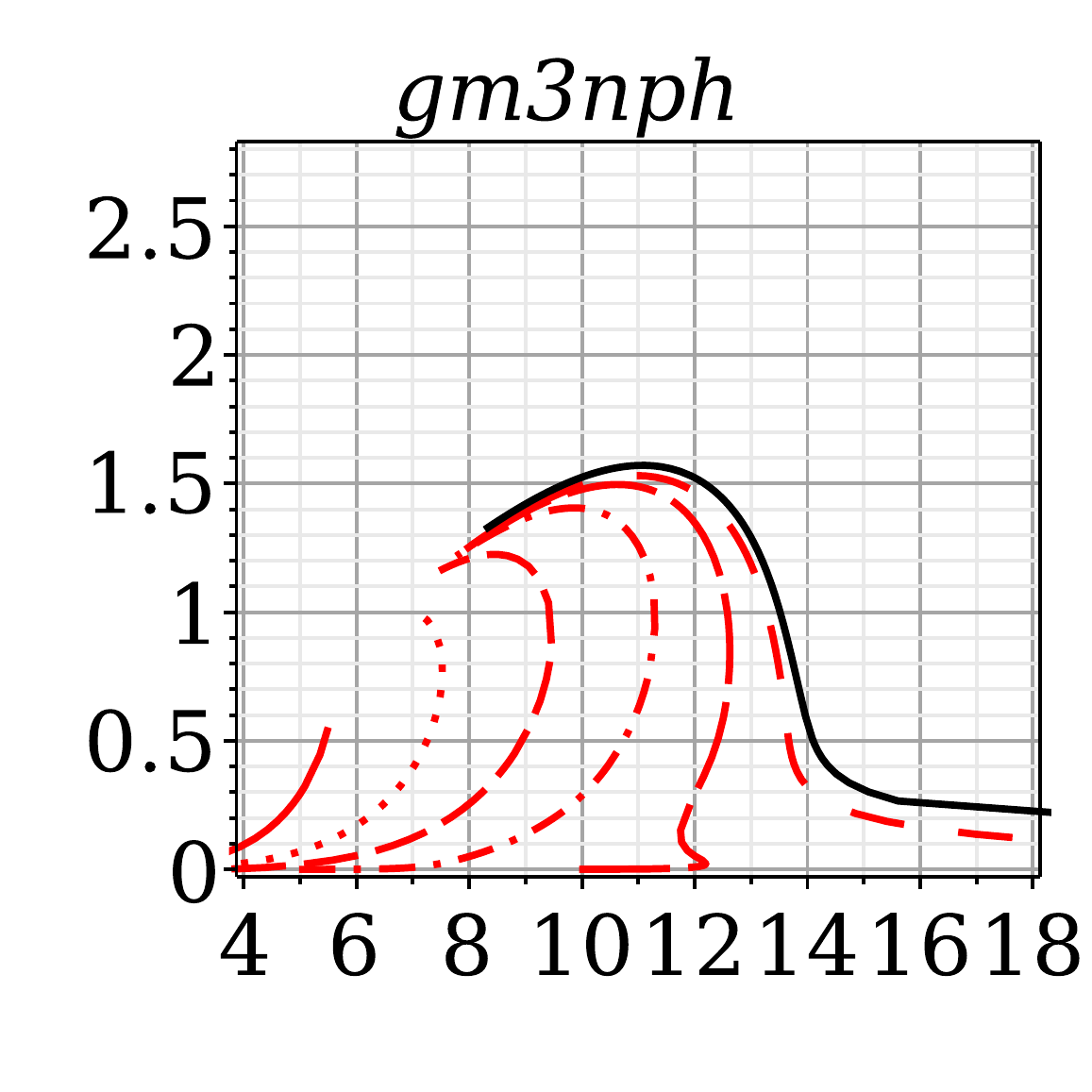} 
\includegraphics[scale=0.15]{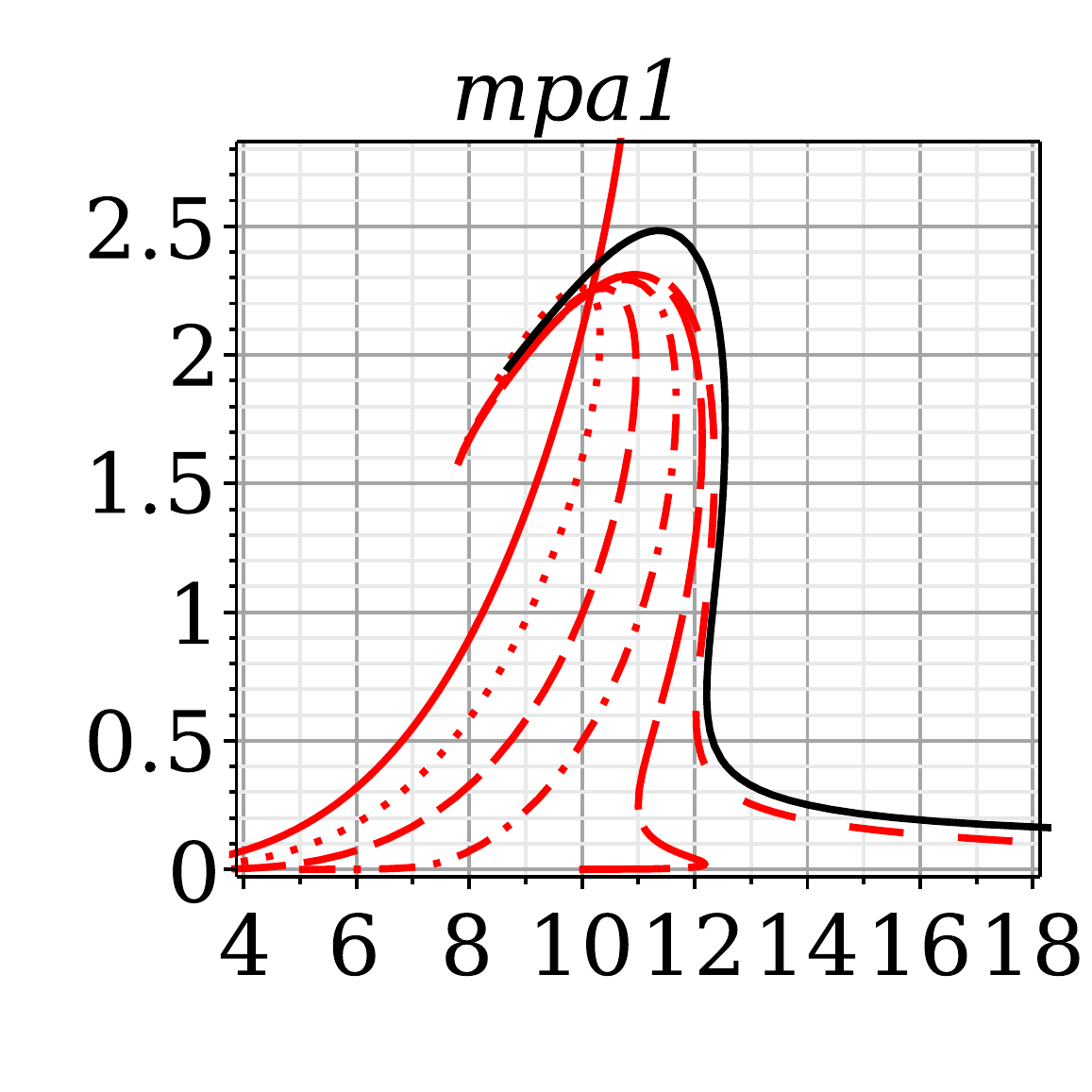}  	\includegraphics[scale=0.15]{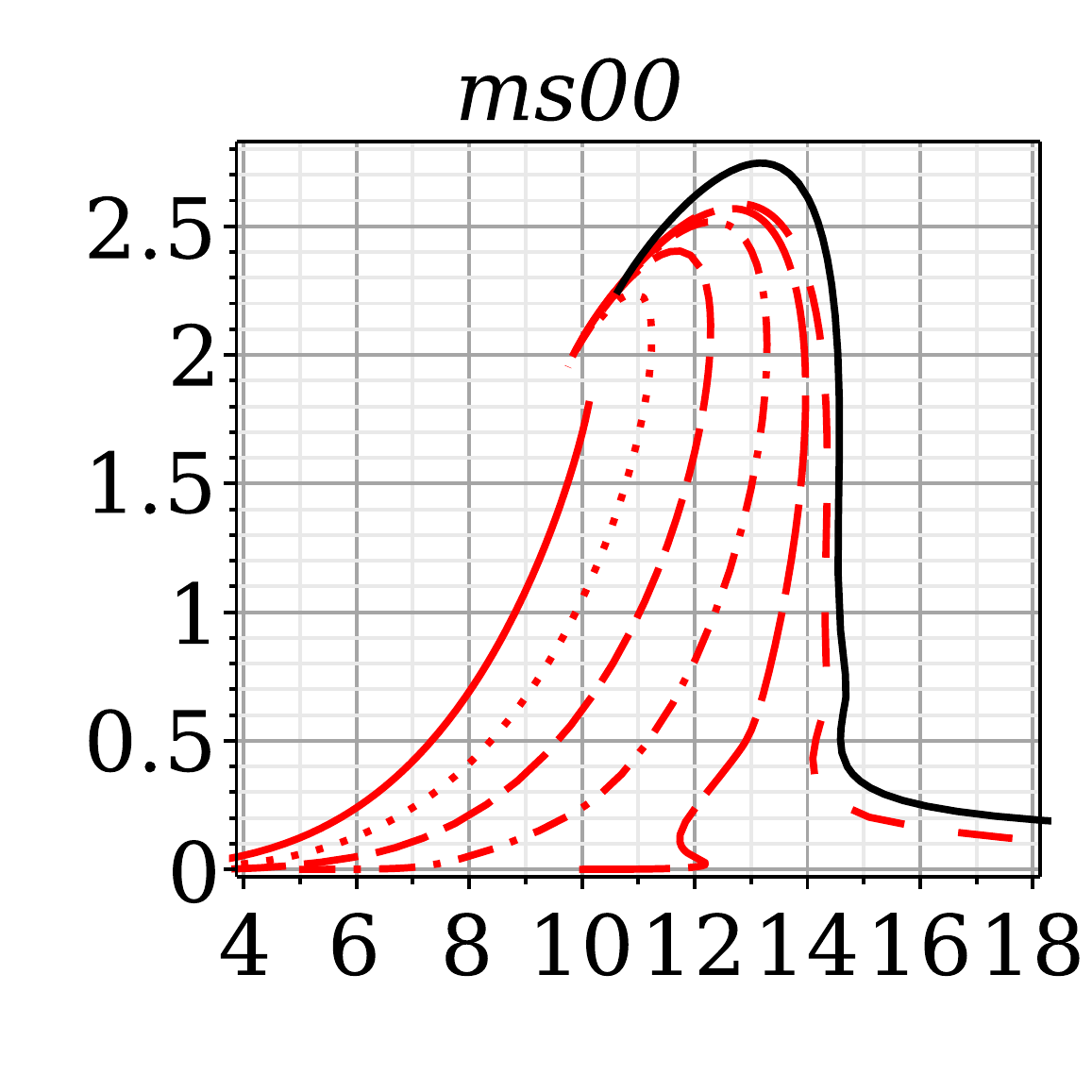} \\  \vspace{-0.4cm}
\includegraphics[scale=0.15]{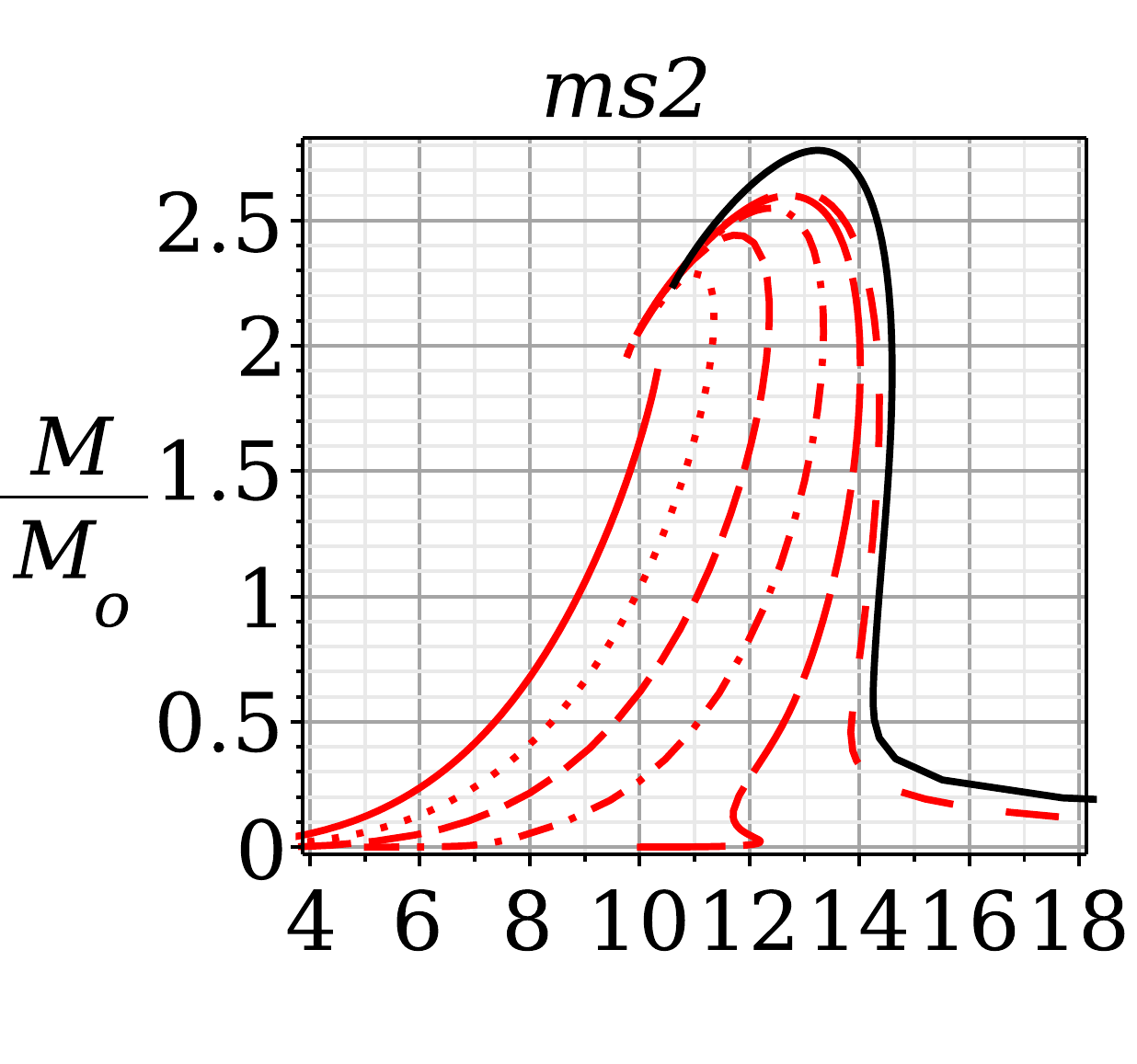} 	\includegraphics[scale=0.15]{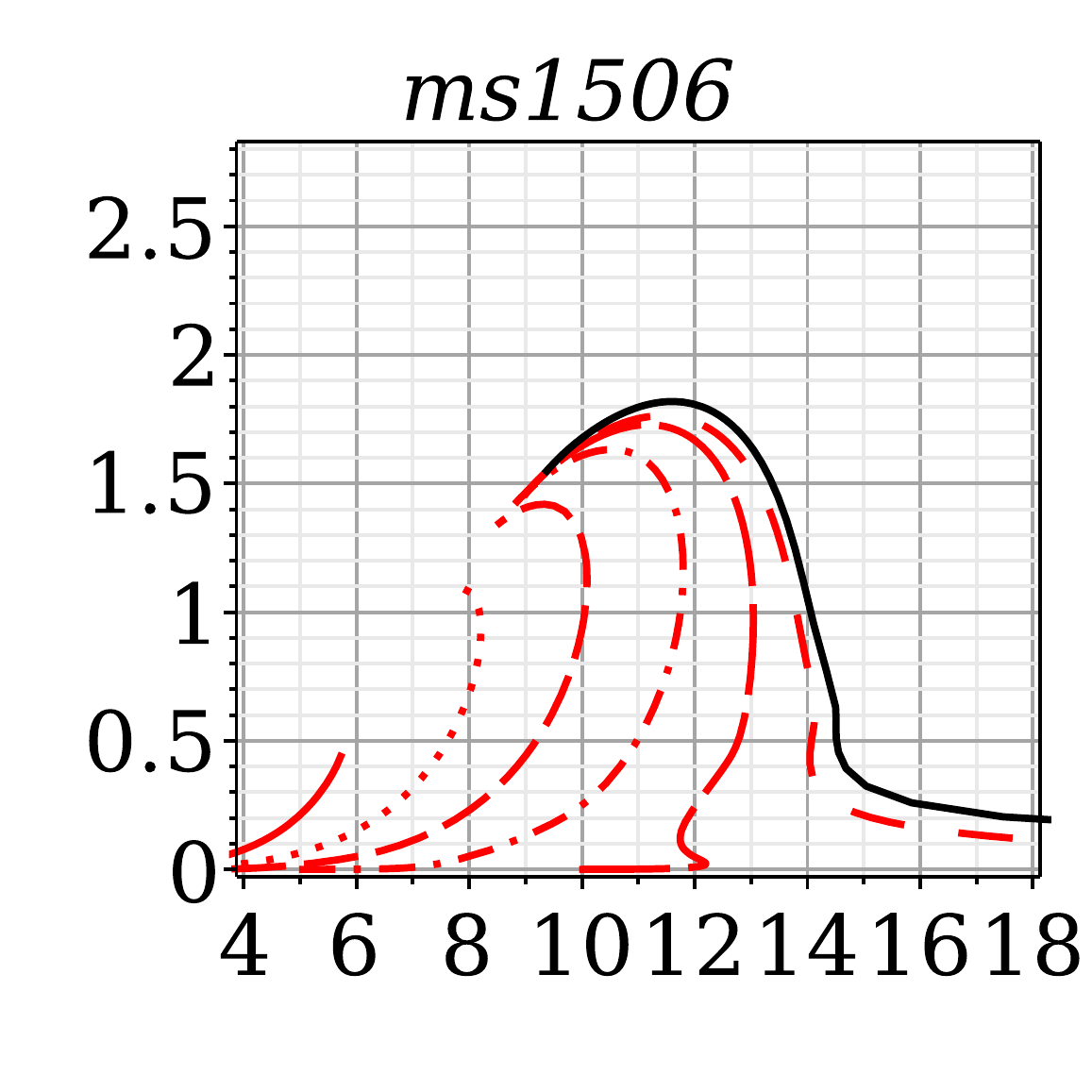} \includegraphics[scale=0.15]{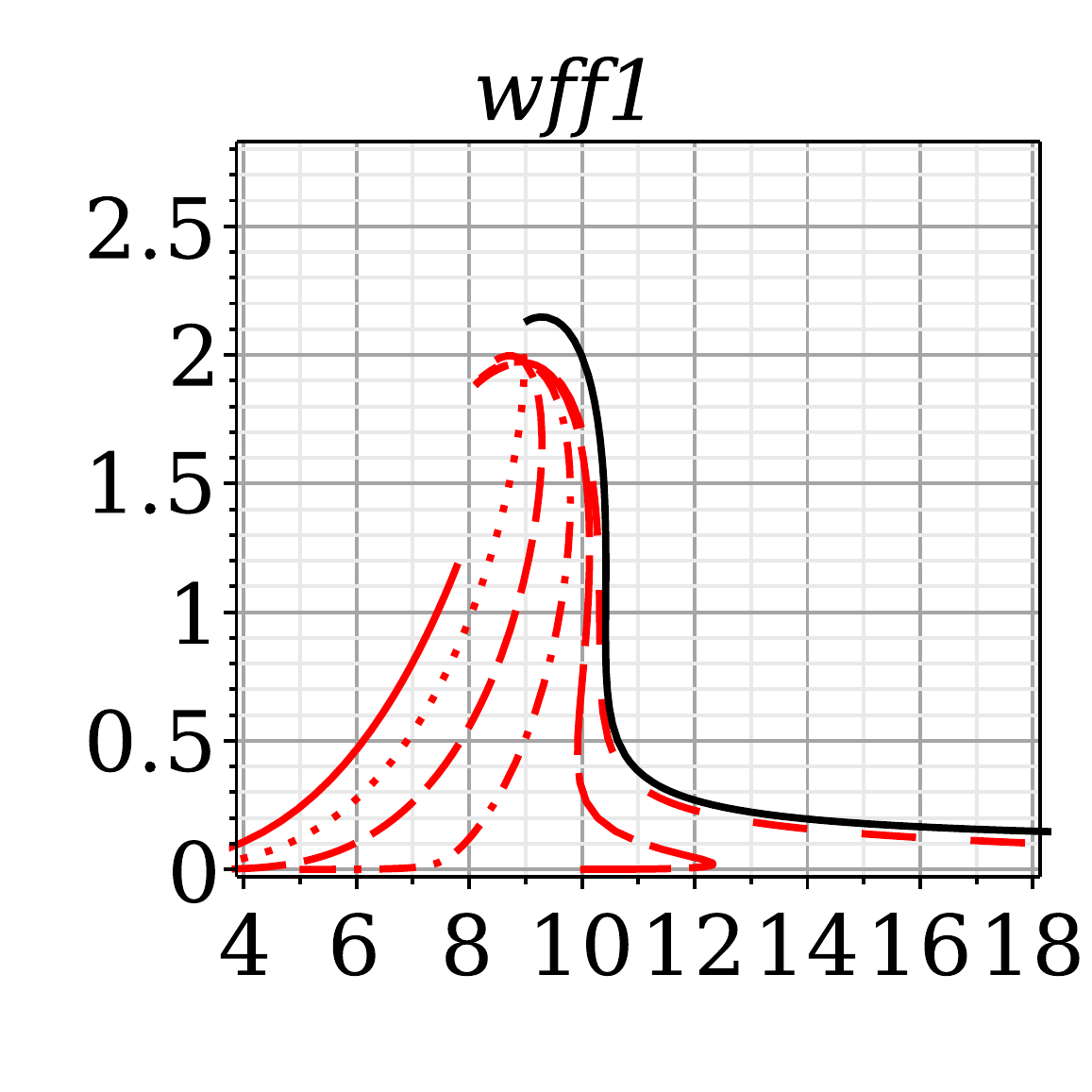}  
\includegraphics[scale=0.15]{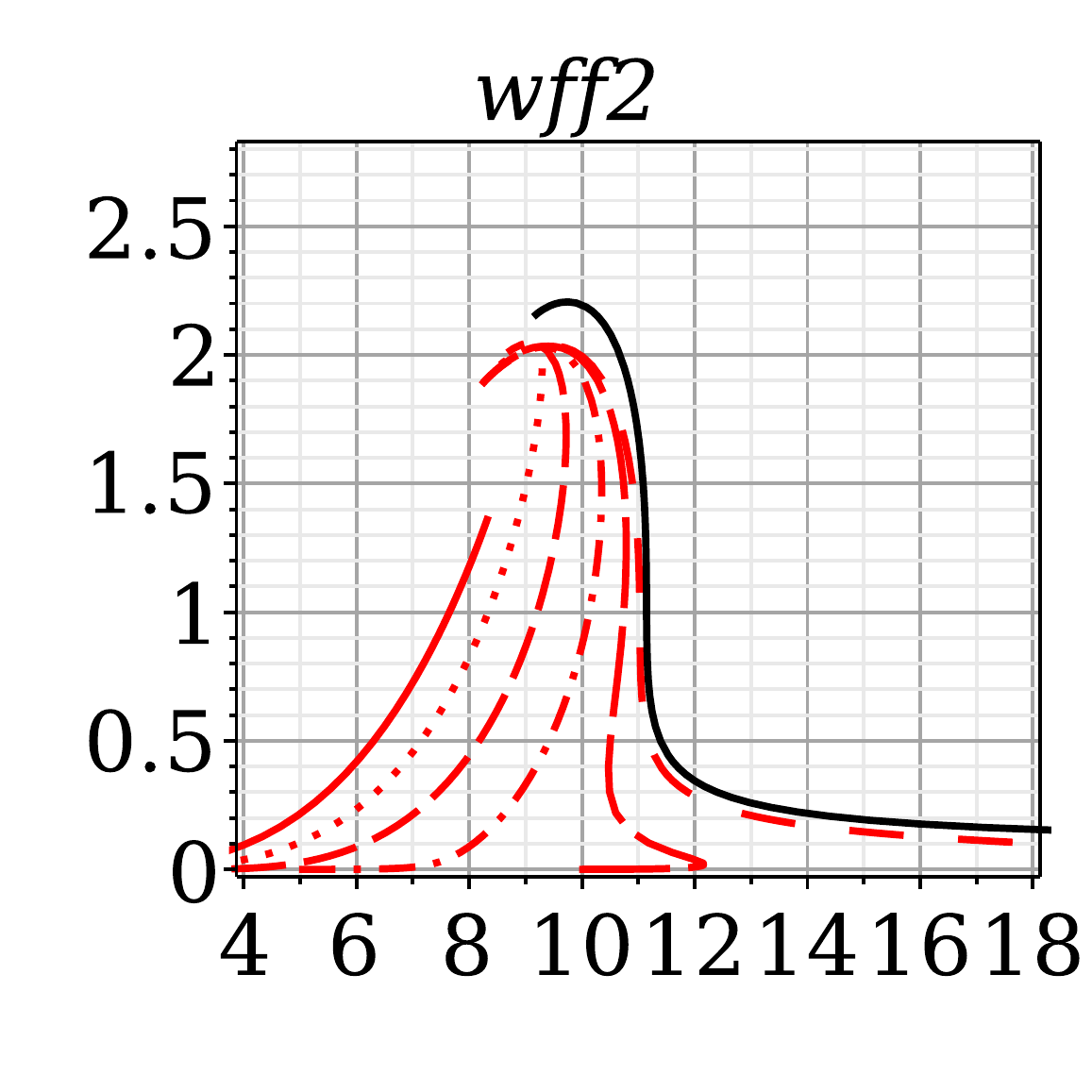}  \includegraphics[scale=0.15]{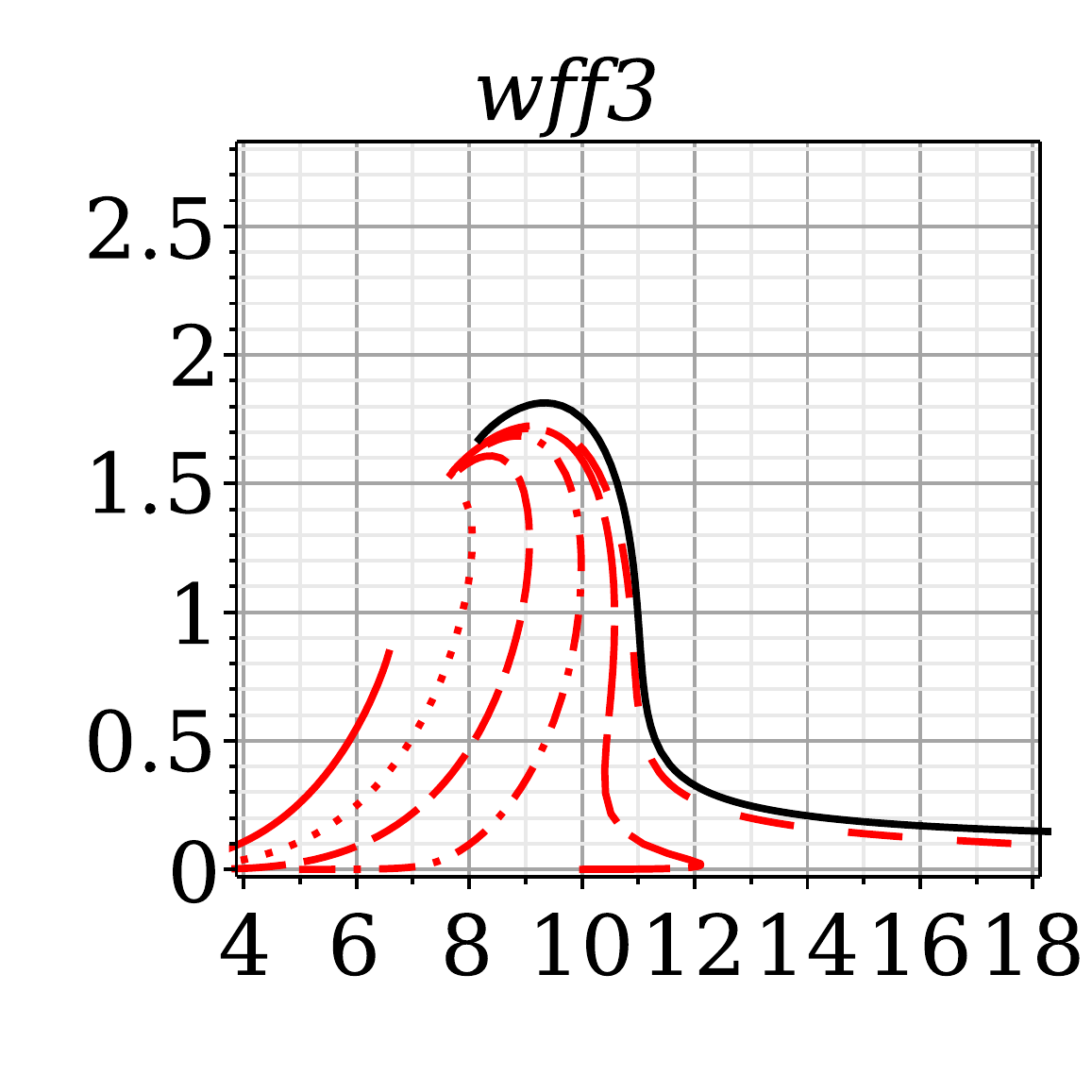} \\ \vspace{-0.4cm}
\includegraphics[scale=0.15]{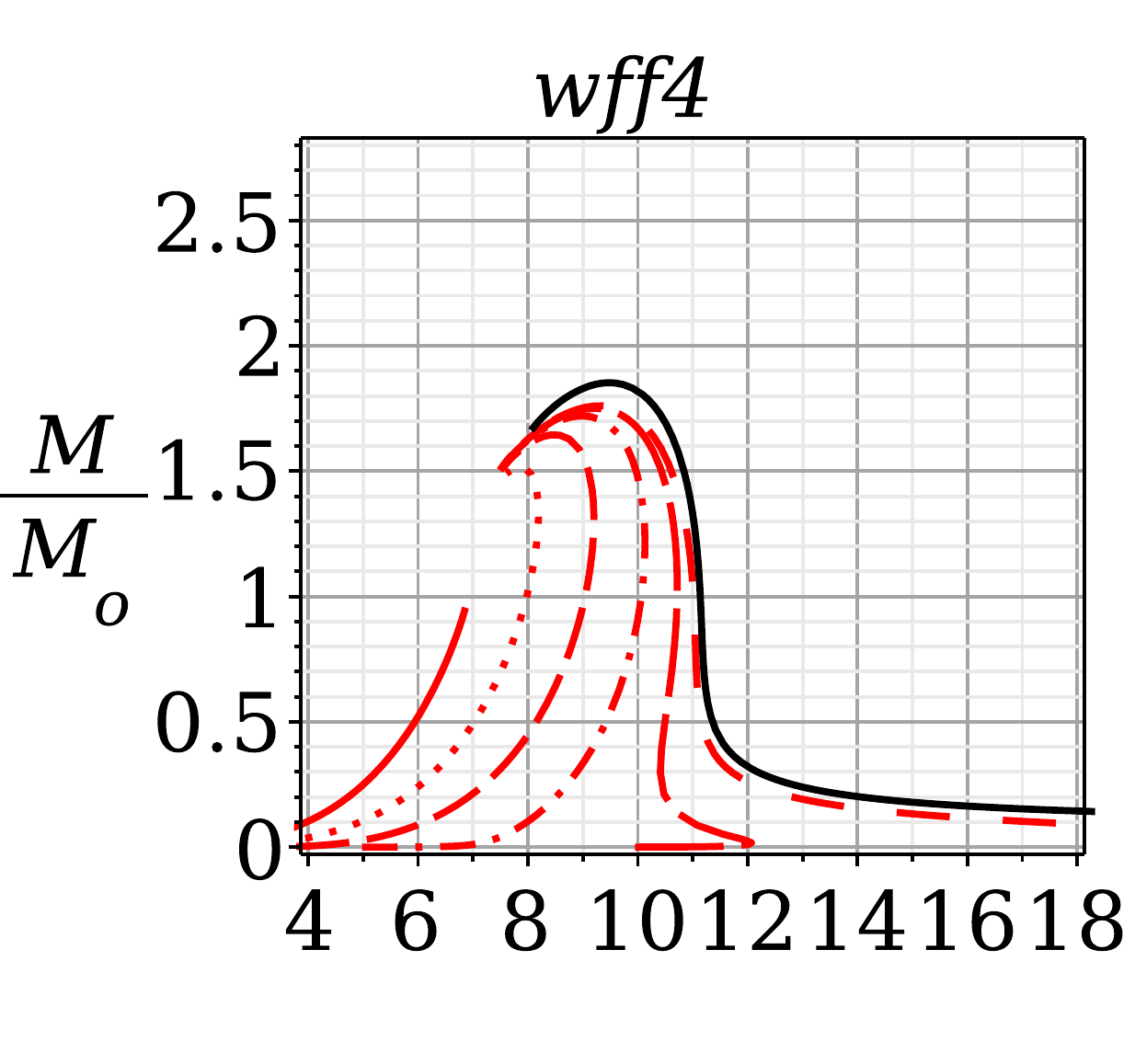} \includegraphics[scale=0.15]{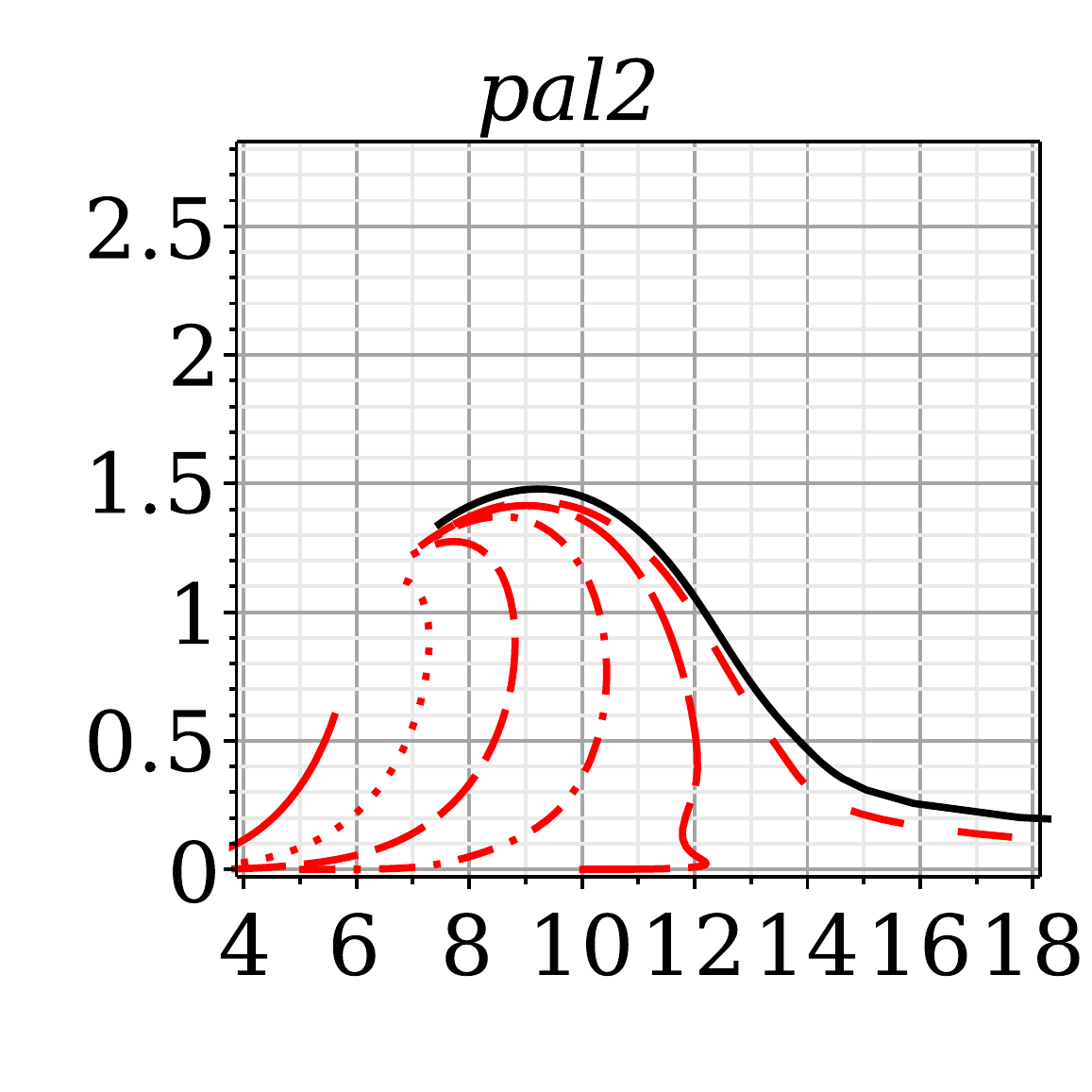}  \includegraphics[scale=0.15]{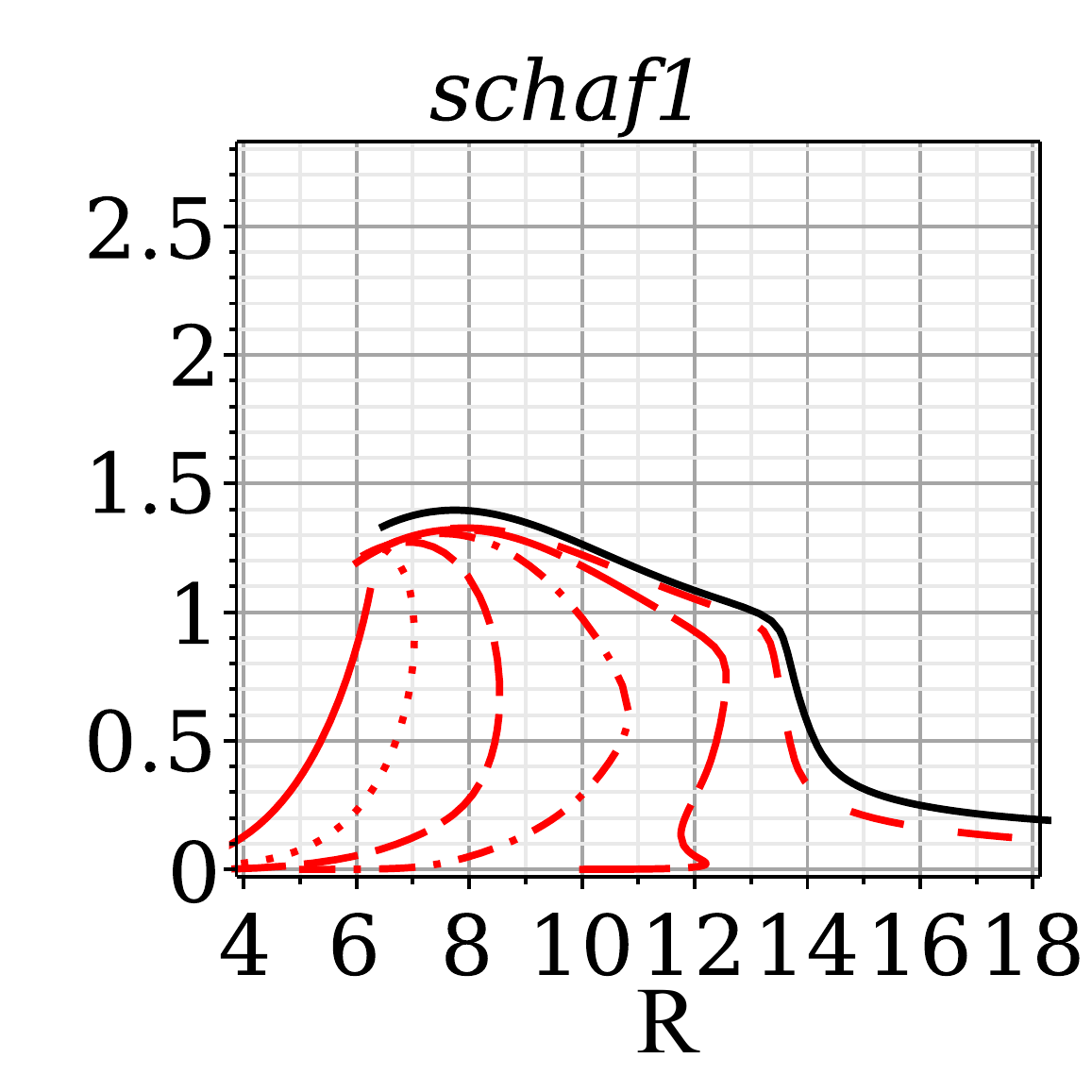} 
 \includegraphics[scale=0.15]{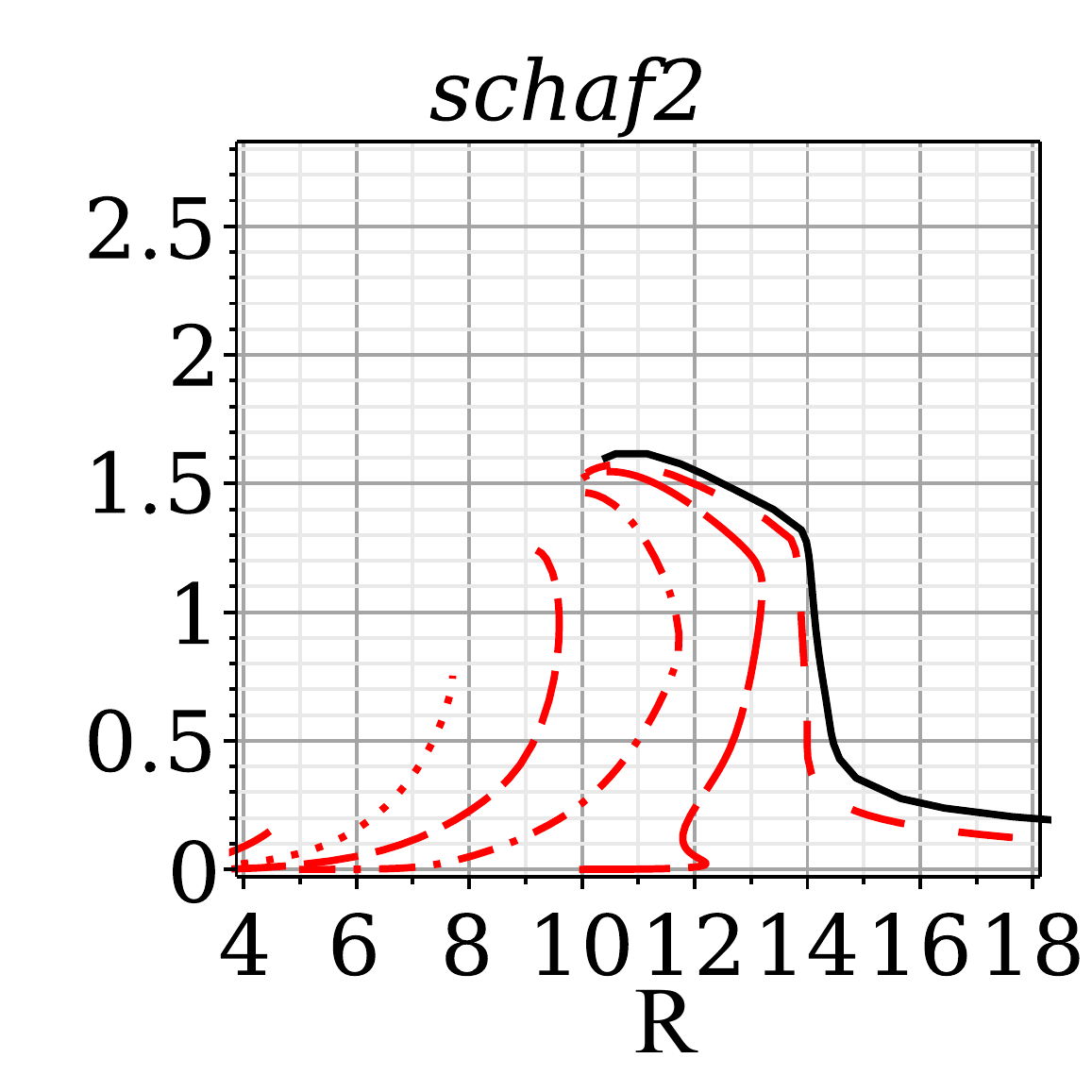} 	\includegraphics[scale=0.15]{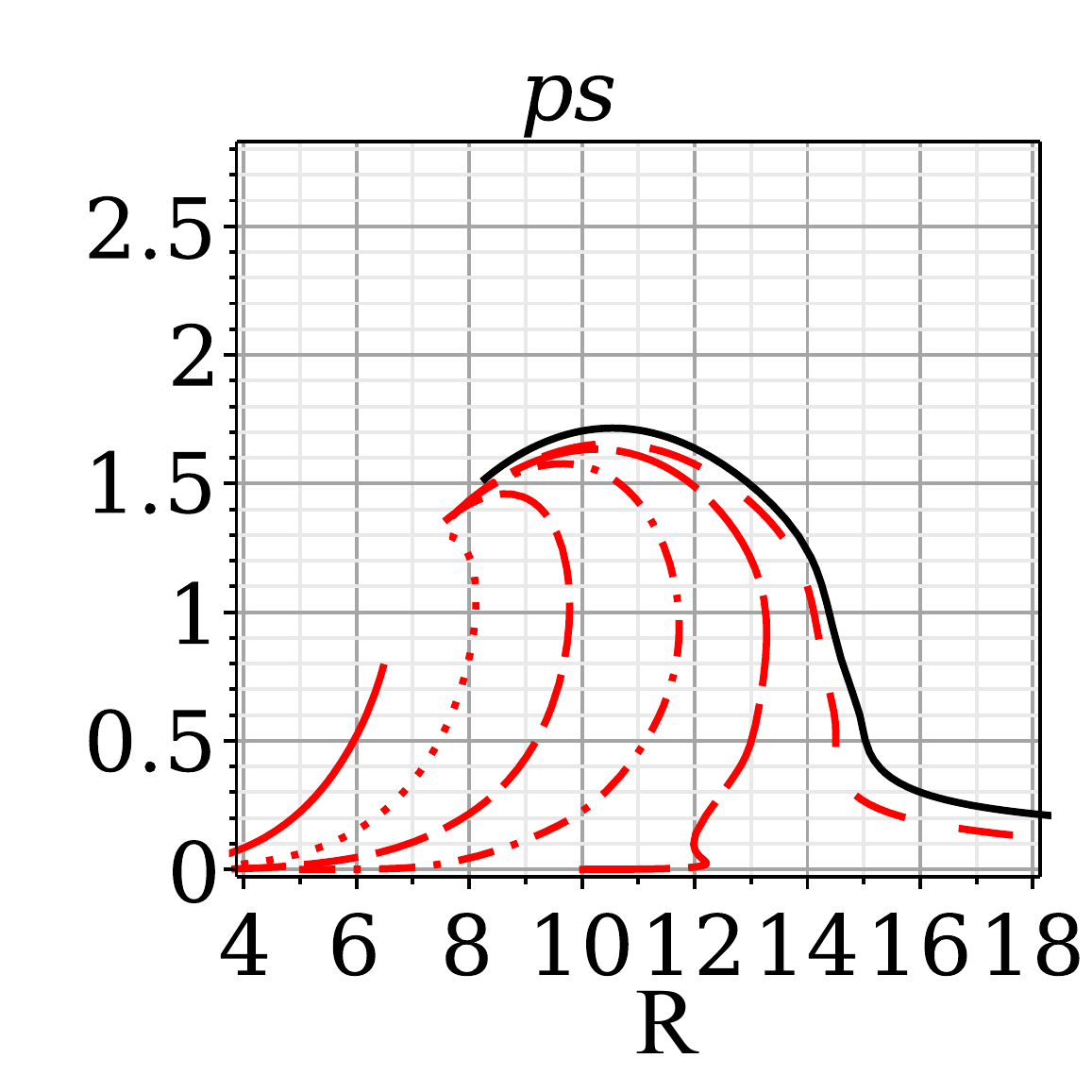} \\  \vspace{-0.4cm}
\vspace{-0.4cm}\includegraphics[scale=0.15]{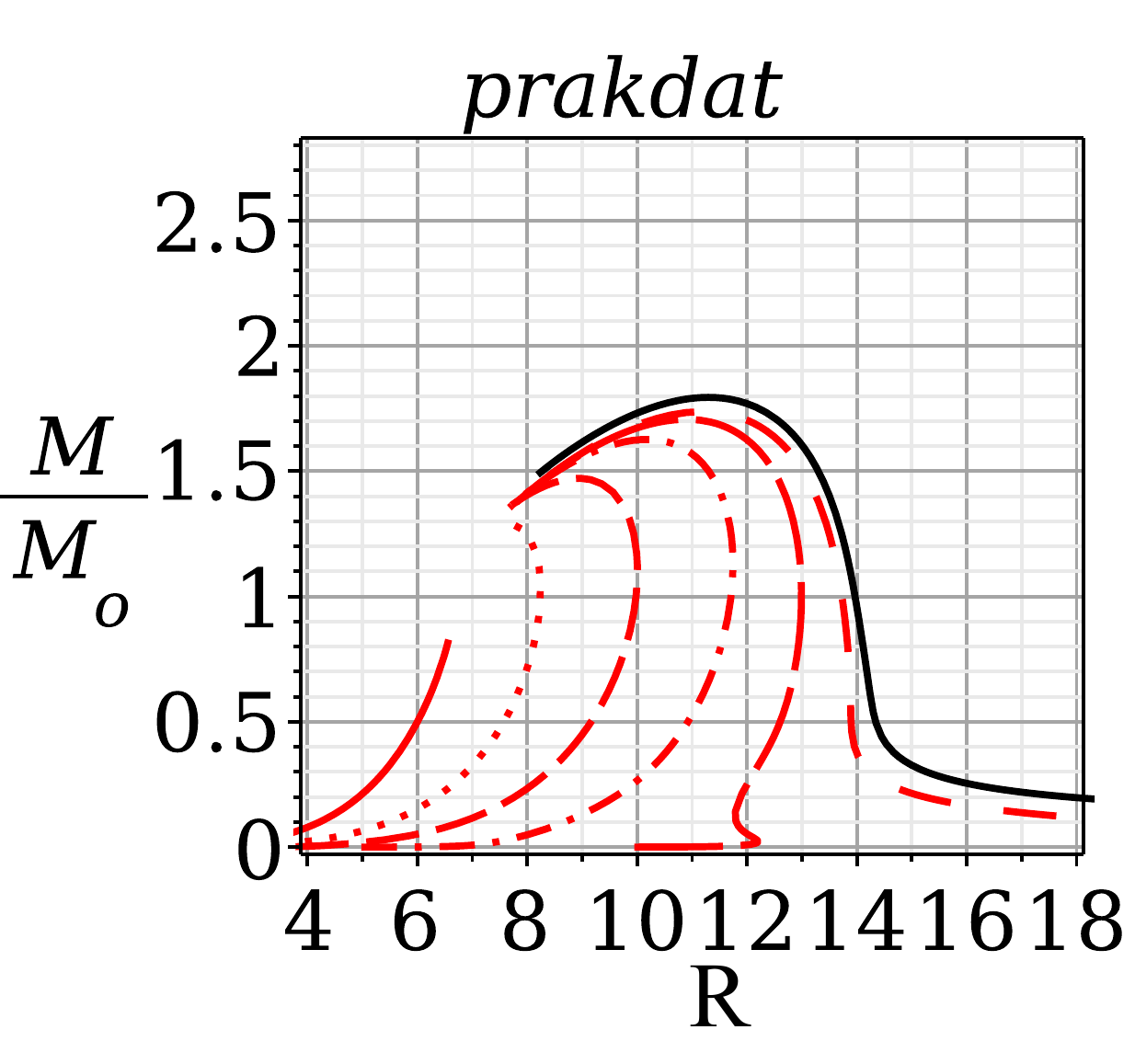} \includegraphics[scale=0.15]{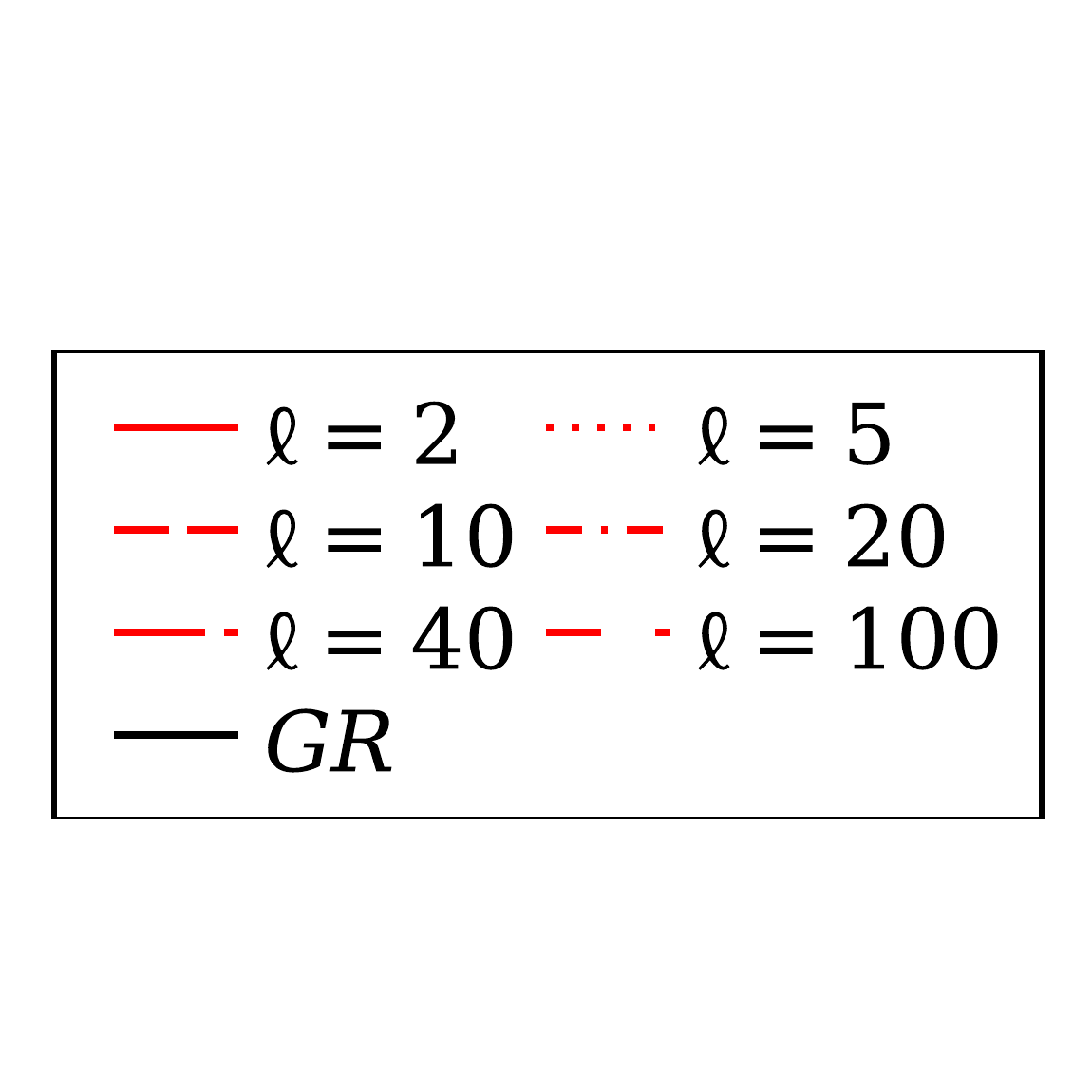} 
	\end{tabular} \end{center}	
\caption{\label{NEWMRxi1}The  mass-radius diagram in the case $\xi=-1$ and $\ell=2$, 5, 10, 20, 40, 100 km  for different equations of state. 
The black curves correspond to the unmodified theory of gravity (GR). 
The red curves of different styles correspond to different values $\ell$.}
\end{figure}

First let us consider in detail the case $\xi=-1$. This case corresponds to the special choice of model parameters of the theory (\ref{action}), such that $\Lambda_{Ads}=-\ell^{-2}$, and in this case the vacuum solution has the particularly simple form (\ref{stealthSS}) corresponding to the Schwarzschild-anti de Sitter black hole. Also, it follows from Eq. (\ref{_eqPsi_dl}) that $\varepsilon_2=-1$ and, since $\varepsilon=\varepsilon_1\varepsilon_2=-1$, one has $\varepsilon_1=+1$. This choice of signs means that we consider the theory (\ref{action1}) with an ordinary positive kinetic term, $\alpha>0$, and a negative nonminimal derivative coupling, $\beta<0$. Results of numerical integration are given in Figs. \ref{ABrhoPsi1}-\ref{NEWMRxi1}.   

In Figs.~\ref{ABrhoPsi1}, \ref{ABrhoPsi_ell} we demonstrate the typical behavior of the functions $A(r)$, $B(r)$, $\rho(r)$ and $\Psi^2(r)$ inside and outside the star. 
The Fig.~\ref{ABrhoPsi1} show the $r$ dependence of this functions in the case $\ell=20$ km and the central baryonic mass density $\rho_{c}=10^{15}\ {\rm g/cm^3}$ for different equations of state. 
For convenience 31 equations of state were divided to three groups, each column of the figures shows functions for this set of equations of state. The correspondence between the curve color and the equation of state is clarified at the  legends at the top of the figure.  
The solid line corresponds to the internal solution, the dotted line corresponds to the external vacuum solution. The internal and vacuum solutions are continuously glued at the boundary of the  star $r=R$. The matter density monotonically decreases and equals zero at the boundary of the star. Solution of $\Psi^2$ equals zero too outside the star. As already mentioned, in the case $\xi=-1$ the vacuum solutions of $A(r)$ and $B(r)$ coincide, but the internal solutions differ. 

Despite the diversity of equations of state, many conclusions will be common for each of them. 
Therefore we will consider one illustrative example to demonstrate these common properties. Next we chose BSk equations only for the sake of demonstrativeness. 
Fig.~\ref{ABrhoPsi_ell} shows  the $r$ dependence of the functions $A(r)$, $B(r)$, $\rho(r)$ and $\psi^2(r)$ in the case BSk19 equation of state, $\rho_{c}=10^{15}\ {\rm g/cm^3}$ and $\ell=10$, 20 and 40 km. The correspondence between the curve style and $\ell$ is given in the legend, black curve corresponds to the unmodified theory of gravity (GR). 
As the parameter $\ell$ increases the curves shift towards to the unmodified solution (GR), the scalar field tend to zero, radius of the star increases. Another equations of state have a similar appearance.

More detailed information about neutron star configurations can be extracted from a mass-radius diagram. 
As an example we will choose the BSk24 equation of state. 
For this purpose in Fig.~\ref{bsk24} for BSk24 equation of state and different values $\ell$ are presented: a) the dependence of the neutron star mass $M$ on the central density $\rho_c$, b) the dependence of the neutron star radius $R$ on the central density $\rho_c$, c) the mass-radius diagram. The correspondence between the curve style and the  $\ell$ value is given in the legend, the black curve corresponds to the unmodified theory of gravity (GR). Also Fig.~\ref{bsk24}d shows the dependence of the sound speed $v$ on the density $\rho$ for BSk24 equation of state. Densities at which the sound speed exceeds the speed of light are marked in gray in the Figs.~\ref{bsk24}a,~\ref{bsk24}b,~\ref{bsk24}d.

We consider observable range of neutron stars masses $M$ and physically permissible densities $\rho$. As noted earlier astronomical observations give a neutron star masses within a range of 1-3 \ms. The permissible values of $\rho$ are limited by the area of applicability equation of state and its analytical representation. In particular, for the BSk24 equation, the authors consider the following range of densities $10^6 \ {\rm g/cm^3}\leqslant\rho\leqslant 10^{16}  \ {\rm g/cm^3}$. Also some equations of state give the superluminal speed of sound  above critical density $\rho\geqslant\rho_{stab}$. Fig.~\ref{bsk24}d shows the dependence of the sound speed on density for BSk24 equation of state. At the density $\rho>\rho_{stab}\approx 2.69\times 10^{15}\,{\rm g/cm^3}$ the equation of state becomes superluminal ($v/c>1$) and causality breaks down \cite{Pearson2018}. Apparently in the case $\rho\geqslant\rho_{stab}$  equation of state do not give a complete description, in particular it was discussed in \cite{Pearson2018}. 

The case of a large values of non-minimal coupling parameter $\ell$ we have discussed later, now we will consider the case of a small values $\ell$ in more detail. 
The case of a small values of the coupling parameter $\ell < 1$ km cannot satisfy the constraint on the stellar masses and central densities simultaneously. Let us consider curves with small values $\ell < 1$ km in the Fig.~\ref{bsk24}a. Indeed the part of the curve corresponding to the range of 1-3 \ms is located at densities significantly exceeding the permissible range of central densities. Vice versa, the permissible range of central densities coresponds to masses significantly smaller mass of the Sun \ms. Thus the case of a small values of the non-minimal coupling parameter $\ell$ takes us away from the applicability area of the equation of state or observable range of stellar masses. 
In the case of small values $\ell < 1$ km the curves shift towards to masses significantly smaller than solar masses or towards higher densities, at which the equation of state is no longer applicable. 
The case $\ell <1$ km is beyond the scope of the work and set of used equations of state. Therefore in the figures we have limited ourselves to the case $\ell > 1$ km. Here BSk24 equation of state was considered, but for other equations the results are approximately the same.

For hydrostatic stability it is necessary to fulfill the condition $dM/d\rho>0$, therefore the unstable configurations correspond to the parts of the curves to the right of the maxima in Fig.\ref{bsk24}a and to the left of the maxima in Fig.~\ref{bsk24}c (and also Figs.~\ref{MRxi1},~\ref{NEWMRxi1}). 
In unmodified theory of gravity superluminal configurations ($\rho_c>\rho_{crit}$) are often unstable, i.e. $dM/d\rho<0$ \cite{Pearson2018}; in particular for the BS24 equation of state the gray area at the Fig.\ref{bsk24}a is located to the right of the maximum of the black curve. A similar situation occurs for modified theory in the case $\ell>1$ km; at the Fig.\ref{bsk24}a the superluminal configurations are unstable, i.e the gray area ($\rho_c>\rho_{crit}$) located to the right part of the curves maxima ($dM/d\rho<0$). The stability of this model will be studied in the next work.

In Figs.~\ref{MRxi1},~\ref{NEWMRxi1} we present the mass-radius diagrams obtained for different values of the nonminimal derivative coupling parameter $\ell$ in the case $\xi=-1$ for 31 equations of state.
%
\begin{figure}[h!]
\begin{center}
\begin{tabular}{c} 
\includegraphics[scale=0.3]{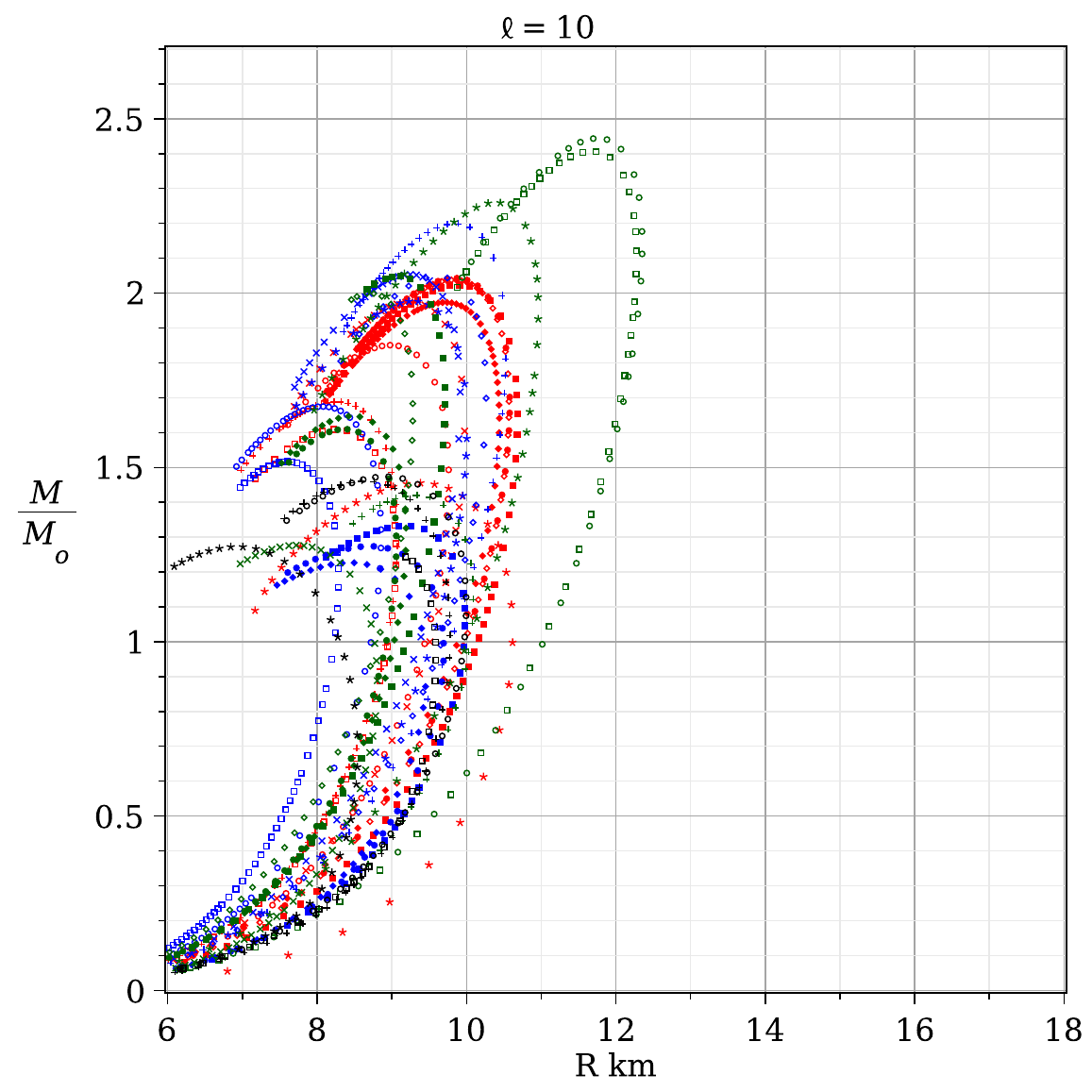}\includegraphics[scale=0.3]{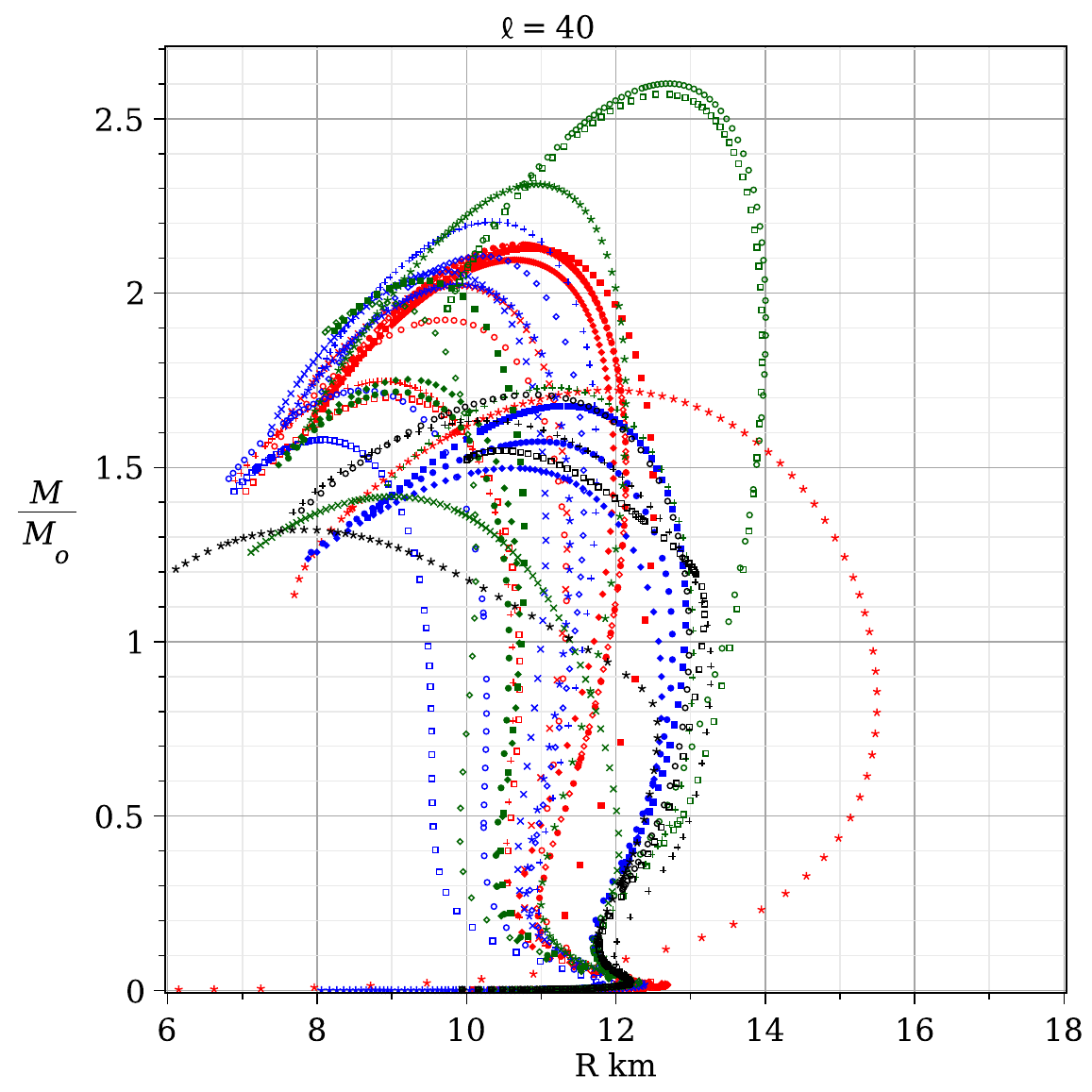}\includegraphics[scale=0.3]{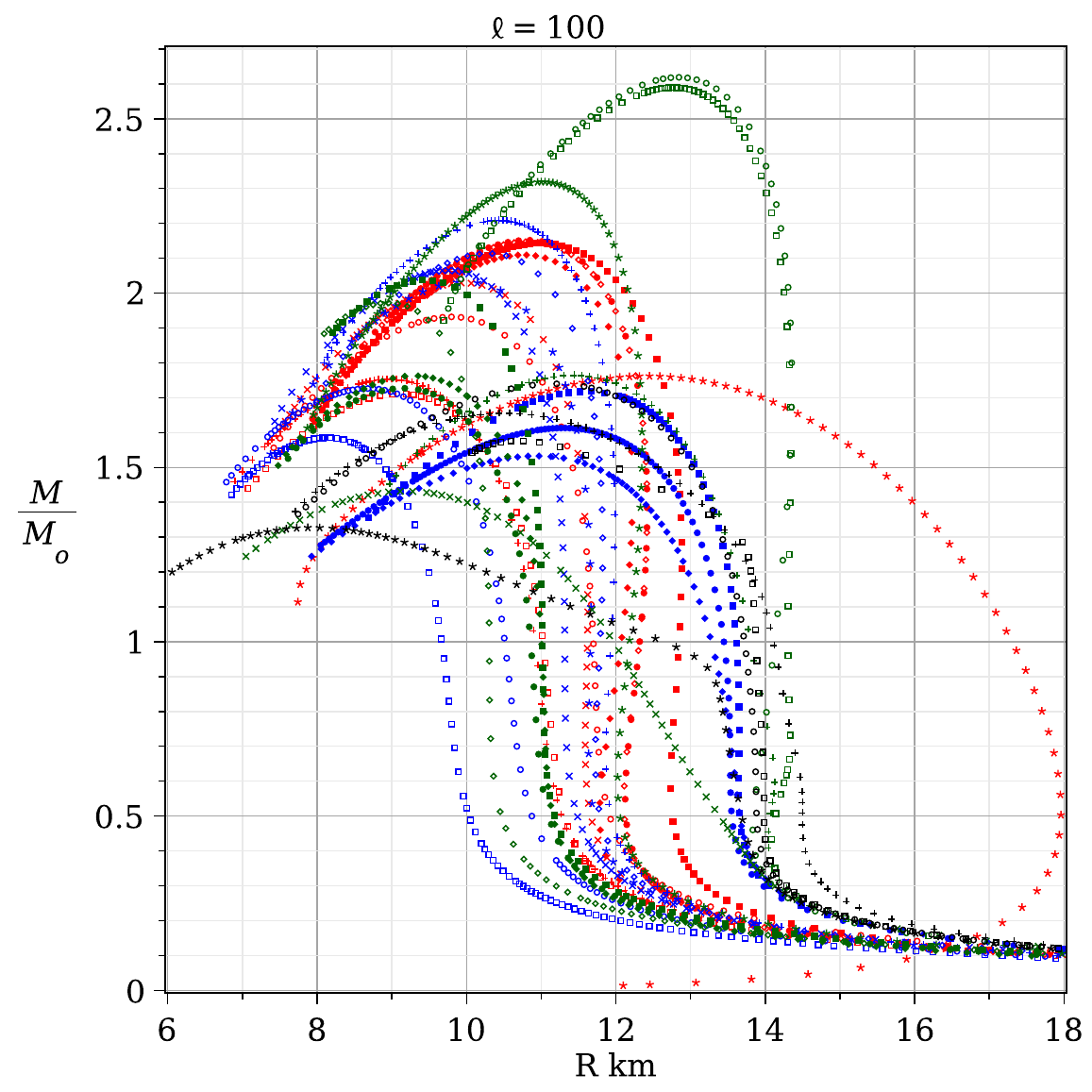}\\ \includegraphics[scale=0.9]{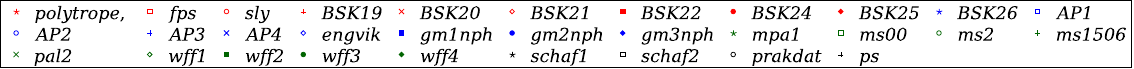}
\end{tabular} \end{center}	
\caption{\label{MR}The  mass-radius diagram in the case $\xi=-1$ and $\ell=10$, 20 and 100 km  for different equations of state. \label{all_eos}}
\end{figure}
For indicativeness the diagrams are presented in two versions: at the Fig.~\ref{MRxi1} the diagrams are grouped by the values of parameter $\ell$; at the Fig.~\ref{NEWMRxi1} the diagrams are grouped by the equation of state. Each curve corresponds to the following range of central densities: $10^{14} {\rm \ g/cm^3}\leqslant \rho_c\leqslant 10^{16} {\rm \ g/cm^3}$. 

At the Fig.~\ref{MRxi1} for convenience 31 equations of state were divided into three subgroups. 
The top row of pictures corresponds to the polytrope, FPS, SLY, BSk19-21, BSk22, BSk24-26 equations of state, 
the central row corresponds to AP1-4 engvik, gm1nph, gm2nph, gm3nph, mpa1, ms00, 
the bottom row corresponds to  ms2, ms1506, pal2, wff1, wff2, wff3, wff4, schaf1, schaf2, prakdat, ps. 
The left column of pictures at the Fig.~\ref{MRxi1} corresponds to $\ell=20$ km, the middle column corresponds to $\ell=40$ km and the right  column corresponds to  $\ell=100$ km. 
The correspondence between the equation of state and the curve style is indicated at the left.

At the Fig.~\ref{NEWMRxi1} the  mass-radius diagram are presented in the case $\xi=-1$ and $\ell=2$, 5, 10, 20, 40, 100 km  for different equations of state. 
The black curves correspond to the unmodified theory of gravity (GR). The red curves of different styles correspond to different values $\ell$.
Finally the diagrams for all equations of state are given in the following Fig.~\ref{all_eos}.

The mass-radius diagram obtained for our model has an essential difference comparing with that in GR. 
Neutron stars in the case of the modified theory of gravity have smaller masses and radii compared to stars in the unmodified theory. 
As the parameter $\ell$ increases, the diagrams shift towards to larger masses and larger radii.
One can see that in the general case the relation of mass and radius tends to that obtained for the case of the unmodified theory in the limit of large values of $\ell$.
There is a qualitative difference between the diagrams for large values $\ell\geqslant 40$ km and small values $\ell\leqslant 20$ km. 
In the case of a small $\ell\leqslant 20$ km the mass of the star increases with increasing radius. This behavior is typical for the so-called bare strange stars or quark stars (see the excellent book \cite{Book_HPY} and references therein). The main feature of bare strange stars is that their radius decreases monotonically with decreasing $M$, with $R \propto M^{1/3}$ for small enough masses of the star. Such the property of strange stars is explained with using the Bag Model for describing quark matter \cite{Book_HPY}. 
However, as it was discussed in our previous work \cite{Kashargin2022}, the specific `strange' relation between mass and radius in our case is forming due to the negative cosmological constant $\Lambda_{AdS}$ given by Eq.~\cite{Kashargin2022}.
In the case of a large $\ell\geqslant 40$ km the form of the diagrams changes, and a vertical section appears at the diagram, where the mass of the star increases practically without increasing the radius. 
The values of the mass and radii significantly differ not only from the case of the unmodified gravity theory, but also strongly depend from the choise of an equation of state. In general, 
realistic equations of state give larger masses compared to the polytropic equation we considered earlier in \cite{Kashargin2022}. 
The largest mass are given by
ms00 and ms2 ($M\approx 2.5$ \ms), 
BSk22, BSk24 and AP3 ($M\approx 2$ \ms).   
The smallest mass are given by
schaf1 ($M< 1.5$ \ms).

Note that, using the mass-radius diagram given in Figs.~\ref{MRxi1},~\ref{NEWMRxi1}, one can in principle restrict possible values of the nonminimal derivative coupling parameter $\ell$. 
As the result, applying observable restrictions for mass and radius of the neutron star, we get the following estimation: 
$1.1 M_\odot\le M_{NS}\le 2.9M_\odot$, and radius, $10\ km\le R_{NS}\le 16\ km$. In general the neutron star model in the theory of gravity with a non-minimal derivative coupling does not contradict astronomical data and is viable in the case $\ell>1$ km.

\subsubsection{The case $\xi\neq-1$ }
\begin{figure}[h]
\begin{center}\begin{tabular}{c}
 	\includegraphics[scale=0.2]{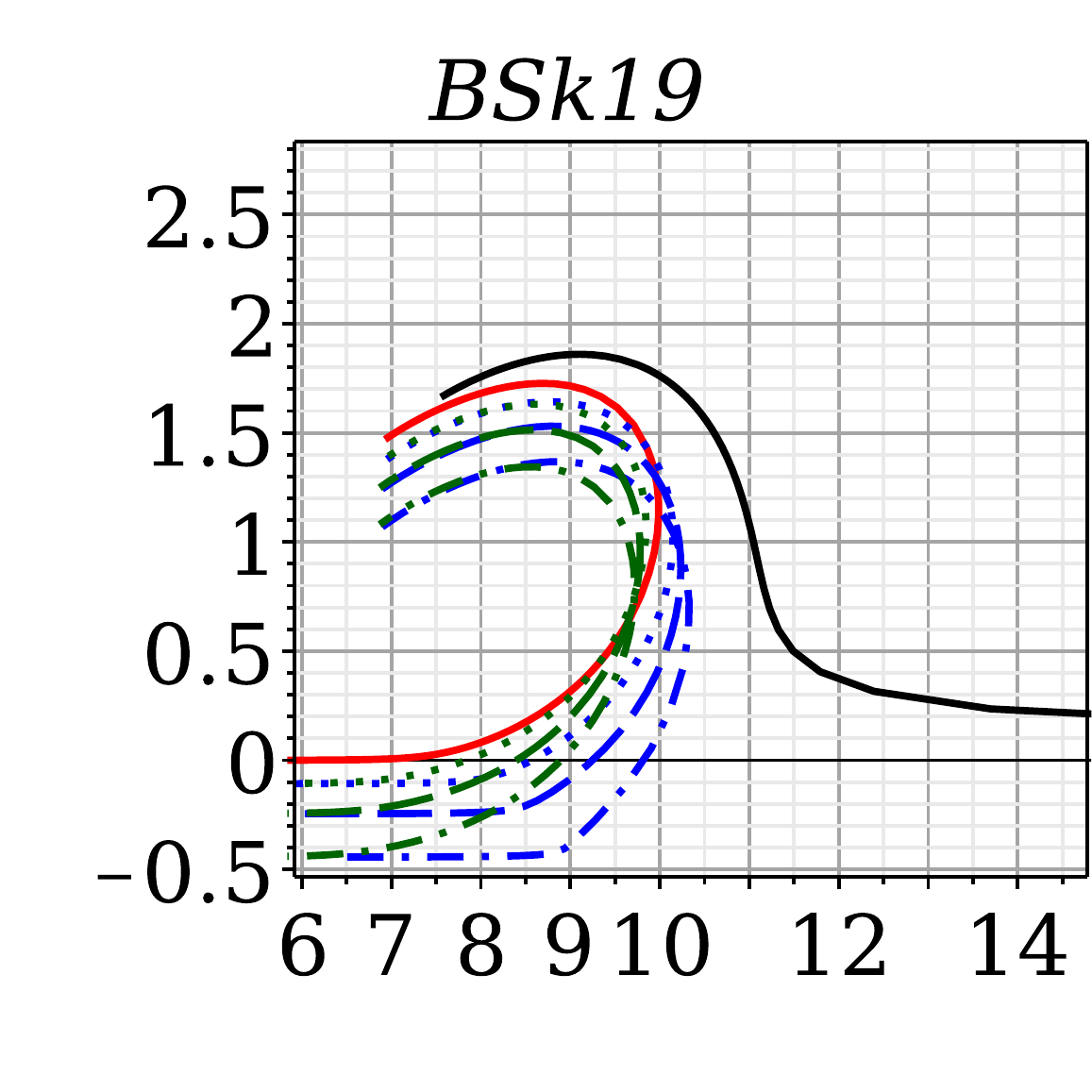} 	\includegraphics[scale=0.2]{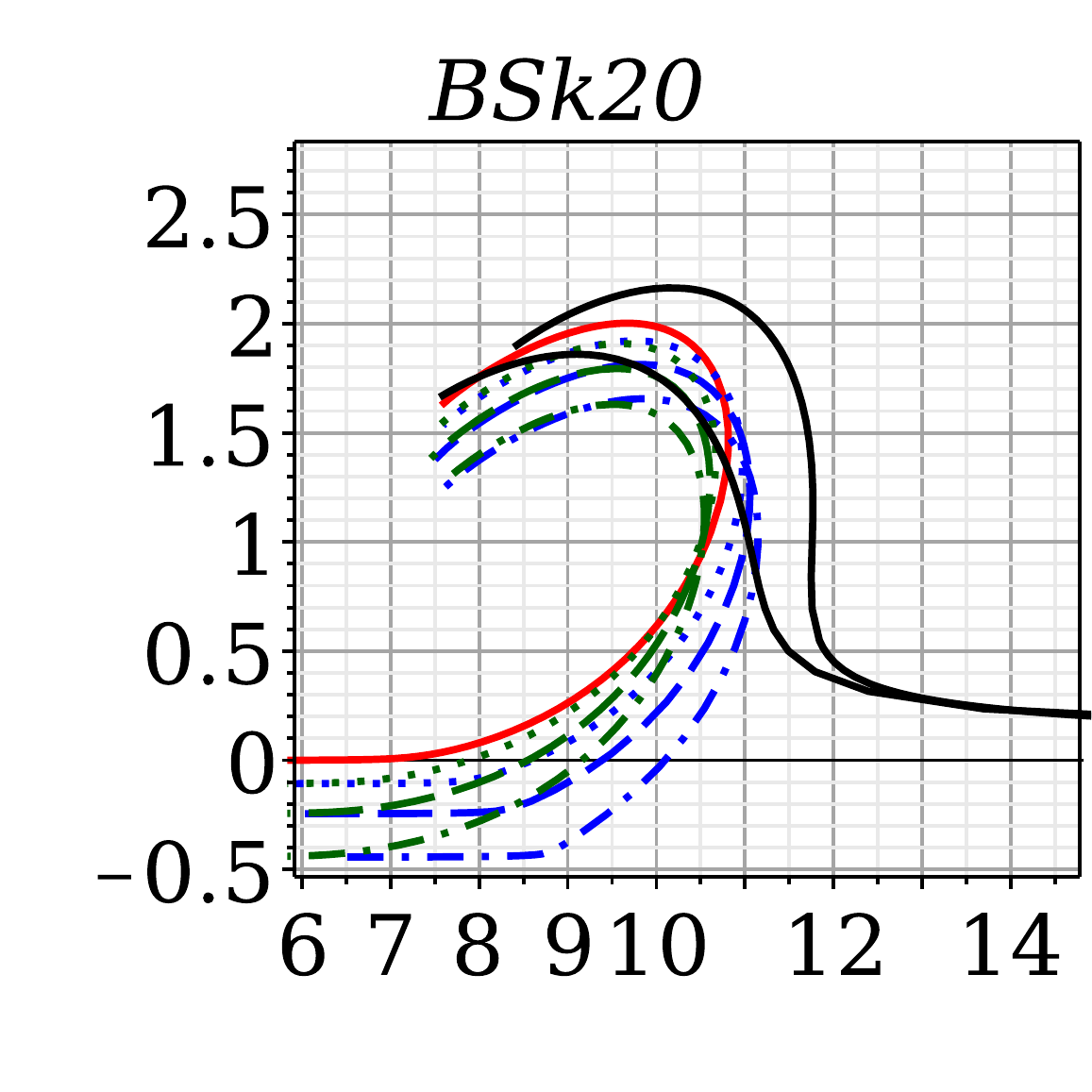} \includegraphics[scale=0.2]{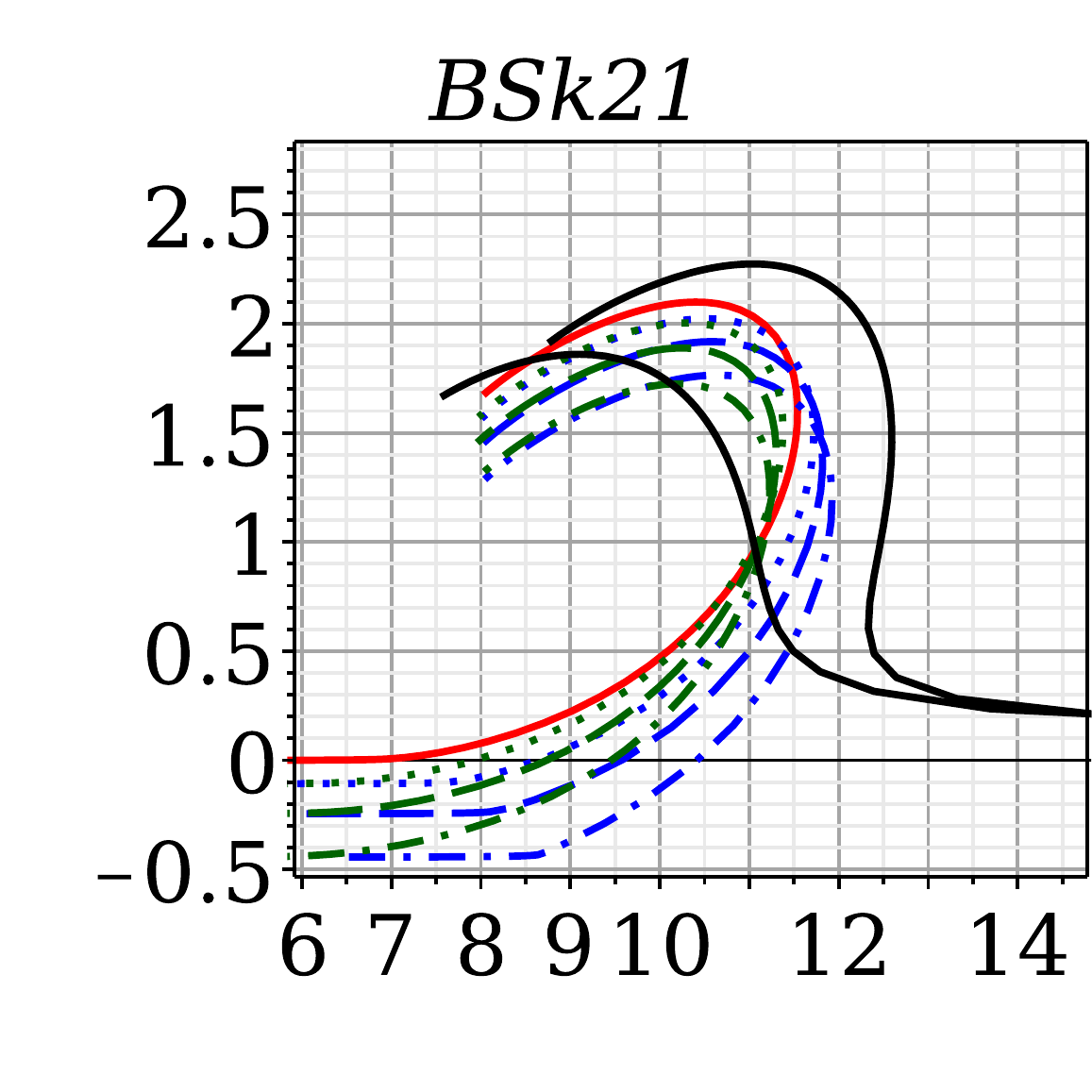}  \includegraphics[scale=0.2]{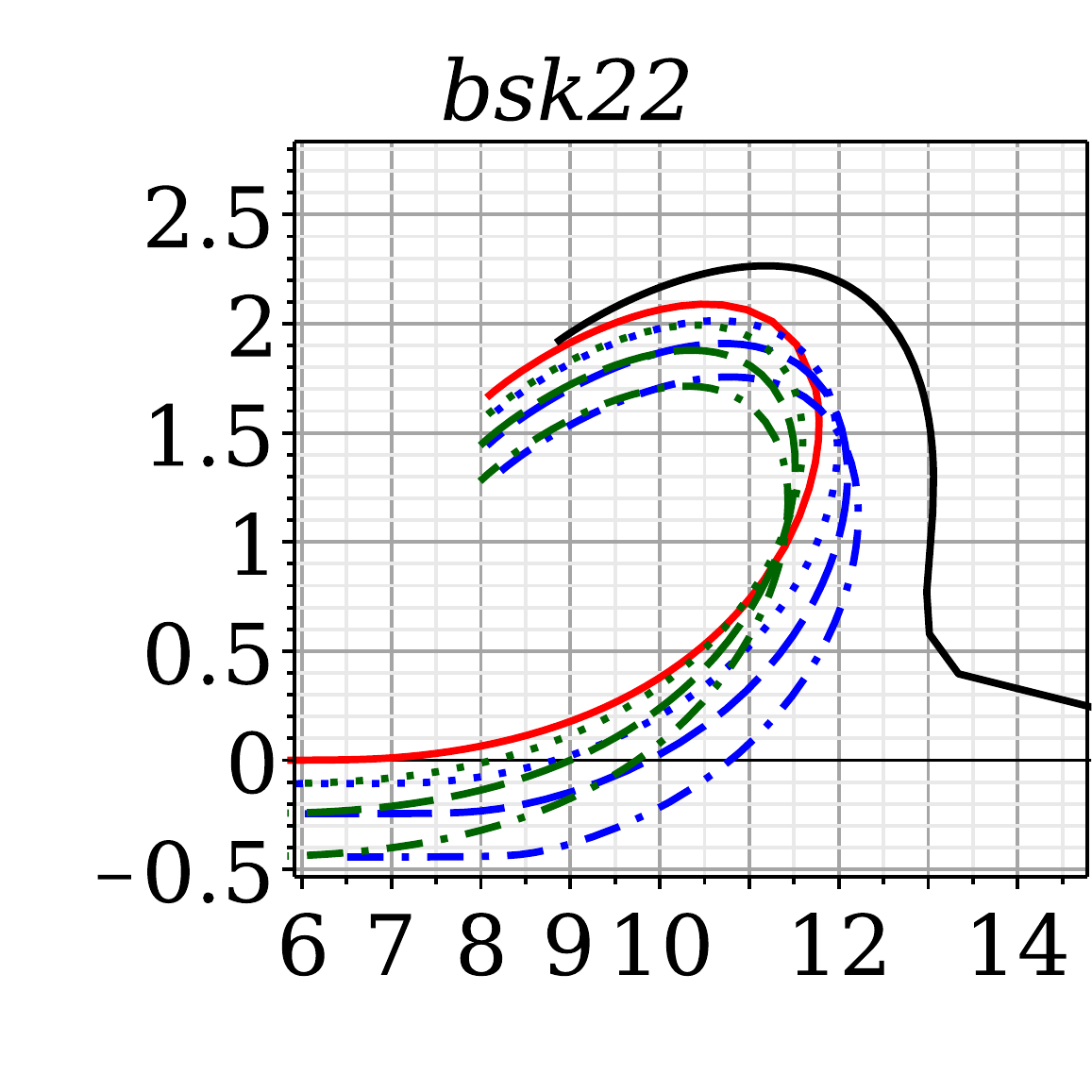} \\
		\includegraphics[scale=0.2]{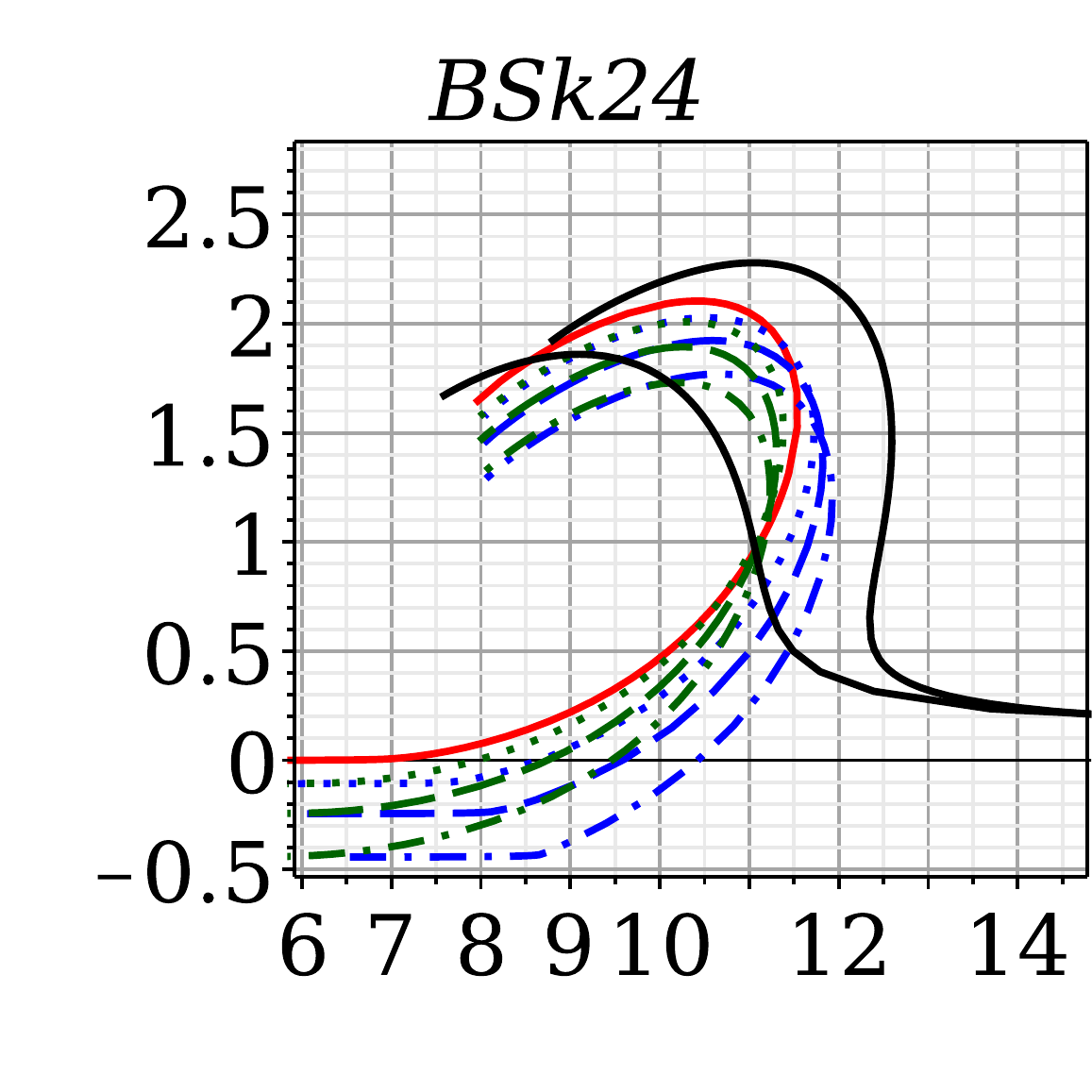} \includegraphics[scale=0.2]{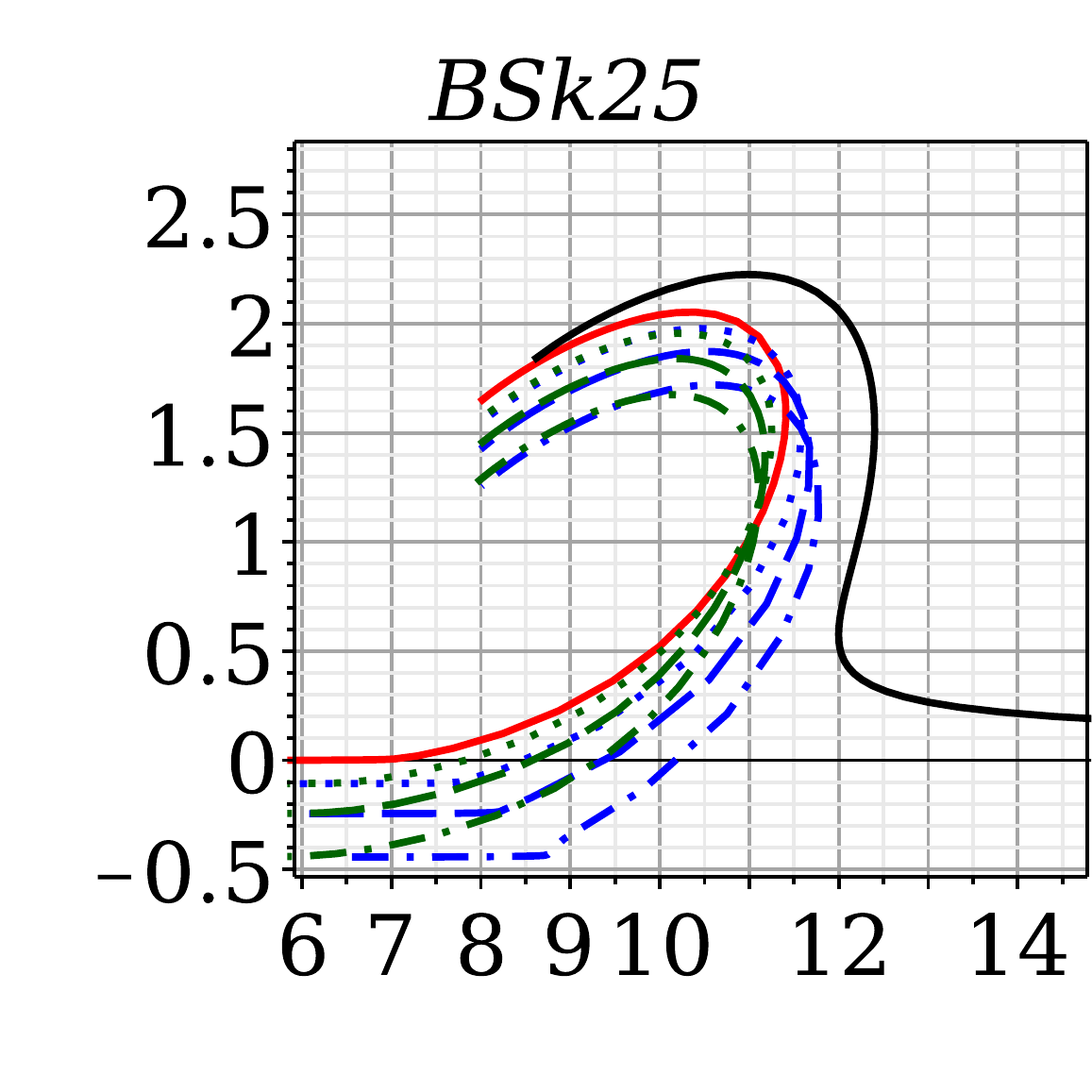} \includegraphics[scale=0.2]{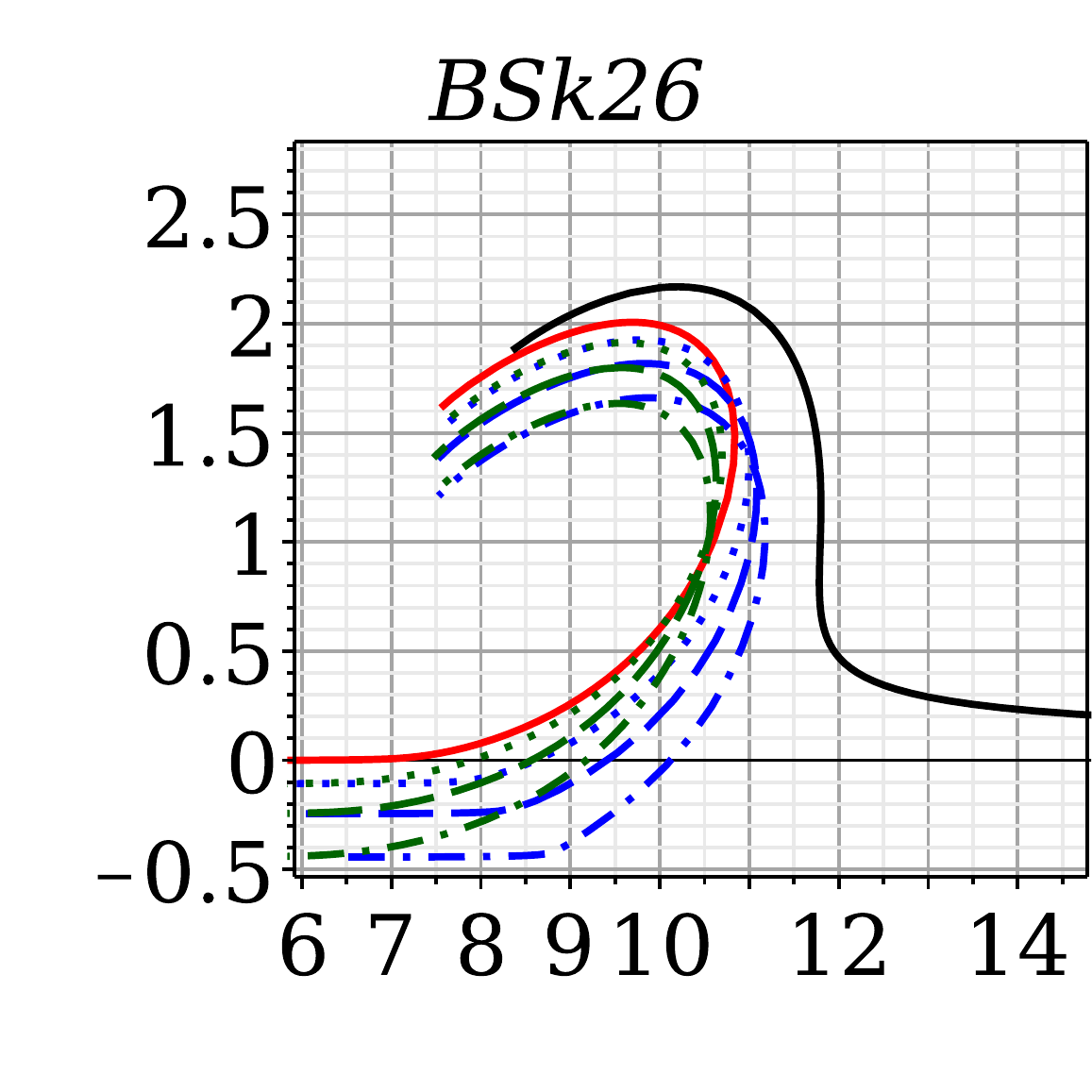} \includegraphics[scale=0.2]{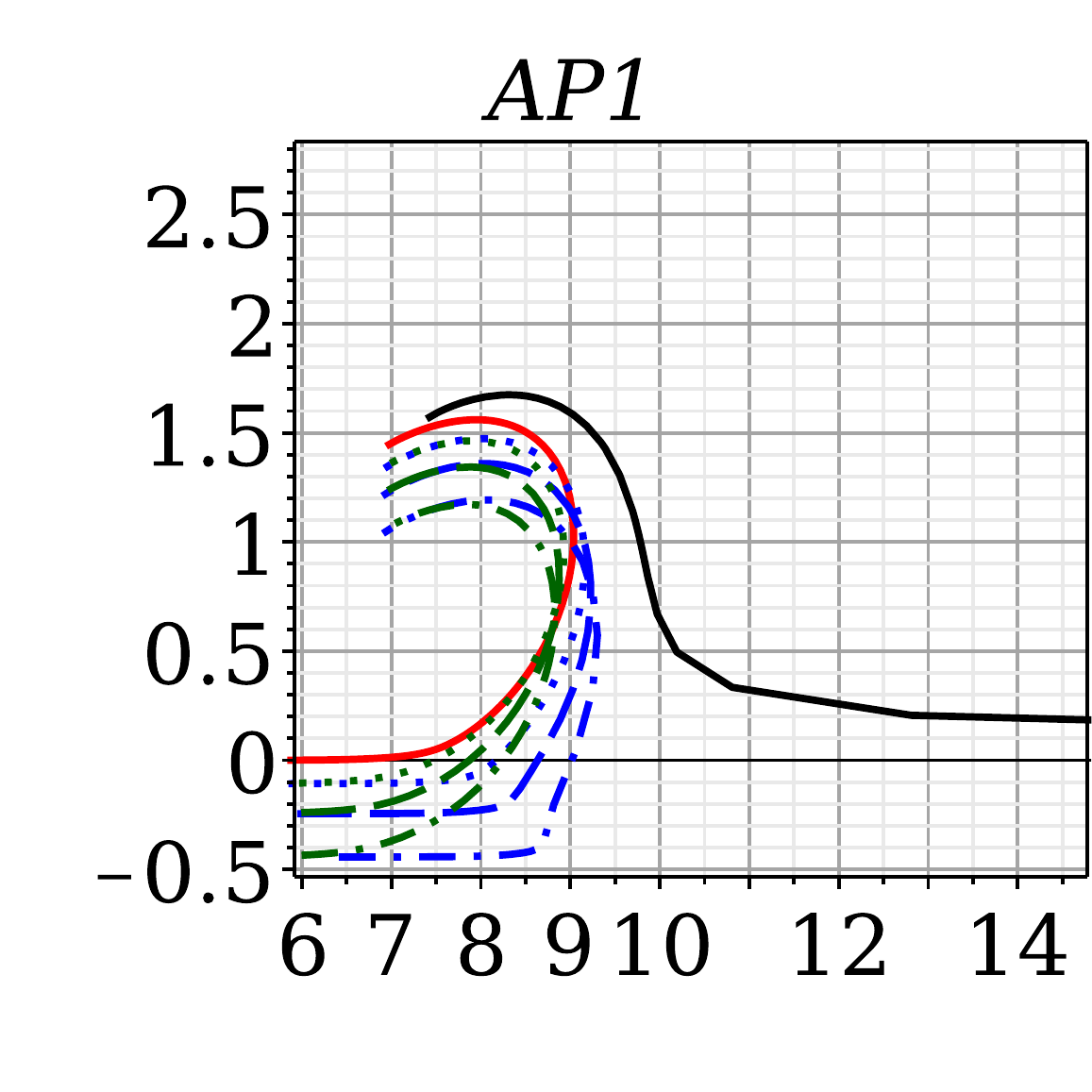}  \\
		 	\includegraphics[scale=0.2]{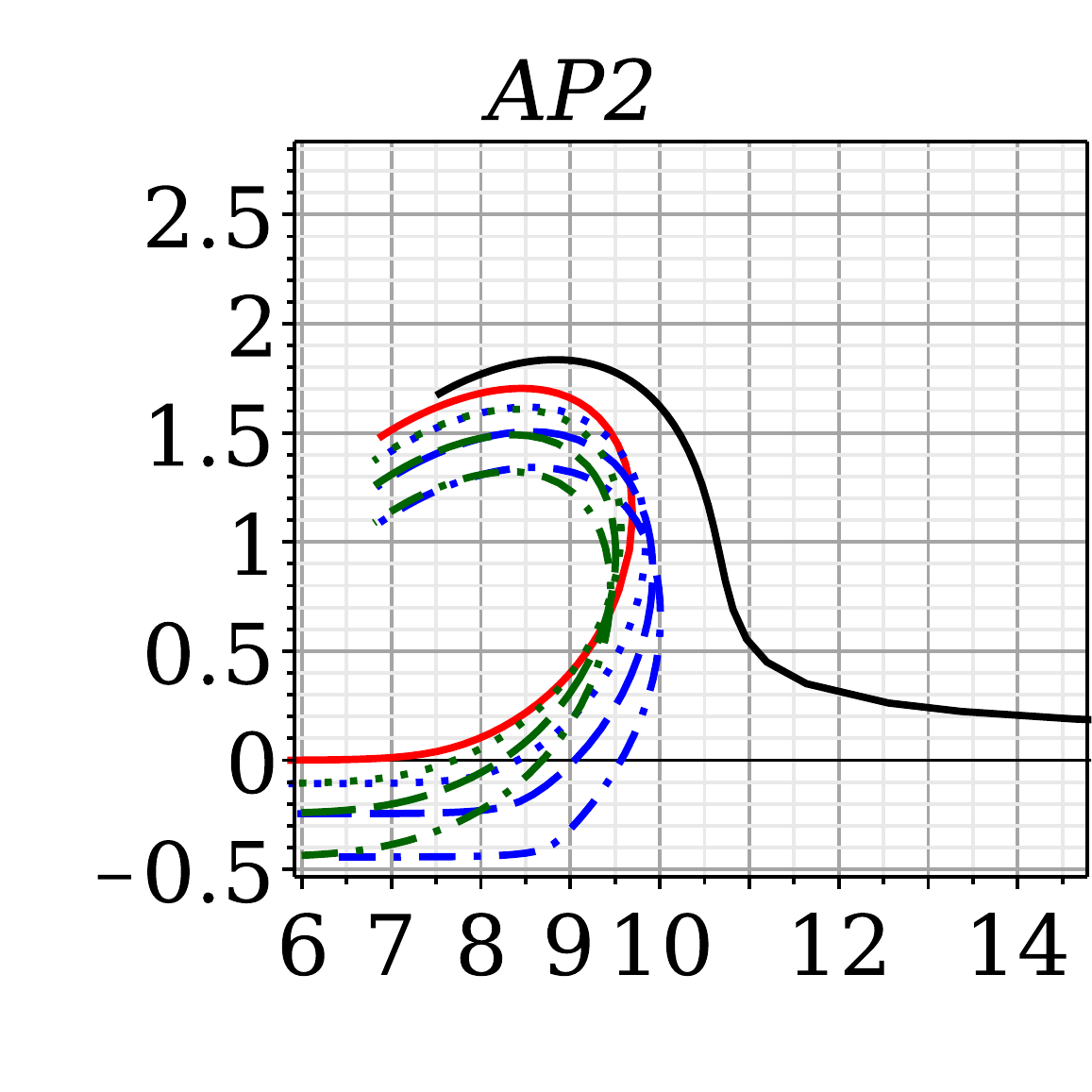} \includegraphics[scale=0.2]{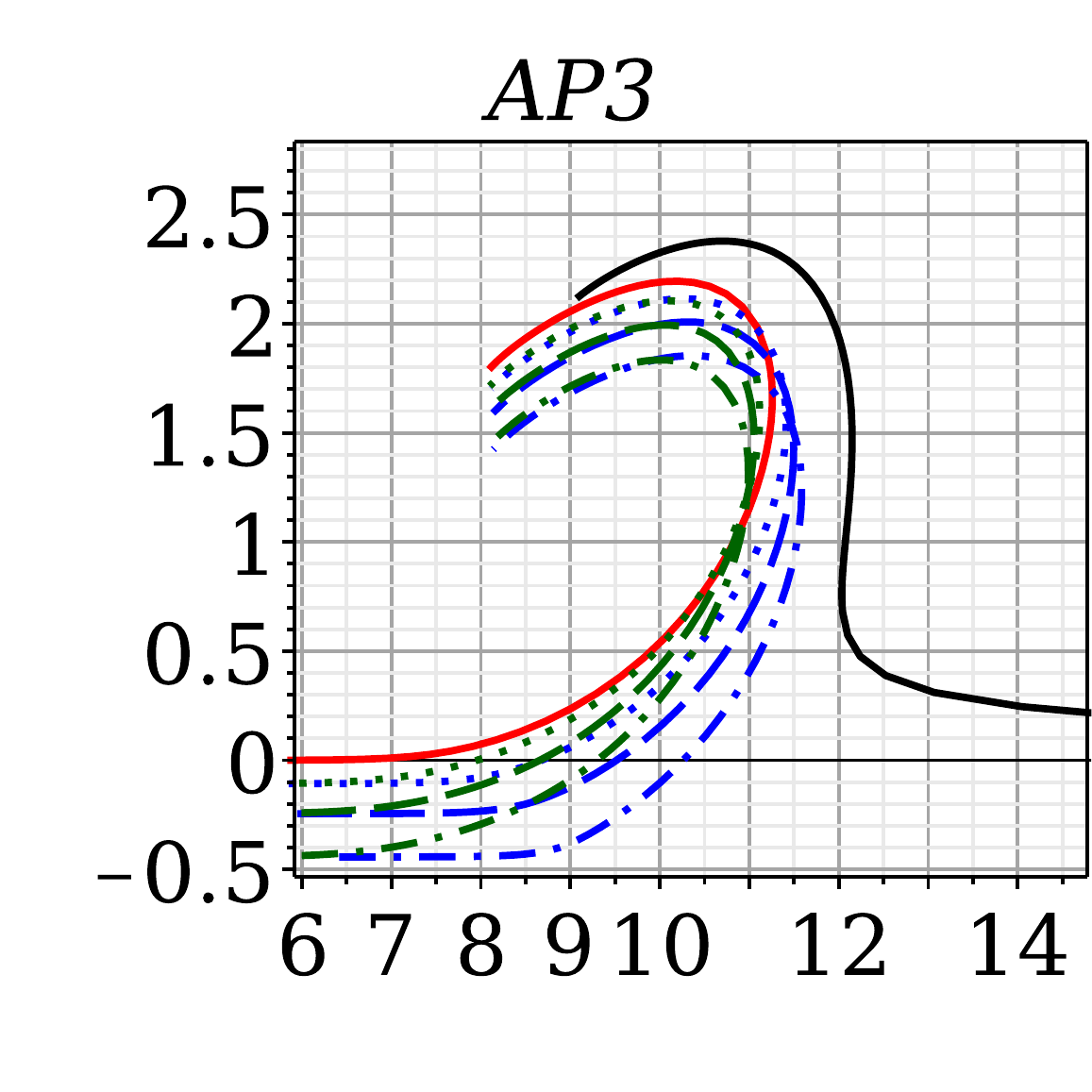} \includegraphics[scale=0.2]{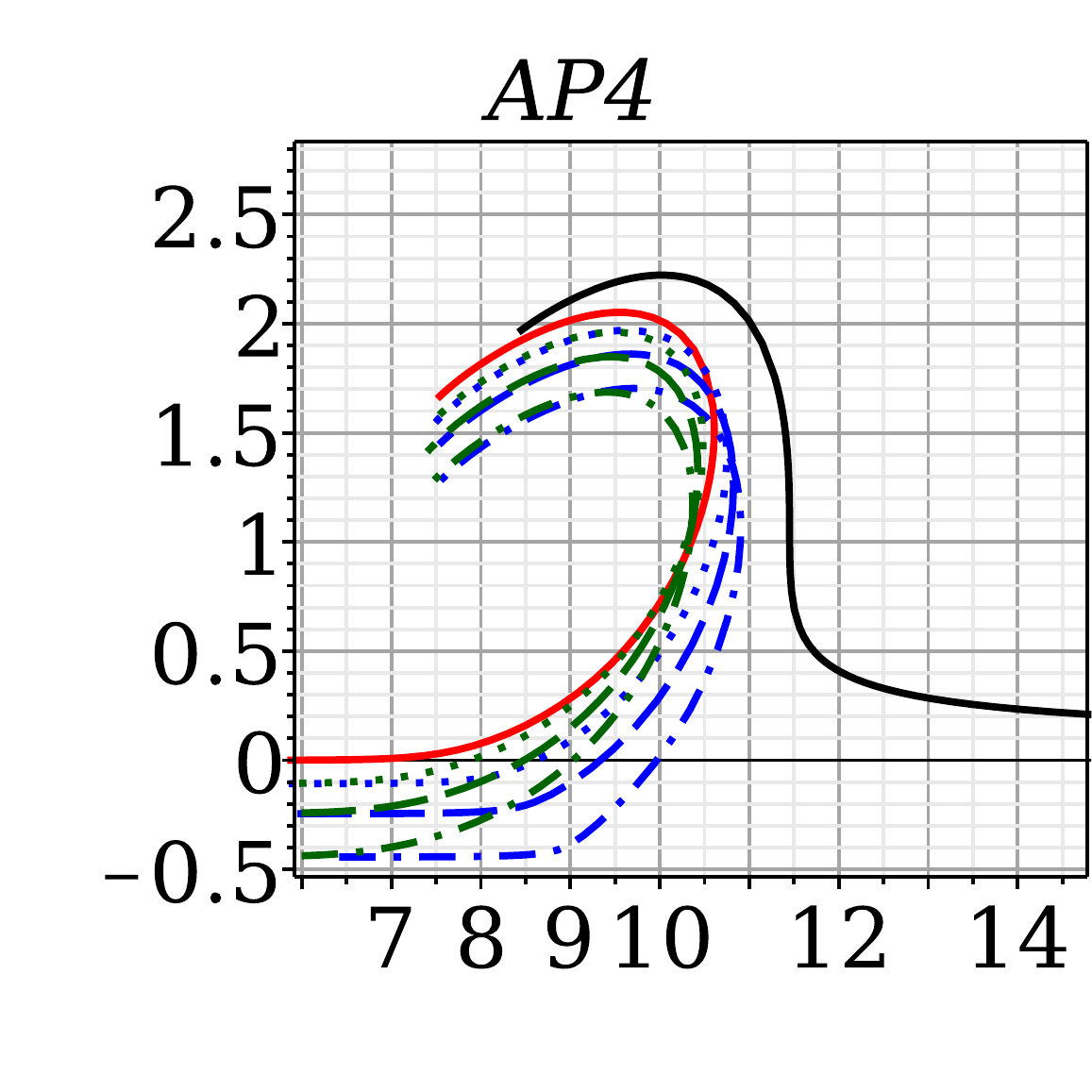} \includegraphics[scale=0.2]{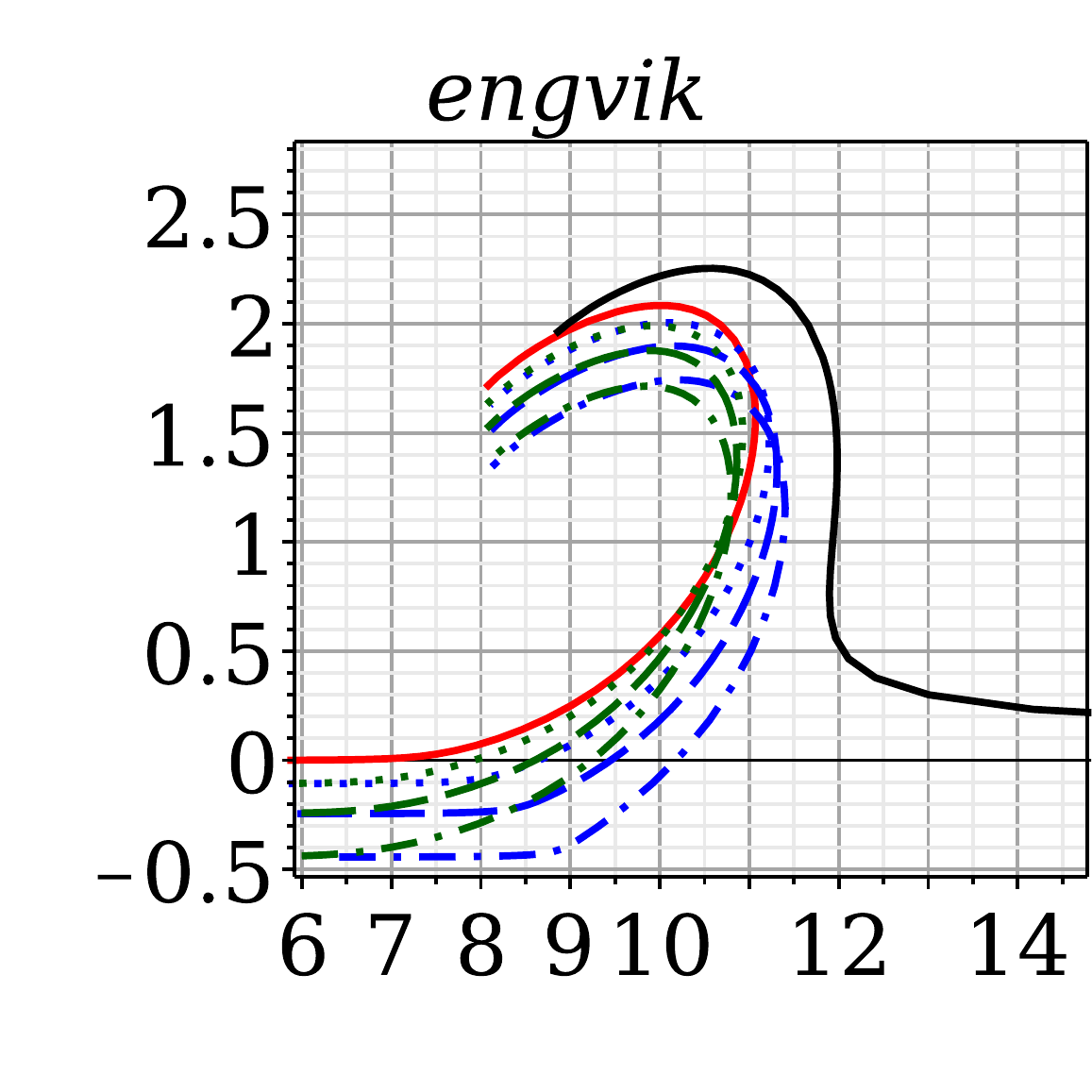} \\ 
			\includegraphics[scale=0.2]{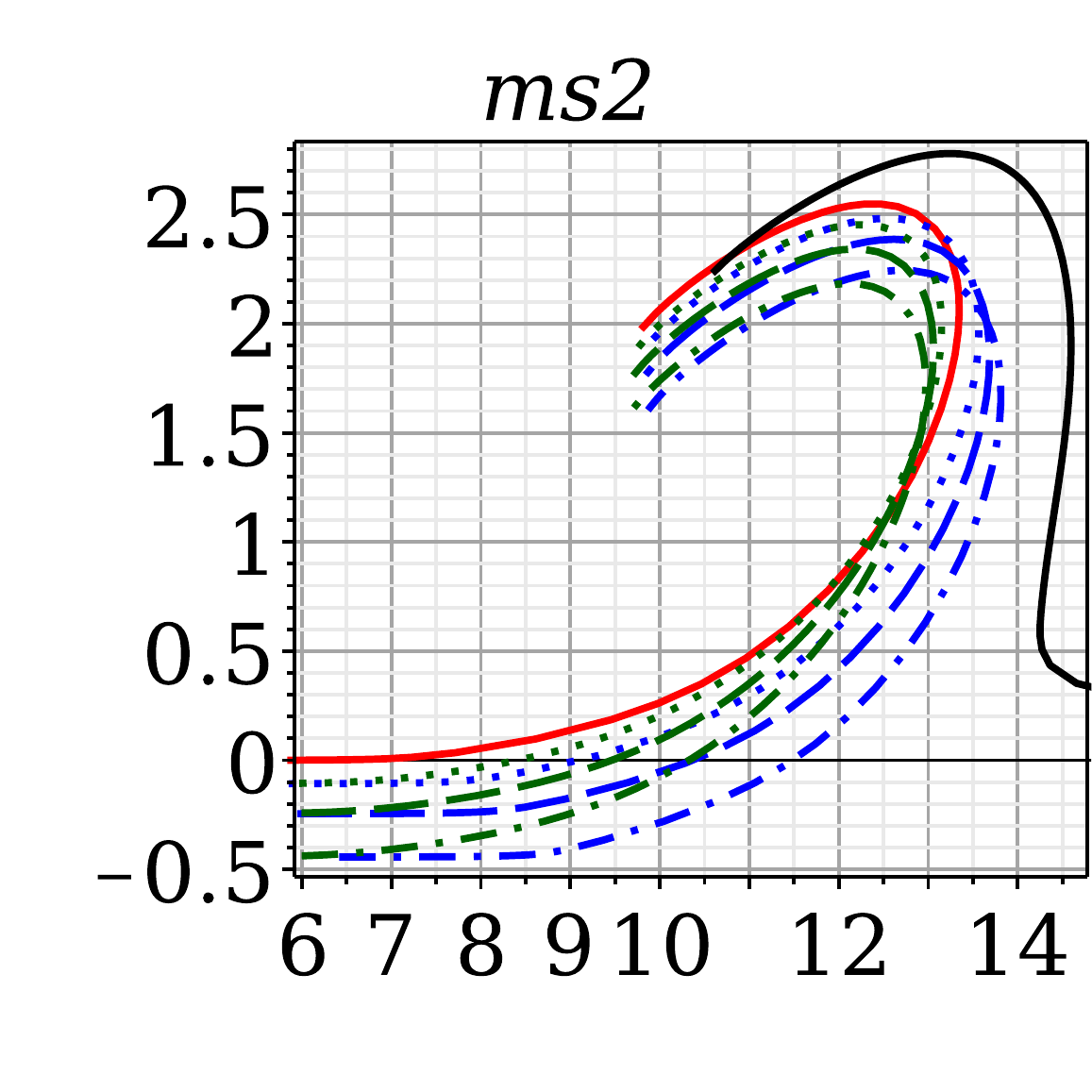}  \includegraphics[scale=0.2]{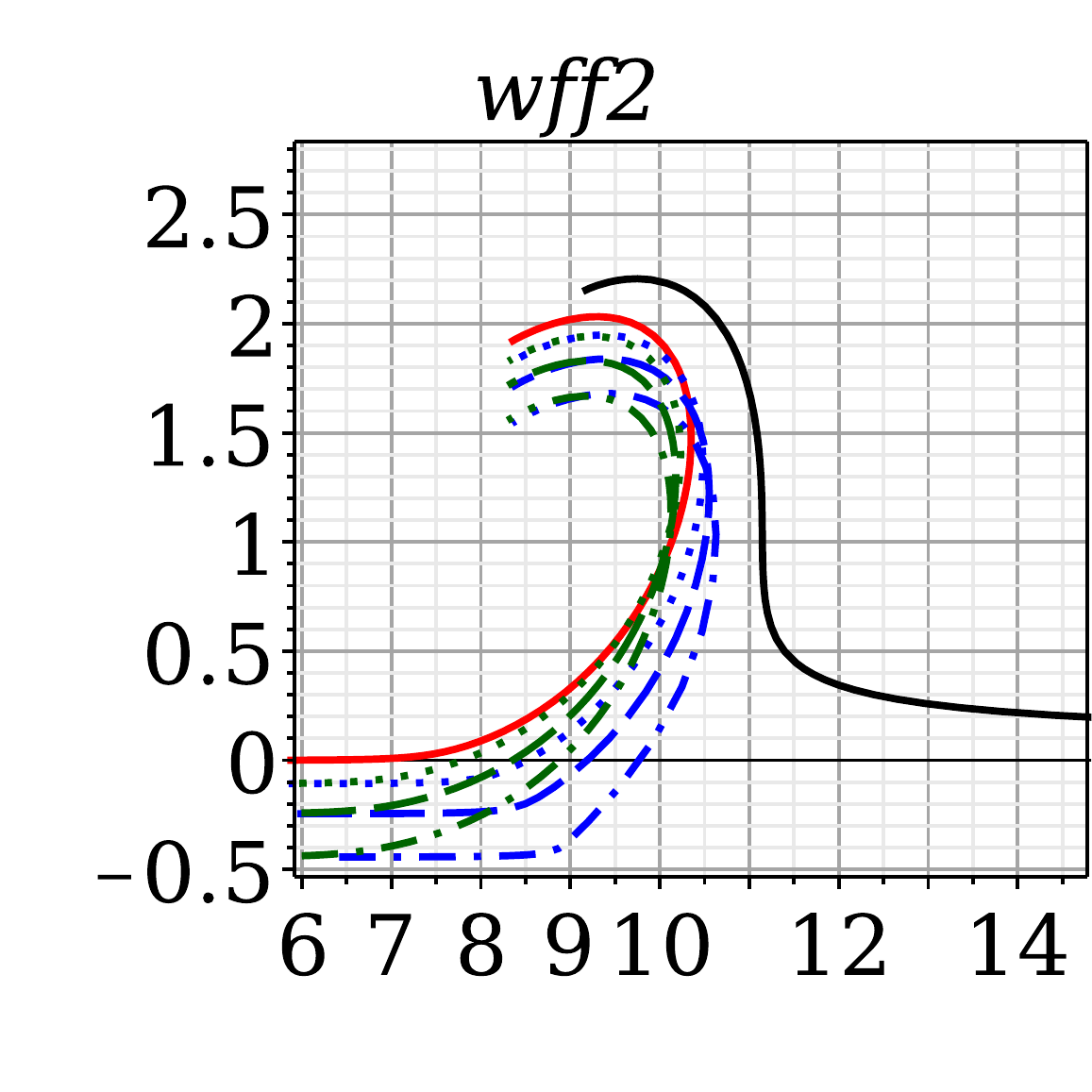}	 		\includegraphics[scale=0.2]{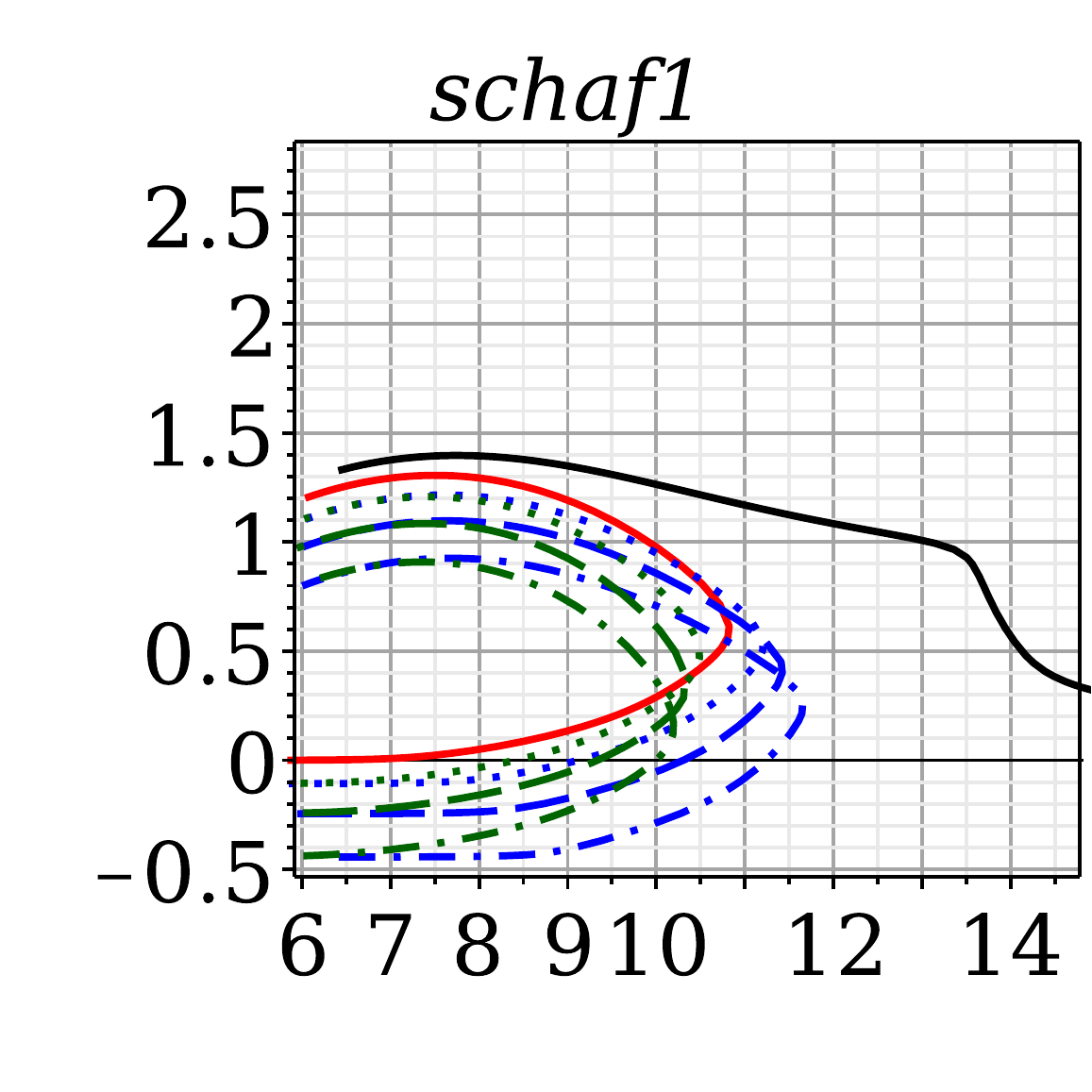}	\includegraphics[scale=0.2]{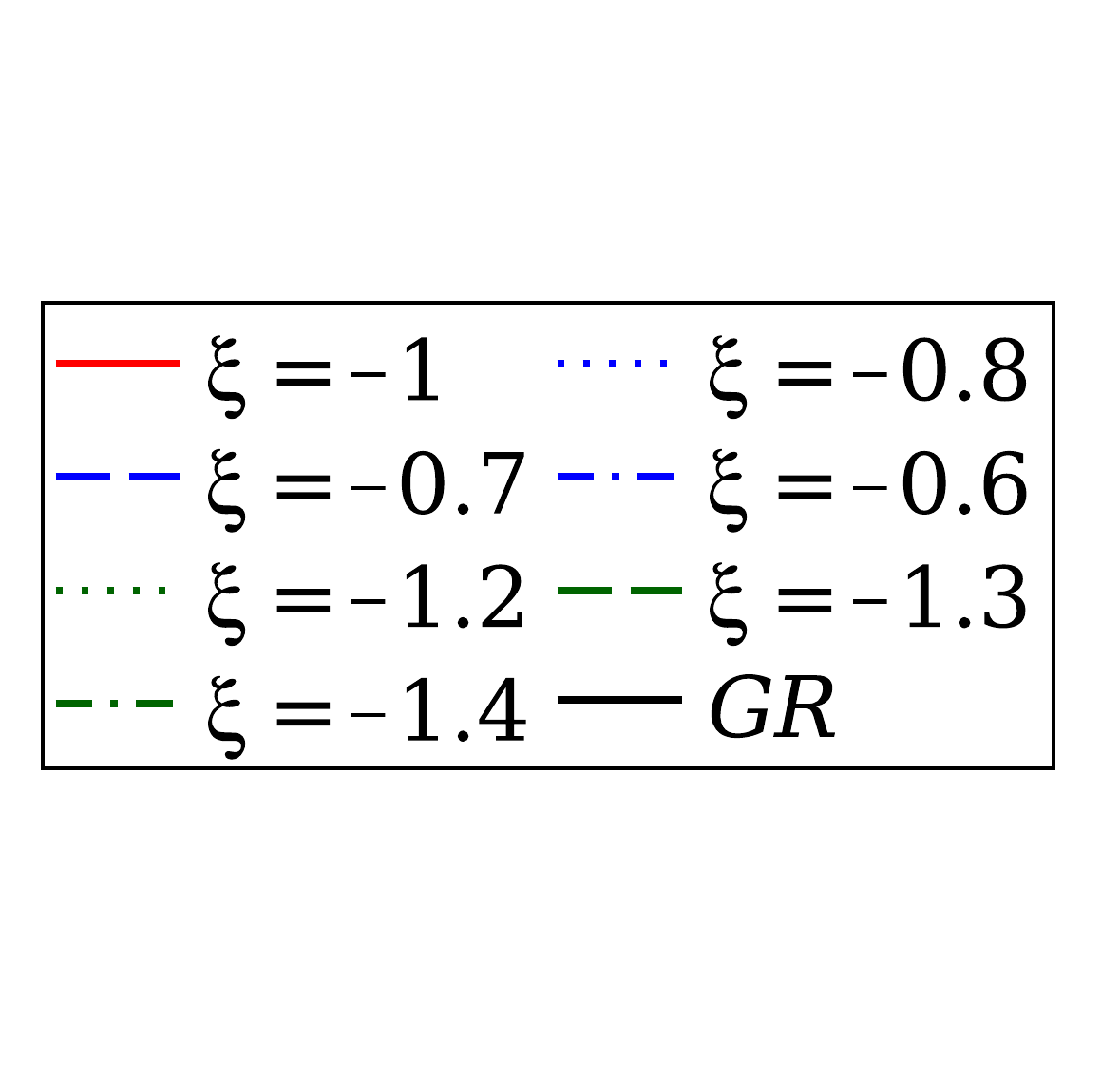} 
	\end{tabular} \end{center}	
\caption{\label{xineq1} The  mass-radius diagram in the case $\ell=20$ km and $\xi=-1.4$, $-1.3$, $-1.2$, $-1$, $-0.8$, $-0.7$, $-0.6$  for different equations of state. 
The black curves correspond to the unmodified theory of gravity (GR);  the red curves correspond to the case $\xi=-1$, the green curves correspond to the case $\xi<-1$, the blue curves -- to the case $\xi>-1$.}
\end{figure}

In this section we will consider case $\xi\not=-1$. Generally $\xi$ takes a value in the interval $-3<\xi<1$. Mass-radius diagrams for various values of $\xi\not=-1$ are shown in Fig. \ref{xineq1}. One can see that in comparison with the case $\xi=-1$  mass-radius diagrams are shifted down and left in case $\xi<-1$, and down and right in case $\xi>-1$. It is also necessary to note that mass-radius diagrams are shifted in the region of negative masses. Of course the baryonic mass (\ref{M_0}) of the star remains to be positive. 
However, it turns out that the asymptotic anti-de Sitter mass given by Eqs. (\ref{A_AdS})-(\ref{Leff}) becomes negative for some values the central baryonic mass density $\rho_{0c}$.

\section{Summary}
In this paper we considered the subclass of Horndeski gravity represented by models with a nonminimal derivative coupling 
of a scalar field with the Einstein tensor and the cosmological constant $\Lambda_0$. 
We numerically constructed neutron star configurations. 
The matter had a form of a perfect fluid and obeyed the equation of state of neutron matter. 
In the previous work \cite{Kashargin2022} the matter obeyed the polytropic equation of state with the adiabatic index $\Gamma=2$. 
The polytropic equation of state does not describe the complexity of matter in neutron stars. 
In the present work besides the polytropic equation  plenty realistic equations of state of neutron star matter were considered. 
We used the analytical representations of following unified equations of state: 
polytrope equation, FPS, SLY, BSk19-22, BSk24-26, AP1-4, engvik, gm1nph, gm2nph, gm3nph, mpa1, ms00, ms2, ms1506, pal2, wff1, wff2, wff3, wff4, schaf1, schaf2, prakdat, ps. 
The model contains three parameters: $\ell$, $\varepsilon$ and $\xi$. 
Parameter $\ell$ is a characteristic length which characterizes the nonminimal derivative coupling between the scalar field and
curvature.  
$\xi=\Lambda_0\ell^2$ is a dimensionless cosmological constant and takes the values $-3<\xi<1$, which provides the required metric signature.  
Note that $\Lambda_0$ is a `bare' (i.e. unobserved) cosmological constant, and an observed  effective negative cosmological constant $\Lambda_{AdS}$ appears as a certain combination of $\Lambda_0$ and the parameter of nonminimal derivative coupling $\ell$.  
In order to provide a regularity of solutions $\varepsilon$ must be equals $-1$ and it
means that the usual kinetic term of scalar field and the modified term with Einstein tensor enter into the Lagrangian with different signs. 
Results of numerical integration demonstrate the behavior of the metric functions, scalar field, baryonic density and the mass-radius diagram for 
different equations of state and different values of the model parameters $\ell$ and $\xi$. 

The case $\xi=-1$ was analyzed in more detail, because in this case the vacuum solution has the particularly simple form of the Schwarzschild-anti de Sitter black hole. 
In the case of a small $\ell\leqslant 20$ km the mass-radius diagram has an essential difference comparing with that in general relativity, namely, radius decreases monotonically with decreasing mass. Such the mass-radius relation corresponds to the so-called bare strange stars or quark stars in general relativity. The specific ‘strange’ relation between mass and radius in our case is forming due to the effective negative cosmological constant $\Lambda_{AdS}$. 
In the case of a large $\ell\geqslant 40$ km the form of the diagrams changes, and a vertical section appears at the diagram, where the mass of the star increases practically without increasing the radius.
As the parameter $\ell$ increases, the diagrams shift towards to larger masses and larger radii. In the general case the relation of mass and radius tends to that obtained for the case of the unmodified theory in the limit of large values of $\ell$. 

The case of a small values of the non-minimal coupling parameter $\ell<1$ km takes us away from the applicability area of the realistic equation of state or observable range of stellar masses. In this case stars with masses more then the mass of the Sun \ms  correspond to densities significantly exceeding the permissible range of central densities for selected equation of state.  Vice versa, the permissible range of central densities corresponds to masses significantly smaller than the mass of the Sun \ms. 

Applying observable restrictions for mass and radius of the neutron star, we get the restrictions on the parameter $\ell$, specifically
$10\leqslant\ell\leqslant50$ $km$. 
In particular, in the case $\ell\gg 1\,km$  the mass-radius diagram approaches the results \cite{Tooper:1965, Rinaldi:2015} without kinetic term of scalar field in the Lagrangian, 
in this case the contribution of the kinetic term much less than the one of the modified term with Einstein tensor in the Lagrangian, 
while the effective negative cosmological constant $|\Lambda_{Ads}|=\ell^{-2}$ is small. 

In the general case $\xi\neq-1$ the mass-radius diagrams are shifted down and left in case $\xi<-1$, and down and right in case $\xi>-1$. It is also necessary to note that mass-radius diagrams are shifted in the region of negative  asymptotic masses. Of course the baryonic mass of the star remains to be positive. 


\acknowledgments

The work is supported by the Russian Science Foundation grant 
No.~25-22-00163. 



\end{document}